\documentclass[a4paper,12pt]{report}
 \usepackage{inputenc}
 \usepackage{graphicx}
\usepackage[T1]{fontenc}
 \usepackage[english]{babel}
\usepackage{latexsym}

\usepackage{hyperref}
\begin{document}

\fontfamily{ptm}
 \large

\section*{Preface}

Modern  physics includes two main theories: general relativity
(the leading gravity theory) and quantum field theory
 implemented as the
Standard Model. These two are based on quite different principles
and symmetries; the Lagrangian approach is their common
denominator. Some people  try to obtain general covariance
(relativity) as emergent phenomenon. In my opinion, general
covariance is too beautiful and important to be emergent. This
book purports to explain quantum description as an emergent
phenomenology (too).

There is a unique variant of Absolute Parallelism (AP embraces
symmetries of both Special and General Relativity) which is both
beautiful (the Little Prince's Principle: true beauty should be
single) and simple (Kolmogorov theory of algorithms' complexity):
no free parameters, nothing (nor $D{=}5$;
 \href{http://arxiv.org/abs/0812.1344}{arXiv: 0812.1344})
 can be changed if to keep the theory safe from emerging
singularities of solutions.

 In linearity, there are 15
polarizations; three of them (electromagnetic-like $f$-waves)
cause linear growth of three other polarizations, so the trivial
solution is linearly unstable; but Riemannian curvature does not
grow, is stable.
 This feature gives an answer to the
 long-standing question:
``Why is there something rather than nothing?'' (They name
Leibnitz, Gr\"{u}nbaum, Parmenides.)
 Or:
``Why are we so lucky that our `solution' differs so drastically
from the trivial one, from `nothing'? All solutions are on equal
footing, are not they?" The answer is: ``Nothing is unstable".

There is also the longitudinal polarization; and
$SO_4$-symmetrical expanding solutions, delivering an example of
L-wave, can serve as an expanding cosmological background, a
`wave-guide' for other polarizations, $f$-waves first of all. If
$f$-amplitudes decrease as $1/\!\sqrt{t}$ (reddening of
thermalized ensemble), the unstable waves grow as $\sqrt{t}$; so
the nonlinear terms should be accounted for (`growing' implies an
`arrow of time').
 The expanding wave itself breaks time reversibility (a kind of
spontaneous symmetry breaking -- solutions' symmetry is always
smaller than the equations' symmetry); it gives anti-Milne model
($a\propto t,k=+1$) which seemingly describes SNe Ia and GRB data
even better than the $\Lambda$CDM-model.

Nonlinear spatially localized field configurations can carry
digital information -- topological charges and quasicharges; an
emerging phenomenology of `topological quanta' on the cosmological
expanding background ($S^3$-spherical shell, of thickness $L$,
filled with waves giving non-linear field fluctuations) is
interesting to explore. The 4-th order prolongation of the
system's symmetric part can be written as a modified gravity with
EM-like energy-momentum tensor: energy is positive and only
$f$-waves (move alone Riemannian geodesics; no gradient
invariance) contribute there. This equation also follows from the
least action principle: the Lagrangian is quadratic in symmetrical
part,  and hence is trivial. After exclusion of covariant
divergences, the main, quadratic terms look like a modified
gravity, $R_{\mu\nu}G^{\mu\nu} + f_{\mu\nu}^2$
 (Ricci and
Einstein tensors). The tangible waves move almost tangentially to
the spherical shell of expanding cosmological waveguide -- that to
be trapped inside it; this feature can potentially explain the
principle of superposition for secondary, proxy $4D$ fields aiming
to describe how topological quanta (extended along the extra
dimension; their parts move along different paths, in agreement
with quantum theory's math) scatter $f$-waves and contribute to
$f^2$-term, giving a derived $4D$ `quantum' Lagrangian.


This book is just my PhD thesis; the original (or more close to
that) Russian version is available at
\href{http://www.arXiv.org/abs/gr-qc/0412130}%
{arXiv: gr-qc/0412130}; the author's summary,
 with the date of
defence (at Tomsk university),
  names of opponents,
 and so on, is also there.

The footnotes marked with asterisk, `*', are `fresh': sometimes my
willingness to add a comment or a reference to my more recent
arXiv-preprint (the thesis was written 15 years ago! yes, it was
the last century) has become irresistible!

 And the retro-picture below shows a few my NSU classmates (and
 me); this is a weak attempt to increase a bit the number of readers!

Ivan L Zhogin, Novosibirsk

\begin{figure}[hbt]
\begin{center}   
\includegraphics*[width=135mm]{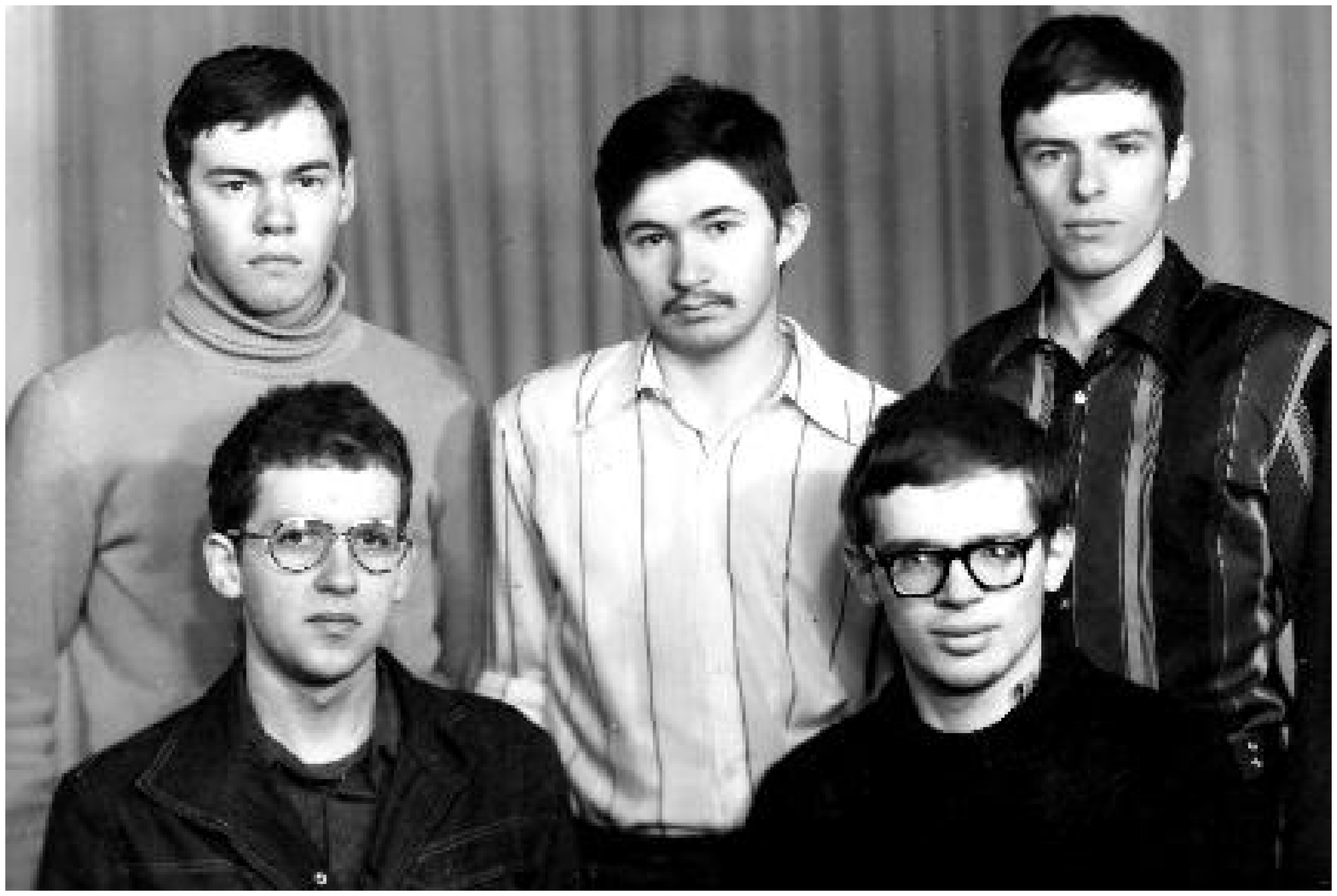} \\
{\normalsize `Life is serious, life is earnest..' Just graduated
from NSU; Novosibirsk, 1981.}
\end{center}
\end{figure}

\newpage \vspace*{2mm}

 \centerline{\bf CONTENTS}
   \vspace{3mm}
 \begin{tabbing} Chapter 0.011\=\hspace{135mm}\= \kill
{\bf
Introduction}\> \>\bf ~5 \\[4mm]
 \> \parbox[t]{125mm}{Test for singularities
\dotfill}\>~9\\
\> \parbox[t]{125mm}{Solution set; topological charges and
quasi-charges\dotfill}\>11 \\
\> \parbox[t]{125mm}{Defended statements
\dotfill}\>13\\[5mm]
{\bf Chapter 1.}
\> \parbox[t]{125mm} {\bf Field equations
}\> \bf 15\\[4mm]
 \> 1.1~\' \parbox[t]{125mm}{ Generally covariant
 frame field equations \dotfill}\>16\\
\> 1.2~\' \parbox[t]{125mm}{ Necessary information on the
compatibility theory; involutive symbols \dotfill}\>
 \raisebox{-9mm}{22}\\[2mm]
\> 1.3~\'\parbox[t]{125mm}{ Einstein--Mayer's classification of
compatible equations of Absolute Parallelism \dotfill}\>
\raisebox{-9mm}{28} \\[2mm]
\>1.4~\'\parbox[t]{125mm}{ Chances for unification of gravitation
and electromagnetism, and the {\em optimum equation} \dotfill}\>
\raisebox{-9mm}{32} \\[2mm]
\>1.5~\' \parbox[t]{125mm}{Updated (overdetermined)
equations  \dotfill}\>37\\
\>1.6~\' \parbox[t]{125mm}{Energy-momentum
tensor and conservation
 laws \dotfill}\>39\\
\>1.7~\'\parbox[t]{125mm}{Post-Newtonian effects
 \dotfill}\>43\\[5mm]
{\bf
Chapter 2.}\>\parbox[t]{125mm}
{\bf Singularities of solutions }\>\bf 47\\[4mm]
\>
2.1~\' \parbox[t]{125mm}{ Gradient catastrophe and co-singularities
\dotfill}\>48\\
\> 2.2~\' \parbox[t]{125mm}{ Field equations free from
co-singularities of solutions \dotfill}\> 51 \\
\> 2.3~\' \parbox[t]{125mm}{
Contra-singularities \dotfill}\>56\\
\> 2.4~\' \parbox[t]{125mm}{Measure of solutions with arising
contra-singularity and dimensionality $D$ \dotfill}\>
\raisebox{-9mm}{59} \\[2mm]
\> 2.5~\' \parbox[t]{125mm}{ Spherical symmetry;
examples of solutions with arising contra-singularity \dotfill}\>
\raisebox{-9mm}{63} \\[2mm]
\> 2.6~\' \parbox[t]{125mm}{ Another co-ordinate
choice; Cosmological solution \dotfill}\>67
\\[5mm] \bf Chapter 3. \>
\parbox[t]{125mm}{ \bf Topological quasi-solitons }\>\bf 70\\[4mm]
\> 3.1~\'
\parbox[t]{125mm}{ Topological charges
 of localized field configurations
\dotfill}\>72\\ \> 3.2~\' \parbox[t]{125mm}{ Absolute, relative
and $k-$ad
homotopy groups \dotfill}\>75\\  
\> 3.3~\'
\parbox[t]{125mm}{ Topological quasi-charges for symmetrical
 field configurations
\dotfill}\> \raisebox{-9mm}{78} \\[2mm]
\> 3.4~\' \parbox[t]{125mm}{ Dimension
$1+4$; SO$_{2}$-symmetrical
 quasi-charges \dotfill}\>82 \\[2mm]
 \> 3.5~\'
\parbox[t]{125mm}{ Phenomenology of quasi-solitons
 (topological quanta);
secondary (proxy) fields and 'flavors'
 \dotfill}\>\raisebox{-9mm}{87} \\[5mm]
 \bf
Conclusion \> \>\bf 95\\[5mm]
\bf Bibliography \>\>\bf 99 \end{tabbing} \newpage

\chapter*{ Introduction}

The well-known merits of the General
Relativity theory (GR) \cite{sing,ligh,lan,pau} 
-- elegance and beauty, reduction of gravitation phenomena to
geometry, the right prediction of post-Newtonian effects -- are
supplemented (or a bit spoiled) with severe demerits or drawbacks.
Firstly, the electromagnetic field was not embraced by geometry,
and hence its existence find no clear explanation \cite{pau,berg}.
In the second place, as it was demonstrated by many authors
(Hawking, Penrose, et al.), in general case, solutions of GR are
spoiled with arising singularities -- points of spacetime where
the curvature invariants are turned into infinity
\cite{lan,pen,how,ger,blh}.
 At last, the {\it metric structure\/} of GR
(i.e., the metric field) corresponds to not the most elementary
representation of the GR symmetry group, I mean coordinate
diffeomorphisms, ${\rm Diff}(D)$.

Topological methods were used (Penrose \cite{pen}, Hawking
\cite{how}, Geroch \cite{ger}) to prove inevitability of
singularity occurrence in solutions of GR (and in the
Branse--Dicke theory as well). In works of Belinsky, Lifshitz,
Khalatnikov \cite {blh, lan} the analytical character of general
GR solution in vicinity of singularity was investigated in the
synchronous coordinate system. (The justification for such a
near-singularity constructions is concerned with the generally
covariant test for singularities which will be offered and
implemented in the second chapter of this work.)

In opinion of many people, the presence of singularities in a
theory designates internal inconsistency of that theory. However,
it is believed that a quantum version of the relativity theory can
be free from singularities (superstring models, which lead in some
limit to $R^2$-gravities, are very popular). In my opinion, it is
quite unlikely, that through just a procedure of quantization one
can transform a bad thing into a good one. The assumption about a
connection between singularities and the {\it arrow of time\/} (an
attempt to benefit from infinity) \cite {pen1} looks even more
fantastic.

So, the main objective of this work is to solve the problem of
singularities (and the problem of gradient catastrophe, or
infinite gradients) of solutions -- staying at a `classical
level', within the framework of the Absolute Parallelism theory
(AP).

Hopes for elimination of singularities are connected with
 different improvements of the
gravitation theory: \\
 1) Complication of structure in theories
 with torsion \cite{obu1, obu2},
etc.; for example, the tetrad theory of gravitation by M{\o}ller
\cite{mell} (a variant of Absolute Parallelism; the frame
equations there, though Lagrangian,  are not good since irregular
(in the first jets, near the trivial
solution); \\
 2) Higher derivatives, $R^{2}$-gravity (the formal test
for the singularities, developed in this work, can be easily
applied to $R^{2}$-gravity \cite{z5}).
 However, as a rule, only the case of particular
symmetric solutions is considered, while the problem of
 singularities of a general solution
remains, in fact, unresolved.

In the works by Vargas et al.\ \cite{var1, var2}, alongside with
the metric, the frame field  is used as the second structure. The
frame structure, however, can not be {\it Lorentzian\/}:
 the global
Lorentz rotations $O(1, D-1)$ acting on a scalar index (as it is
the case in the `classical' absolute parallelism) can be
inadmissible -- it depends on the kind of `currents' in
\cite{var1}); and the second structure is not connected directly
with the Riemann structure (the metric is defined independently
from the frame field). The theme of
 singularities in \cite{var1,var2} is not mentioned at all.

The second structure was introduced as an attempt to solve the
problem of geometrization of  electromagnetic field. However, as
it will be shown in the first chapter, the Lorentzian structure
(or the frame structure) of absolute parallelism, a frame field
$h_a{}^\mu(x)$ (see also \cite{z2, z5}) is large enough for this
problem.
 And for the {\it `optimum'\/}  equations of AP, quite reasonable
 answers to the questions about energy-momentum tensor
 and about post-Newtonian effects are possible.
The presence of only one
 structure\footnote{* So, it is a single
 field theory (irreducibility).}
 is more preferable  from the
aesthetic point of view; also this simplifies very much the
analysis of singularities in general solutions of compatible AP
equations.

Absolute parallelism was formulated and investigated rather
 in detail  in a number of Einstein's works
 (some works had been accomplished in co-operation with Mayer)
 relating to the period
from 1928  to 1931  \cite {ei2, ei3, ei4}.
 That to learn about earlier mathematical works
 of a similar direction,
one can look at \cite{ship}.

As  Pais mentions \cite {pais}, some time Einstein assigned
 the big hopes for this theory, though his colleagues,
 physicists, expressed all doubts.
 Pauli asked about a tensor of  energy-momentum, and about
 the post-Newtonian effects \cite [p. 334]{pais}.

However the main difficulty, as it is possible to assume,
 was the problem of 'search' (or explanation)
 for the electromagnetic field.
The theory contains a natural candidate for this role,
 the antisymmetric tensor of the second rank composed
  from the second derivatives of the frame field,
  $h ^ {a} {} _ \mu $. However, all equations
   of absolute parallelism
(there are plenty compatible field equations in AP),
 with the unique exception (the already mentioned {\it optimum\/}
 system of equations; it doesn't admit
  the usual dimension, $D{=}4$),
do not allow the Coulomb-like asymptotic behavior. (As there is an
Einstein and Mayer's work, \cite {ei5}, devoted to
spherically-symmetric solutions of AP equations, it is possible to
assume, that they have noticed the problem
 of Coulomb asymptotics. Some  noncovariant candidate for the role
 of electromagnetic field was considered as well; that, certainly,
 could not have any success.)
These exclusive equations will be the basic object of our attention.
 They are unique  also with the other remarkable property:
 their solutions of general position are free from co-singularities
(connected with the gradient catastrophe).

Perhaps, due to this  reason (the problem with
 `Coulomb-law'\footnote{*
It seems the Pauli's questions were also the case against.})
 Einstein has left the  absolute parallelism theory.

 For  greater completeness of this historical digression,
 it is necessary to mention  two popular articles written
 by M.P. Bronshtein  and devoted to AP; recently both articles
 have been reprinted in \cite{gore1}.
It is interesting,  in one of these articles,
AP is brought into compare
  with attempts to describe the aether as an elastic environment,
  incompressible or, on the contrary, easy (or super-)
  compressible (for exception of longitudinal
  waves).\footnote{ At the end of the second chapter,
  the spherically-symmetric problem is considered.
  At some choice of coordinates, after integration
  of a part of the equations, one can arrive to a system
  of  equations looking like gas dynamics equations
  for Tchaplygin's gas, where the gradient catastrophe
   is absent.}
 As a matter of fact, in principle,
 one can use this short term 'aether'
  as a brief designation of that `spacetime' manifold inhabited
  with all necessary characteristics, fields.

So, the first question arising at construction of a field theory,
 is how to choose a suitable mathematical or geometrical
 structure of spatial manifold. It is meant, that there
 is a differentiable structure; that is, all points of space
 $M $ can be regularly marked by means of coordinates
 $x ^ {\mu} $. Any regular (nonsingular),
 differentiable replacements of coordinates  are allowed;
 they make up the  group  ${\rm Diff} (M ^ {D}) \
 (D $ -- dimension of space),
 the group of coordinate diffeomorphisms.

The frame field $h^a {} _ \mu (x ^\nu) $ of absolute parallelism
 (a set of $n $ vector fields) is supplied with a symmetric
  constant matrix $ \eta _ {ab} $
  which  specify the signature of space;
  redefining both the frame field and the matrix $ \eta $
  through multiplication on a non-degenerate matrix,
  it it is possible to bring $ \eta _ {ab} $ to a diagonal form,
 with diagonal elements $ \lambda _ {i} = \pm1 $.
 That is, the {\it geometrical structure\/} of AP contains
 the {\it  frame field with the signature;}
 the field $h_a {} ^ \mu {} $ is the elementary object
 realizing representation of the following
  {\it left--right group\/}
 of transformations: \\
1) `Right' coordinate diffeomorphisms (act on the Greek index); \\
2) 'Left' global Lorentz rotations $ O(1, D-1) $ acting on
 the Latin index, and global scale transformations; \\
 AP equations are invariant  with respect to this large group.
In comparison with  Riemann (or metric)
 structure of General Relativity,
 the group of symmetry becomes larger,
 while the representation becomes simpler
 (the vector representation).

\section*{Test for singularities}

The modern theory of compatibility or formal integrability
 \cite{pomm, vino} allows easily to mark out  compatible equations
  of AP; for such equations,
  formal solutions in a form of Taylor's series do exist.
  In the first chapter, as against the second one,
  only non-singular jets of solutions are dealt with.
  That is,  the initial term of series is a non-degenerate matrix,
  and one can transform, using a left-right (or only right)
  transformation, this term into the unit matrix.

Extending the test of compatibility for the cases of degeneration of
either the co-frame matrix $h^a{}_\mu$ (co-singularities)
or the (contra-)frame $h_a{}^{\mu}$
 (more exactly, the frame density of some weight),
 one can obtain a generally-covariant formal test for existence
  of singularities in solutions: the necessary condition
  of existence of formal solutions,  solution jets growing from
  (or arbitrarily close to) singularity, is
retention of regularity and compatibility of the senior terms of
 a system (of  equations). This is the case if the
symbol of a system remains involutive.
The symbol can be related to that degenerated matrix itself
 (contra-singularities), or to its minors (co-singularities)
 of as large co-rank as possible. A detailed analysis and
 application of this formal test for singularities
 is carried out in the second chapter.

The possible approach that to judge on sufficiency of this test is
to study how the left-right symmetry group of  AP equations acts
near singularities (acts on  jets of singular solutions), and how
to choose orbits of `physical' formal solutions (which would have
space-like Cauchy surface `approaching' the surface of
singularities) and make estimation of their prevalence, or
measure.\footnote { The certain analogies and parallels to
questions of the book \cite{arno} are possible.}

So, the overall objective of the second chapter  is  to demonstrate
 that there is the unique frame field equation (system of equations;
 with unique choice of space dimension, $D=5 $), which solutions
 (of general position) are free from singularities.
 It turns out that the singularity-free equations of
 the second chapter coincide with the `optimum' system of
 the first chapter.

It is necessary to note, that  the least action principle and
Lagrangian equations are usually considered as the only possible
way (of doing physics).  In actual fact, the Lagrangian approach
yet does not guarantee compatibility of equations (it  provides
some identity, but sometimes this is not sufficient). From the
`Lagrangian point of view',  perspectives to unite both
$g_{\mu\nu}$ and $f_{\mu\nu} $ within a framework of a single
structure seems rather improbable, because gravitation and
electromagnetism should give  similar (of the same order)
contributions to the Noether $D$-momentum.

 The requirement of
absence of singularities, however, seems to be more important and
fundamental than the least action principle.
 The singularity-free  equations (i.e., the optimum equations
 mentioned earlier) are not of Lagrangian kind.
There exist only an approximate Lagrangian (for the linearized
equations)
 and an approximate `weak' Lagrangian
of  electromagnetic kind (`weak Lagrangian' is the term coined
 by Ibragimov \cite{ibra} for the case
when prolonged equations are also required for minimization
 (of the `weak action');
either (of two)  includes only a part of all covariants
 (type $h' $ or $h'' $), therefore even quantization of linearized
 equations seems problematic. The second prolongation of
 the equations leads to an energy-momentum tensor of electromagnetic
 form \cite{z5}.\footnote{* In  fact, some combination of
  equations (symmetrical part)  squared can serve as an exact
  (although trivial) `weak' Lagrangian, and the starting point
  for a Lagrangian phenomenology of topological quanta.}

Furthermore, there is a close relation between
 the existence of minimizing
 functional and  Lyapunov stability.
 That this relation was working, it is desirable to assume
 the following scenario. The exact solution of non-Lagrangian
 equations has extremely complex, stochastic character
 (practically non-repeatable).
 Trying to keep an eye on a behavior (noise-averaged, on some scale)
 of stable (hence, presumably symmetric) nonlinear
 field configurations with topological (quasi-)charges,
 one may be forced to introduce secondary phenomenological fields,
 which evolution already conforms to some Lagrangian rules.
 The exact details of this scenario for 5-dimensional
 absolute parallelism depends also on the mechanism of reduction
 of the extra  dimension (with possible `relativistic character'
 of this extra dimension, as in the case of spherically-symmetric
 expanding cosmological solution).

 \section*{Set of solutions; topological charges and quasi-charges}

The main issue of the third chapter is to  analyze qualitatively
 the set of singularity-free solutions of AP equations
 ($D=5$, the topology of space is trivial), as well as the subsets
 of symmetric (and, probably, stable in some sense) solutions.
 These problems lead us to the means and methods
 of homotopy topology.

Deforming a  frame field $h^{a}{}_{\mu} (x)$ on a Cauchy surface
 $R^m ~ (m=D-1) $  in such a way that the metric becomes trivial,
 then making  boosts vanishing, we come to a field
 of rotation matrices,
  $s (x)\in SO_ {m} $ (`chiral' field).
The set of the localized maps
 $s(x): ~ R ^ {m} \to { SO}_ {m} $
 is, generally, disconnected, that is,
consists of a number of connected subsets, or components  --
 homotopy classes; these classes form a homotopy group,
 denoted as $ \pi _ {m} (SO _ {m}) $.
 For example, $ \pi _ {4} (SO _ {4}) =Z_2+Z_2$.
Affiliation with different homotopy classes corresponds
 to different values of  topological charge.

Large {\sl symmetry of AP equations} (left-right covariance)
 gives the possibility of {\sl symmetric solutions}:
 in this case, some left-right transformations do not change
 such a solution at all (a kind of stationary `point').
Classification of symmetric configurations of
 $h $-field can be similarly reduced  to classification
 of symmetric configurations of SO-chiral field,
 and to calculation of {\sl relative\/} homotopy groups
 (or dyad groups;
 for the definition see \cite{dubr, fuks})
 in the case of simple symmetries, and to calculation
 of $k$-ad homotopy groups
 $$ \pi_{l} (A_0; A_1, \ldots, A_{k-1}) ~
 (A_i\subset A_0\subset SO_m) $$
 in general case; and this time we have topological quasi-charges.
 In the case $m=4$,
all quasi-charge groups can be found using
the exact $k$-ad homotopy sequence.

Homotopy groups and topological charges are widely used
 in theoretical works (see for example \cite {ryba, olsh}),
 but the role of relative and $k$-ad homotopy groups
 in classification of symmetric configurations
of `topological' (or chiral) fields, apparently, was not noted yet.
 Triad homotopy groups are introduced in the
  series of papers of Blakers  and Massey \cite{blak},
  and remarkably exposed by  M.M.  Postnikov in his lectures
  on algebraic topology  \cite {post}.
  In the same lectures, there  was an unsuccessful attempt
  to define tetrad homotopy groups; this can be a little
  corrected and generalized  that to define by induction
  the $k $-ad homotopy groups.

The problem how  to describe the morphism of a set
of homotopy classes
 of solutions of higher symmetry into the set of
 homotopy classes of configurations of smaller symmetry
  requires a detailed description of representatives of
  the first nonzero  class  of both homotopy sets (groups);
  for this purpose one can use the means of symmetric framed
  (sub)manifolds (sub-manifolds equipped  with vectors and
  having some symmetry).
  The connection  between differentiable mappings and
  framed manifolds is expounded in \cite{dubr, pon};
  \cite{dubr}  contains other useful information like
 quaternion representation of $SO_ {4} $ group:
\vspace {-2mm} \[ { SO}_4=S ^ {3} _ {(l)} \times S ^ {3} _
{(r)}/\pm ~; \] at the reversal of one co-ordinate (at  the
orientation change) the left and right spheres of unit quaternions
 ($S ^ {3} _ {(l)} $ and $S ^ {3} _ {(r)} $) switch places;
 it is rather important for perspectives to introduce
 `phenomenological' chiral (spinor) fields as a means
 to describe topological charges, elements of
 the topological charge group
 $\pi_{4} ({ SO}_4) =Z_2 +Z_2$.
The very possibility that, in a classical field theory,
 there appear some features  of a quantum field theory,
 at some  cosmological environment), looks
 very interesting and noteworthy.

The hypothesis about  topological nature of elementary charges
 was put forward by Wheeler \cite{whee} in the context
  of Riemannian geometry of GR. A more refined attempt of
  geometrization (or topologization) of elementary charges, and also
 `geometrization' of spinors, was undertaken
 by Sakharov \cite{sakh}. This approach is  based
 on a complication of topology of Wheeler's `wormholes'
 (or Weyl's 'handles'), with usage  of topological
 parameters of three-dimensional knots.
 However, a changing of space topology
 (creation of a 'handle') should be accompanied by singularities,
 and that is unattractive; it is not clear how to  describe
 such a process in a  'regular', consistent manner.

In the {\bf Conclusion}, some interesting problems suitable
 for future research, and possible approaches to  solving
 these problems, are briefly indicated.

The author is grateful to V.A. Aleksandrov and V.M. Khatsimovski
 for useful discussions of this work.

\section*{Defended statements}

1. Within the framework of the theory of absolute parallelism (AP),
 with the single structure
(frame field  and the Minkowski signature), the unification
of gravitational and electromagnetic fields
looks possible (at increase of space dimension): \\
- \ the energy-momentum tensor exists and has
 the electromagnetic form; \\
- \ classical post-Newtonian effects for the `optimum' equation
 are the same as in the GR theory, while spin-dependent effects
 can differ from predictions of General Relativity.

2. Generalization of the compatibility test  on the cases
 of degenerate co-frame or contra-frame  matrix
gives the generally-covariant formal test (existence condition)
for singularities of solutions. The requirement of the
absence of
singularities
 let one to unambiguously choose the compatible system
 of field equations (left-right covariant, second order,
 with well-posed Cauchy problem);  space-time
 dimension is fixed simultaneously
 ($D=5 $;  the `singularity-free', or `optimum' equation of AP).

3. The spherically-symmetrical problem for this equation is
reduced
 to one second order equation
 (at the certain choice of coordinates),
 or (in other coordinates) to a system of two first order equations,
 which is  similar to the  gas dynamics equations for Chaplygin gas.
 There are no stationary spherically-symmetric solutions
 (except for trivial one), but non-stationary solutions
 are possible, such as a  single wave running along the radius;
 this fact is very interesting for possible
 cosmological applications.

4. The problem of homotopy classification  of (localized)
solutions
 and symmetric solutions  is reduced to calculation of absolute
 and relative (if symmetry is simple; generally -- $k$-ad)
 homotopy groups of
 rotation groups;\footnote{* And their quotients.}
  these correspond respectively to topological charges
  and quasi-charges.

5. For $D=5 $, the quasi-charge groups are calculated  for symmetry
 groups containing a continuous subgroup (special attention
 are payed to symmetries which are a part of the symmetries
 of the cosmological solution).
 The morphisms of quasi-charge groups,
 induced by embedding of symmetries, are received through
 the analysis of corresponding symmetric framed (sub)manifolds.

6. At certain `cosmological conditions' (nonlinearity of noise,
 energy stability of [soliton-less] background),
secondary (auxiliary)  phenomenological (3+1)-dimensional fields
 can be introduced to describe
 stable, averaged along the additional dimension, reproducible
  features and parameters of evolution of topological
   quasi-solitons (* or topological quanta).

\chapter*{Chapter 1. \ Field Equations}

 \addtocounter {chapter} {1}

At first,  a set of notations and definitions of covariant objects,
 tensors, which are necessary to compose
covariant equations of Absolute Parallelism (AP), is entering.
 The general form of AP equations, generally-covariant
 (more precisely, left-right covariant; and not underdetermined)
 systems of equations of the second order, is discussed.
 Then, the necessary information (definitions and theorems)
 from the formal theory of
compatibility of partial differential equations \cite{pomm}
 is to be adduced.
The main  stages in compatibility analysis of some
system of equations are:  first, to check whether the symbol
 of a system (of  differential equations) is involutive or not;
 and second, to check whether the all necessary identities
 exist or not. Both stages can be implemented in a finite
  number of steps.
  The symmetry  of  AP equations (including general covariance)
  simplifies  very much these actions.

Then, the results of  Einstein and Mayer's work \cite{ei4} with
classification of compatible AP equations
 are discussed. The number of compatible equations is quite great
 (as against the general theory of relativity
 where the requirements
 of compatibility and general covariance unequivocally determine
 the form of equations), therefore some additional reasons or
 restrictions are necessary to choose the most attractive and
 appropriate system of equations. In the second chapter the
 aprioristic approach is used: the requirement of absence
 or rarity of solutions with a birth of singularity is put forward;
 whereas in this chapter the reasons connected to experiment
 are actively involved: the opportunity of Coulomb
 (or Yukawa) asymptotics
 (that is, an opportunity to compare the antisymmetric tensor
 composed of the second derivatives of the frame field  with
 the tensor of electromagnetic field); correct results for
 the `classical' post-Newtonian effects. Both these requirements
 are satisfied for the unique variant of AP
 (the `optimum' equations), for which the extra space
 dimension(s) should be added
 (in case $D=1+3 $ the equations  lose their regularity).

\section*{1.1 \ Generally-covariant frame field equations}
In the theory of absolute parallelism, the properties of space,
its geometry,  is defined by a $D$-frame field (a set of vector
fields), or co-frame matrix (co-vector fields):
\[h_a {} ^ \mu (x ^ {\nu}), \ \ \ h^a {} _ \mu (x) \, ; \  \
\ h_a {} ^ \mu h^b {}_\mu = \delta ^ {b} _a \,. \]
 The metric is
not independent, but can be expressed through $h$-field and the
matrix of Lorentz (or Minkowski) signature $ \eta _ {ab} $:
 \begin{equation} \label {metr}
 g _ {\mu \nu} = \eta _ {ab} h^a {} _ \mu h^b {} _ \nu \,,
 \ \mbox {where \ \ }
\eta _ {ab} = \eta ^ {ab} = {\rm diag\,} (-1,1, \ldots, 1) \ .
\end{equation}
Concerning coordinate transformations, Latin indexes have scalar
 character, and Greek ones -- vector character.

One can perceive the frame field $h (x) $ as a
 {\it Lorentzian  structure}, because it admits,
 together with coordinate diffeomorphisms
 (acting on the Greek index),
  the global transformations of `expanded' Lorentz group,
  $ O ^ {*} (1, D-1) $,
(the scale transformation is added  to the Lorentz
 group\footnote{This is just the  group of transformations
 of inertial coordinates.})
  acting on the Latin index:
\begin{equation} \label{lrsy} h ^ {*a} {}_\mu (y) =
\kappa s^a {} _b h^b {} _ \nu (x)
\partial x ^\nu\!/\partial y ^\mu;
  \  \kappa> 0,   \ s^a {} _b\in {O} (1, D-1);
  \ \kappa, s ={\rm const}.
\end{equation}
The  AP equations of frame field should be invariant (or
covariant) with respect to the left-right transformations (\ref
{lrsy}); therefore, they should be composed through covariant
entities, i.e. tensors with indexes of both sorts, whose
transformations are similar  to (\ref {lrsy}). Tensors with Latin
indexes will be named $L$-tensors (i.e., Lorentzian; or,
concerning coordinate transformations, $D$-scalars), and tensors
with Greek ones are $D$-tensors (tensors of the diffeomorphism
group).

This requirement of {\it left-right} covariance
 will be automatically
fulfilled if one follows the next easy rules: \\
1. Usual covariant differentiation [with symmetric connection,
which preserves  the  metric (\ref{metr})] is used;
$g_{\mu\nu;\lambda} \equiv 0 $. \\
2. For operations with Latin indexes (contraction,
 raising, lowering), one  uses
$ \eta _ {ab},  \ \eta ^ {ab} $; with Greek ones --
 the metric $g ^ {\mu\nu} $
or $g _ {\mu\nu} $; at last, to change the kind of an index
 (change the ABC), the frame
 $h_a {} ^ \mu $ or co-frame $h^a {}_\mu$ should be used.\\
3. One can coin the notion of {\it mathematical dimension}
 for any covariant $ \Psi $ -- according to the power of constant
 $ \kappa $  which multiply $ \Psi $  at the {\sl scale}
 transformations (\ref {lrsy}); for example:
\[h_a {} ^ \mu {} \sim \kappa ^ {-1} ;  \
 h^a {} _ \mu \sim\kappa ; \ \eta_ {ab} \sim
\kappa ^ {0} , \  h^a {}_ {\mu; \nu} \sim \kappa . \] The usual
physical dimension arises if the scale transformation is
accompanied by the analogous replacement of coordinates:
\[x^{*\mu} = \lambda x ^ {\mu} , ~~\mbox {where \ }
 \lambda =\kappa ;
\mbox {\ \ so} \]
\[h_a{}^\mu{} \sim h^a{}_ \mu \sim\lambda^{0} , ~~~
\eta_{ab} \sim g _ {\mu\nu} \sim \lambda^{0} , ~~~
h^a{}_{\mu;\nu}\sim \lambda^{-1} , \mbox {\, etc.} \]
 The third rule states: field equations should contain terms
 of the same ('mathematical' or 'physical') dimension; certainly,
 it it is not supposed to introduce any
 `dimensional fundamental constants'.

As \ \ $ \eta_{ab} = \eta^{ab} = {\rm const,} \ \ g_{\mu\nu;
\lambda} \equiv 0 \,$, \ in covariant ($LD-$covariant) expressions
(where, in particular, only covariant differentiation is used) one
may do not distinguish upper and lower indexes and to omit
$\eta^{ab} $ and $g^{\mu\nu}$ in contractions,\footnote{At such
agreement it is necessary to use the `physical' dimension.}
understanding that
\[
\Psi_{.. a.. a..} = \Psi_{.. a.. b..} \eta^{ab}, \ \ \
\Psi_{..\mu.. \mu..} = \Psi_{.. \mu.. \nu..}g^{\mu\nu} . \]

The first derivatives of frame field give the tensors
\begin{equation}
\label{gala} \gamma_{a\mu\nu} = h_{a\mu; \nu} \,, \ \ ~
\Lambda_{a\mu\nu} = 2h _ {a [\mu; \nu]} =h _ {a\mu, \nu} - h _
{a\nu, \mu} \,.
\end {equation}
Here, as usually \cite {lan, how, dubr}, square brackets mean
alternation of indexes (and round ones -- symmetrization).
 Note  the symmetry properties
 of $ \Lambda $- and $ \gamma $-tensors:
\[\Lambda_{abc} =-\Lambda_{acb},
 \ \ \ \gamma_{abc} = - \gamma_{bac}, \]
where, certainly, $ \Lambda_{abc} = \Lambda_{a\mu\nu} h_b {}^ \mu
{} h_c {} ^ \nu {} $.

The irreducible  parts of these tensors  are denoted as follows:
\begin{equation}
\label{sdef} S _ {abc} =3\Lambda _ {[abc]} = \Lambda _ {abc} +
\Lambda _ {bca} + \Lambda _ {cab} =6\gamma _ {[abc]} \,;
\end{equation}
\begin{equation}
\label{phi} \Phi_a = \eta^{bc} \Lambda_{bca} = \Lambda_{bba}
=-\gamma_{abb} ~.
\end{equation}
Taking into account (\ref {sdef}), (\ref {gala}),
 it is easy to express
$ \gamma $ through $ \Lambda$:
\[\gamma_{abc}=\frac12(\Lambda_{abc} +
 \Lambda_{bca}-\Lambda_{cab})
. \]
 The vector $ \Phi _ {\mu} $ is a natural candidate
 to be compared  (maybe,  with some
factor) with the vector-potential of electromagnetic field;
 this explain
the following notation (for a part of the second
 derivatives of frame field):
\begin{equation}
\label {fmn}
 f_{\mu\nu} =2\Phi_{[\mu;\nu]} =
  \Phi_{\mu; \nu}-\Phi_{\nu; \mu} \,,
\end{equation}
\[\mbox {or \ } f_{ab} = \Phi_{a, b}-\Phi_{b, a}
 + \Phi_{c} \Lambda _ {cab} \,.
\]

It is frequently convenient to pass to Latin (scalar)
 indexes (the covariant
differentiation of $D$-scalar entities comes to usual partial
derivative). The last equation uses the following helpful notation:
\begin {equation}
\label {sdif} {} _ {, a} = {} _ {; \mu} h_a {} ^ \mu\,.
\end{equation}
At differentiation of $D$-scalar entities, we have
 a simple rule how to  transpose
 `scalar differentiations' (here $ \Psi $ is a $D$-scalar, that is,
 it can have only Latin indexes):
\begin{equation}
\label {abc}
\begin{array}{rcl}
\Psi_{,a,b}-\Psi_{,b,a} & = &(\Psi_{;\mu}
 h_{a\mu})_{;\nu}h_{b\nu}-(ab)= \\
&= & \Psi_{;\mu}\gamma_{a\mu b} - (ab)=-\Psi_{,c}\Lambda_{cab}\,.
\end{array}
\end{equation}
In presence of Greek indexes, more bulky terms with the
 Riemann curvature tensor appear:
 \[\Psi_{\cdots\mu\cdots, a,b} - (ab) =
\Psi_{\cdots\nu\cdots, a, b} R_{\mu\nu ab} -\Psi_
{\cdots\mu\cdots, c} \Lambda_{cab} \,. \] The curvature tensor can
be easily expressed through $ \gamma $ (or $\Lambda $), by
definition:
\begin {equation}
\label {trim} R _ {a\mu\nu\lambda} =
2h _ {a\mu; [\nu; \lambda]} =2\gamma _ {a\mu [\nu; \lambda]} \,.
\end {equation}

 The well known expressions  follow for the Ricci tensor
 and the scalar curvature
(Ricci scalar; see, for example, \cite [\S 98] {lan}):
\[R _ {ab} = - \Lambda _ {(ab) c, c} - \Phi _ {(a, b)}-\frac12
\Lambda _ {cda} (\Lambda _ {cdb} + \Lambda _ {dcb}) + \frac14
\Lambda _ {acd} \Lambda _ {bcd}-\Lambda _ {(ab) c} \Phi_c \,, \]
\begin {equation}
\label {sca} R =-2\Phi _ {c, c}-\frac12\Lambda _ {abc}
 \Lambda _ {abc}
+ \frac {1} {12} S _ {abc} S _ {abc}-\Phi_a\Phi_a\,.
\end {equation}

As we know, the definition of Riemannian curvature tensor (\ref
{trim})\footnote{* That is, the fact that curvature is a
`derivative entity' relative to the metric.} leads to
 the Bianchi
identities; in a similar way, the definitions
 (\ref {gala}),
(\ref {sdef}), (\ref {fmn}) lead to the following identities:
\begin{equation}
\label{iden} \Lambda_{a [\mu\nu; \lambda]} \equiv 0 ~ \mbox { \ \
\ [from
 (\ref{gala})]}
\end{equation}
\hspace*{35mm} \ or \ \ \ $ \Lambda _ {abc, d} + \Lambda _ {adq}
\Lambda_{qbc} + (bcd) \equiv 0 \,; $ \hfill (\ref{iden}$'$)
\begin{equation}
\label{ides}
 S_{[\mu \nu \lambda; \varepsilon]} \equiv
3/2\Lambda_{a [\mu\nu} \Lambda ^ {a} {} _ {\lambda \varepsilon]} ~
\mbox {\ \ \ \ \ [from (\ref {sdef})]};
\end{equation}
\[f_{[\mu\nu;\lambda]} \equiv 0 ~ \mbox {\ \ \ [from (\ref {fmn})]}.
\]
Contraction of two indexes in (\ref {iden}$ '$)
 gives the next identity:
\begin{equation}
\label{idef}
 \Lambda_{abc, a} +f_{bc} \equiv 0 \,.
\end{equation}

Let us consider $LD$-covariant second order equations of a frame
field. The Cauchy problem can have unique solution if the number
(of components) of  equations exceeds the number of field
components, $D^{2}$, minus $D$ (this takes into account the
freedom  of  coordinate choice), but this condition
 is not sufficient. For example the equation
 \  $R_{abcd} =0 \mbox {\ [see \ (\ref{trim})]} $ \
does not result, obviously, in a well-posed Cauchy problem (though
at big $D $ it has a lot of equations).

Let us consider systems of $D^{2}$  equations; so, the equations
have a form of a  second rank tensor which can be divided on the
symmetric and antisymmetric parts. The `uniqueness'  of Cauchy
problem depends on the principal terms (with the highest
derivatives) of equations, or on the {\it symbol\/} of a system:
it is necessary, that  the equations (and identities) would
provide
 (as differential consequences, as prolong equations)
the  `covariant evolution equations' for  the basic covariant,
tensor $ \Lambda_{abc} $, that is, the equations of the next
form:\footnote{* Further remark: the middle equation here does not
contain  terms $\Phi''$ because of symmetry considerations.}
\[\Phi_{a, b, b} =\Lambda' \Lambda + \Lambda^3, \  \
S _ {abc, d, d} = \Lambda' \Lambda + \Lambda^3,
\ \  \Lambda_{abc, d, d} = \Phi'' +\Lambda' \Lambda + \Lambda^3. \]
Taking into account all definitions, (\ref {sdef}),
(\ref {phi}), (\ref {fmn}), and the identity (\ref {idef}),
one can write down the symmetric and antisymmetric parts of
 covariant AP equations:
\[{\bf G}_{\mu\nu}=
2\Lambda_{(\mu\nu)\lambda;\lambda}+\sigma
(\Phi_{\mu;\nu}+\Phi_{\nu;\mu}-2 g_{\mu\nu} \Phi_{\lambda ;\lambda
})+(\Lambda^2)=
\]
\begin{equation}
\label{systa} \mbox{}=
 -2G_{\mu\nu}+(2\sigma-2)(\Phi_{(\mu;\nu)}-g_{\mu\nu}
\Phi_{\lambda;\lambda})+V_{(\mu\nu)}(\Lambda^{2})=0,
\end{equation}
\begin{equation}
\label{systb} {\bf H}_{\mu\nu}=
 S_{\mu\nu\lambda;\lambda}+\tau f_{\mu\nu}
+V_{[\mu\nu]}(\Lambda^2)=0,
\end{equation}
here the `quadratic terms', $ V_{\mu\nu}(\Lambda ^2)$,
 should be determined from the requirement of compatibility;
  \ $
G _ {\mu\nu} =R _ {\mu\nu}-\frac12 g _ {\mu\nu} R $
 \ is the Einstein tensor \cite {ligh, lan}.
The first (and most important) terms in
 (\ref {systa}), (\ref {systb})
 are taken with factor unit as their disappearance breaks
 the well-posedness  of  Cauchy problem.
 The other higher-order terms depend on two free parameters,
 $ \sigma $ and $ \tau $; the choice (by  replacement \ $
{\bf G} _ {\mu\nu} \to {\bf G} _ {\mu\nu} + k g _ {\mu\nu} {\bf G}
_ {\lambda\lambda} $) of the factor at $g _ {\mu\nu}
 \Phi _ {\lambda, \lambda} $
will become clear some later [see \ (\ref {maxa})].

The differential consequences from (\ref {systa}), (\ref {systb})
(the first {\it prolongations}) give a couple of Maxwell-like
equations:
\[ {\bf G}_{\mu\nu;\nu}=(\sigma-1)f_{\mu\nu;\nu}+2(\sigma-1)
R_{\mu\nu}\Phi_{\nu}+V_{(\mu\nu);\nu}=0~,\]
 \[
{\bf H}_{\mu\nu;\nu}=\tau f_{\mu\nu;\nu}+ V_{[\mu\nu];\nu}=0~;
\]
they should be consistent with each other; that is, there should
 exist an identity of the next form:
\begin{equation}
\label{maxa} \tau{\bf G}_{\mu\nu;\nu}+
 (1-\sigma){\bf H}_{\mu\nu}\equiv (\Lambda
{\bf G}, \Lambda {\bf H}).
\end{equation}
In Einstein and Mayer's work \cite {ei4}, the equations possessing
such an identity have been listed.

If $ \tau\neq0 $,  the `electromagnetic current' $J_\mu $ is
trivial, i.e., the equation $J_{\mu; \mu}=0$ comes to identity
automatically:
\begin{equation}
\label{ide2}
 J_{\mu}(\Lambda\Lambda^{\prime})
 \propto V_{[\mu\nu];\nu}\,,~~J_{\mu;\mu} \propto
  V_{[\mu\nu];\nu;\mu}\equiv 0\,.
\end{equation}
In the case $\tau=0$, the second identity is steel necessary for
compatibility of the system, and this requires  more work
 \cite{z2}; this case is considered in detail  in \S 1.3 (it is
remarkable also by the absence of co-singularities and gradient
catastrophe, \S 2.2).

 It is also necessary to add here some comments about
 the notations accepted in the works of Einstein (and Mayer)
 on AP \cite {ei2, ei3, ei4}. The other covariant differentiation
 was used there: with asymmetric connection, which is defined
 by the requirement of `frame preserving'
 (to denote it, we use the overlined  symbol `$ \overline {;} $'):
\begin{equation}
\label{asco} h_{a\mu\overline{;}\nu}=0,
 \mbox{ \ i.e., \ } h_{a\mu,\nu}-
\overline{\Gamma}^{\lambda{}} _{\mu\nu}h_{a\lambda{}}=0~, \ \ \
\overline{\Gamma}^{\lambda{}}_{\mu\nu}=
 h_a{}^\lambda{} h_{a\mu,\nu}~.
\end{equation}
The antisymmetric part of asymmetric connection,  $\overline
{\Gamma} $, forms a  tensor,  torsion tensor, which coincide with
our tensor $\Lambda $:
\[  \overline{\Gamma}^{\lambda{}}_{\mu\nu}-
 \overline{\Gamma}^{\lambda{}}_{\nu\mu}=
 \Lambda^{\lambda}{}_{\mu\nu}\,.\]
As it is easy to understand, the `new' (or `frame') curvature
tensor, $ \overline{R} $, composed from
 $\overline{\Gamma}$-connection, is identically equal to zero.
 Really, in view of identity
(or definition) $h_{a\mu\overline{;}\nu} =0$, it is possible to
replace $D-$tensors on $L-$tensors (one can  carry $h _ {a\mu} $
through  $ \overline {;} $-differentiation), and the rule of
rearrangement (or permutation) of differentiation indexes will be
reduced to expression (\ref {abc}), which has no terms with the
curvature (for $D$-scalars, both usual and covariant
differentiation are the same -- for any connection). This
identical equality, $ \overline {R}_{\mu\varepsilon\nu\tau}=0 $,
was reflected in the name of the theory: absolute parallelism.

So, that to switch to our notations in the equations of \cite
{ei2, ei3, ei4}, one can replace the Greek indexes on Latin, and
differentiation (\ref {asco}) (with asymmetric connection), $
\overline {;} \mu $, on the `scalar' differentiation (\ref
{sdif}). Our willingness `to keep' the usual covariant
differentiation (as well as Riemann curvature and Riemannian
geodetics) will find an explanation at the end of this chapter (\S
1.7), during discussion of post-Newtonian effects. In principle,
one can use any combination of symmetric and asymmetric
($\overline{\Gamma}$) connections (that is still connection), and
only further analysis of solutions of  equations (interpretation
of covariants, the energy-momentum tensor), depending on a choice
of AP equations, can give preference  to some connection.

\section*{1.2 \ Necessary information on the
 compatibility theory;\\ \hspace*{11mm}
 involutive symbols}
 The modern theory of {\it compatibility},
 or {\it formal integrability},
of systems of partial differential equations  is usually
 formulated with good deal of complicated mathematical methods
\cite {vino}. There is, however, a more simple and democratic
presentation of this theory: the book of J.F.\,Pommaret
\cite{pomm}; it delivers  a finite recipe or algorithm of
compatibility verification  for an arbitrary system of (nonlinear)
equations.\footnote{* See also
\href{http://en.wikipedia.org/wiki/Jet_bundle}%
{wikipedia.org/wiki/Jet\_bundle}
and reference therein, including
 \href{http://www.arXiv.org/abs/0908.1886}{arxiv: 0908.1886}.}

Compatibility of a system of equations simply means that there
 exist formal solutions in a form of Taylor's series.
 The compatibility theory provides also information on
 a solution arbitrariness:  how many  parameters are free
 to choose in the series terms of some
 order.\footnote{* In other words, some integers
 (called characters), $\alpha_0, \alpha_1,\ldots$,
define how many functions of $D$ variables, $D-1$ variables, et
cet., determine the general solution.}

Let $ \cal E $ is a trivial fiber bundle of dimension $m+n $
 (space $R ^ {m+n} $ with coordinates $ (x, y) $)
  above $n$-dimensional space $X $ with a projection
   $ \pi $, $ \pi: {\cal E} \to X $.
Independent variables $x ^ {\mu} $, coordinates of the base
 $X $ (in this paragraph $ \mu=1, \ldots, n $),
  dependent variables $y^A $ (`field' variables;
   $A $ is a set of indexes) serve as fiber coordinates.
 To deal with sections $y=f (x) $ of bundle $ {\cal E} $,
  they introduce further bundles, bundles of jets
   $J _ {q} ({\cal E}) $ with additional coordinates
  $p ^ {A} _ {\Gamma} $, which are coefficients of
$q$-th order expansion in Teylor series
 ($ \Gamma = \mu_1\cdots\mu_l $
 $ - $ a compound index, $ | \Gamma | = l $) :
\[
f^{A}(z) =y^{A}(x)+ p^A_{\mu}(x)(z^\mu -x^\mu)+
 p^A_{\mu\nu }(x)(z^\mu
-x^\mu)(z^\nu -x^\nu)/2+ \cdots \]  \[ =
 \sum_\Gamma  p^A_{\Gamma}
(z-x)^\Gamma/|\Gamma |!\,; \ \
 (q\ge|\Gamma|\ge0,\ \ p^{A}=y^{A}\,. \]
There is a set of natural projections
\[ \pi ^{q+r}_{q}:  J_{q+r}({\cal E})\to J_q({\cal E}) \ ;
\ \ \ \dim J_q({\cal E})=n+m\frac{(q+n)!}{q!\, n!} \ .\]

A differential equation of an order $q $ (or a system of equations;
 for brevity sake, sometimes we  omit indexes)
\[F^{B} (x; y, y', y",\cdots, y^{(q)})=0 \]
corresponds to  the analogous equation on the jet bundle
$J_q({\cal E})$:
\[F(x; p_{\Gamma})=0 \ \ \ (0\le |\Gamma| \le q); \]
this defines some subset ${\cal R}_q \subset J_q({\cal E})$.

Differentiation of this system, ${\cal R}_q$, defines {\it
$r$-prolongation\/} (prolonged systems of  equations)
\[{\cal R}_{q+r} \subset J_{q+r} ({\cal E}):
 \ \ F(x; p) =0, \ F' =0, \ \cdots, \ F^{(r)} =0, \]
\[\mbox {where, for example, \ \ }
(F')_{\mu} =d_\mu F =\frac {\partial F} {\partial x^{\mu}} + \frac
{\partial F} {\partial p_{\Gamma}} p_{\Gamma \mu} \,. \]

The system ${\cal R}_{q}$ is called {\it regular} if the system
${\cal R}_{q+r}$ at any $r\ge 0 $ is a fiber submanifold  in
$J_{q+r}({\cal E})$.

That to check the regularity of a system, one needs to consider
some auxiliary vector bundles above ${\cal R}_{q}$ and to check
that their ranks do not vary. We will denote tangents [co-tangent]
manifolds as \
 $T({\cal R}_{q}) \subset T(J_q ({\cal E}))$, \ with coordinates
  $(x, p_{\Gamma}; u, v_{\Delta})$,
   $ |\Gamma|, |\Delta|\le q\  $; \
[respectively, $T^{*}(X)$].

The  {\it  symbol\/}  $G_q$ of a system ${\cal R}_{q}$ is
important for  analysis of equations; this is the vector space
above $ {\cal R}_{q}$, $G_q\subset T({\cal R}_q)$, `generated' by
the senior jets $p_{\Gamma}, \ |\Gamma| =q$, and defined by the
following equation:
\begin{equation}
\label{symb} G_q:\ \  \ \frac {\partial F} {\partial p _ {\Gamma
}} (x; p) v_ {\Gamma} =0, \ \ \ | \Gamma | =q, \ \ (x; p) \in
{\cal R} _ {q}.
\end{equation}
Symbol dimension depends on the number of the highest jets
 (the highest order derivatives) and on the number
 $N _ {F} $ of the equations of system $ {\cal R} _ {q} $
 [or system (\ref {symb}) which determines the symbol]:
\begin{equation}
\label{dimg} \dim G_q = m \frac{(q+n-1)!}{q!\,(n-1)!} - N_{F}
\end{equation}

The next statements from \cite{pomm} is of special importance (it
is a kind of the essence of this theory):

The compatibility condition ({\bf  Corollary 4.9} from \cite{pomm}):

{\sl Let
${\cal R}_{q}\subset J_{q}({\cal E})$ is a $q$-th order
 system on a fiber manifold
$\cal E$,so that ${\cal
R}_{q+1}$ is a fiber submanifold in $J_{q+1}({\cal E})$.

If the symbol $G_{q}$ is 2-acyclic and the mapping $\pi^{q+1}_{q}:
{\cal R}_{q+1} \to {\cal R}_{q}$ is surjective, then
 the system ${\cal
R}_{q}$ is formally integrable (compatible).}

The involutivity conditions ({\bf  Corollary 4.11}
 from \cite{pomm}):

{\sl If the symbol $G_{q}$ is involutive and the mapping
$\pi^{q+1}_{q}: {\cal R}_{q+1} \to {\cal R}_{q}$ is surjective,
then  the system  ${\cal R}_{q}$ is involutive.}\footnote{System's
involutivity is some `stronger' than (i.e., includes)
compatibility, but simpler to check.}

This clause `the map $\pi_{q}^{q+1}$ is surjective' means just
that\footnote{That is, this prolongation does not give rise to new
$q$-th order equations.}
\[ \pi^{q+1}_{q}({\cal R}_{q+1})= {\cal R}_{q} \ . \]

 We will use also the next condition of symbol involutivity
 (it is a variant, or reformulation
  of {\bf Definition 2.16} from \cite{pomm}):

{\sl Symbol $G_{q}$ is involutive, if the next relation holds: }
\begin{equation}
\label{inv} \dim  G_{q+1}= \dim G^0_q+\dim G^1_q + \cdots + \dim
G^{n-1}_{q}\ \ (\dim G^n_q\equiv 0) .
\end{equation}
Here, subspaces $G^{i}_{q}\subset G_q$ are defined by
 addition to (\ref{symb}) the next
(noncovariant) equations ($G^{0}_{q}=G_{q})$:
\begin{equation}
\label{giq} v^{A}_{\mu_1\cdots \mu_k  \cdots \mu_q} = 0 \mbox{ \
if \ } \exists k \, \mbox{ that \ }\mu_k  \le i  \,.
\end{equation}

\subsection*{Application of compatibility theory to AP equations}
 In AP the number of field components $m$ is equal to
 $n^{2}$.\footnote{In this subsection we use $n$ instead of $D$
 (and Euclidean signature).}
  Let us take as an example a system ${\cal R}_2$ of the next form:
\begin{equation}
\label{ex} {\bf E}_{a\mu }= \Lambda_{a\mu \nu ;\nu }+ A_{a\mu
}(\Lambda^{2})=0.
\end{equation}
After differentiation of (\ref{ex}), one can try to exclude the
highest derivatives ($h'''$) in some combination of prolonged
equations ${\cal R}_{3}$. If it is possible, the requirement of
surjectivity of the mapping $\pi^{3}_{2}: {\cal R}_{3} \to {\cal
R}_{2}$ means that the other terms should also vanish
 through the use
of equations ${\cal R}_{2}$; in other words, there should exist a
corresponding identity. Otherwise, this gives rise to a new and
irregular (in first jets) second order equation, a kind of
$h'h''=0$ (more exactly, $\Lambda \Lambda '=0$); `irregular (in
first jets)' is because the new symbol of `supplemented' system,
$G^*_2$, will
 depend on the first derivatives $h'$, hence $\dim G^*_q$ will
change at $h'\to0$ (more exactly, $\Lambda \to0$).

The only combination of this kind for system (\ref{ex}) is
equation ${\bf E}_{a\mu ;\mu }=0$; note that in the case
 $A_{a\mu}=0$ there exist an easy identity:
  ${\bf E}_{a\mu ;\mu }\equiv0$.

Let us check that the symbol of system (\ref{ex}) is involutive;
during this analysis, it is convenient to use the Euclidian
signature (this choice does not affect the dimensions of spaces
$G^{i}_q$) and indexes $a,\ldots ,i = 1,\ldots ,n\,$.

The symbol $G_{2}$ (a vector space over $R_{2}$) of the system
(\ref{ex}) can be defined by the linear equation [see
(\ref{symb})]
\begin{equation}
\label{sym1} G_{2}:\ \ \qquad {\bf e}_{a\beta}=\frac{\partial{\bf
E}_{a\beta}}{
\partial h_{c\mu,\nu \lambda }} v_{c\mu ,\nu \lambda }= 0\ .
\end{equation}
In the notation of coordinates $v_{ab,cd}$,  jet-indexes are
separated by coma that to emphasize their symmetry with respect to
permutations. Symbol's involutivity should be checked over all
points
  $(x, h, h^\prime , h^{\prime\prime})\in R_{2}$; however,
only the frame matrix $h^{a}{}_{\mu }$ define (\ref{sym1}), and
this matrix, if it is nondegenerate, can be reduced to the unit
one through a coordinate transformation.
 If $h^{a}{}_{\mu }=\delta^{a}_{\mu }$,
 the equation of symbol, (\ref{sym1}), just coincides
with the linearized equation:
\begin{equation}
\label{sym2} {\bf e}_{ab} = v_{ab,cc} - v_{ac,cb} = 0 \,.
\end{equation}
This equation differs from the Maxwell equation, $a_{b,cc}-
a_{c,cb}=0$, only by the `extra' index `$a$'; the system
(\ref{ex}) can be supplemented (in analogy with the Lorentz gage
$a_{b,b}=0$) with the equation $h_{a\mu ,\nu } g^{\mu \nu }=0$,
which define a coordinate system (fix a gauge). It is therefore
evidently, that the system (\ref{ex}) has unique (up to a gauge)
solutions to the Cauchy problem.

 A general systems, (\ref{systa}), (\ref{systb}), is equally
 suitable (with well-posedness of Cauchy problem) if its symbol
 has the same dimensions: $\dim G_{2+r}$ for all $r$ coincide with
 analogous values of the system (\ref{ex}).

Taking into account (\ref{dimg}), (\ref{giq}), (\ref{sym2}), find
$\dim G_{3}$ and $\dim G^{i}_{2}$ for the system (\ref{ex}):
\begin{equation}
\label{g3} \dim G_{3} = n^{3}(n+1)(n+2)/6 - n^{3} + n ,
\end{equation}
\begin{equation}
\label{gi2} \dim G^{i}_{2} = n^{2}(n-i)(n-i+1)/2 - n^{2} + n^{2}
\delta ^{i}_{n} + n \delta ^{i}_{n-1} .
\end{equation}
The first term in (\ref{g3}) corresponds to the number of
coordinates  $v_{ab,cde}$ (accounting the symmetry of jet
indexes), the second term is the number of equations  ${\bf
E}'=0$, the third  is the number of identities for this system.
The terms in (\ref{gi2}) are: the first item means the number of
coordinates $v_{ab,cd}$ subject to (\ref{giq}) ($q=2$), the second
is the number of equations (\ref{sym2}), the third term expresses
the fact that, in the case of $G^{n}_{2}$, the equations
(\ref{sym2}) turn into identity provided (\ref{giq}). At last, the
last term, $n\delta ^{i}_{n-1}$, is added because, in the case of
space $G^{n-1}_{2}$,  $n$ equations  from (\ref{sym2}) turn into
identity [given (\ref{giq})]:
\[ {\bf e}_{an} = v_{an,nn} - v_{an,nn} \equiv  0
\ \ \ (n \mbox{ is the `fixed value' index}).
\]
Substituting (\ref{gi2}), (\ref{g3}), one can check that the
equality (\ref{inv}) is valid; that is, the symbol $G_2$ of system
(\ref{ex}) is involutive (if $h_a{}^\mu{}$ is nondegenerate).

 If the corresponding identity does exist
(a kind of ${\bf E}_{a\mu;\mu} \equiv {\bf E}\Lambda$), the system
(\ref{ex}) is compatible; it means that all other identities, in
the next orders of differentiation, do exist for sure.

The expressions (\ref{g3}) and (\ref{gi2}) are also valid for any
system (\ref{systa}), (\ref{systb}) excepting the case $\tau=0$
(when the `second' identity (\ref{ide2}) does not follow
`automatically'), when one should add to Eq.~(\ref{gi2}) the
additional term $+\delta ^{i}_{n-2}$, because in this case the
component of equation, ${\bf h}_{nm}= 0\ (m=n-1)$, [see
(\ref{symb}), (\ref{sym1}), (\ref{systa})] from (\ref{symb})
written for $G^{n-2}_{2}$ turns into identity as long as the
additional equations (\ref{giq}) are taken into account:
\[ {\bf h}_{nm} \sim  S_{nmc,c}=0 \ ; \]
here $c\neq n,m$ (note that  $S_{\mu \nu \lambda
}=S_{[\mu\nu\lambda]}$), hence this equation is a part of
(\ref{giq}).

It is not difficult to obtain analogous expressions for  $\dim
G^{i}_{3}$ and $\dim G_{4}$, which are valid both for (\ref{ex})
and for other variants (excepting the case $\tau=0 \ \& \ \sigma
=1$, when the `Maxwell equation' is absent, and the evolution
equation $\Phi_{a,b,b}+ \cdots =0$ necessary for well-posed Cauchy
problem is lacking too):
\[ \dim G^{i}_{3} = n^{2}C_{n-i+2}^{3} - n^{2}(n - i) +
n (1 - \delta ^{i}_{n})\ , \]
\[ \dim G_{4} = n^{2} C_{n+3}^{4} - n^{3}(n+1)/2 + n^{2}\,. \]
Substituting these equations into (\ref{inv}), one can finally
prove that the symbol $G_{3}$ is involutive.

So, in the case $\tau=0, \ \sigma\neq1$, symbol $G_2$ of a system
(\ref{systa}), (\ref{systb}) is not involutive, but its
prolongation, symbol $G_3$, is involutive, and compatibility of
such a system is provided by first two identities (of different
order on differentiation).
\newpage

\section*{1.3 \ Einstein--Mayer's classification\\
 \hspace*{11mm} of compatible equations of Absolute Parallelism}
 The paper \cite{ei4} by Einstein and Mayer
 is devoted to a full search
for compatible second order AP equations. More exactly, the
authors were seeking for equations which obtain `one
four-dimensional identity' (the spacetime dimension were fixed
there, $D=4$, so the identity (\ref{maxa}) is just
four-dimensional vector). There are several rare cases, when the
presence of such an identity is not enough for equation
compatibility; these cases will be explained later.

As a rule, in \cite{ei4}, field equations were not divided
 on the symmetric and antisymmetric parts
 (\ref{systa}), (\ref{systb});
instead a combination of these parts was taken which enters into
identity (\ref{maxa}). Therefore, especial attention should be
payed to the cases when this combination contains either only
symmetric or only skew-symmetric part.

The four classes of compatible (having that identity) field
equations were found in
 \cite{ei4}, and the most simple-looking  is
the case II$_{22112}$. Taking an arbitrary rank three tensor,
linear in $\Lambda_{}$ and antisymmetric
 with respect to the last two
indexes,
\[ K_{abc}=K_{a[bc]}=a\Lambda_{abc}+bS_{abc}+
c \eta_{ab}\Phi_{c}-c \eta_{ac}\Phi_{b}~, \] one can arrive to a
simple second order equation (for frame field $h$)
\begin{equation}
\label{sys1} {\bf A}_{a\mu}=K_{a\mu\nu;\nu}=0,
\end{equation}
which has the evident identity:
  \[ {\bf A}_{a\mu;\mu}\equiv 0
~~(K_{a\mu\nu;\nu;\mu}\equiv 0)~.
\]
The old parameters $\sigma$ and $\tau$ can be linked to $a, b, c$:
\begin{equation}
\label{abcst}
 \sigma= -\frac c a ~, \ \ \ \tau=\frac{a+c}{a+2b}~.
\end{equation}
According to the arguments adduced in the previous paragraph, the
cases $a=0$ and/or $a+2b=0$ are unappropriate and correspond to
incompatible equations (in spite of identity), so one can chose
$a=1$. The other special case is
\[ a+c=0~,\mbox{ \ when \ } \sigma=1,~\tau=0, \]
when compatibility requires two identities (Maxwell equation does
not follow from both symmetric nor antisymmetric part of the
system); one can easily check that the second identity is lacking.

The next evident class of compatible AP equations is the class
 (with two free parameters, as the previous class) of Lagrangian
 equations, which is labeled in
 \cite{ei4} as I$_{12}$.
 Let's take a Lagrangian (density)
  $h{\cal L}$, where $h=\det h^{a}{}_{\mu}$,
 ${\cal L}$ is an arbitrary `full' scalar ($LD$-scalar); that is,
\[ {\cal L}=\frac a 4 \Lambda_{abc}\Lambda_{abc} +
\frac b {12} S_{abc}S_{abc} + \frac c 2 \Phi_{a}\Phi_{a}~.\]
 Taking into account the symmetry properties of
 tensors $\Lambda$ and $S$, transform the differential:
\[d\,{\cal L}= \frac1 2 K_{abc}\, d\Lambda_{abc}=
K_{abc}\, d(h_{a\mu,\nu}h_b{}^\mu{}h_c{}^\nu{}) ~, \]
\[ \mbox{where  \ \ \ }
 K_{abc}=K_{a[bc]}=a\Lambda_{abc}+bS_{abc}+
c \eta_{ab}\Phi_{c}-c \eta_{ac}\Phi_{b} \]
 (as for the previous class of equations).
  Making variation, one should distinguish co- and
 contra-indexes, but the result, naturally, can be written in a
 covariant form:
\begin{equation}
\label{sys2} {\bf B}_{a\mu}=\mbox{} - g_{\mu\nu} \frac{\delta
(h{\cal L})}{h\delta  h^a{}_{\nu }} =K_{a\mu\nu;\nu}
+\Lambda_{bca}K_{bc\mu } -h_{a\mu}{\cal L}=0~.
\end{equation}
 Similar to the General Relativity equation,
  the Lagrangian (generally
 covariant) AP equations have a natural identity which is easily
 checkable (it can be derived in a similar way as the derivation
 of the GR identity in \cite{lan}: a coordinate variation leads
 to a frame variation while the action does not change):
\[ {\bf B}_{a\mu;\mu } -{\bf B}_{bc}\Lambda_{bca} \equiv 0\,. \]
Of course, the relations (\ref{abcst}) are also valid for
equations (\ref{sys2}).

The skew-symmetric  part of equations (\ref{sys2}),
\[
{\bf B}_{[\mu\nu ]}=(\frac a 2+b)(S_{\mu \nu \lambda ;\lambda }
-\Lambda_{\mu ab}\Lambda_{ab\nu } +\Lambda_{\nu  ab}\Lambda_{ab\mu
})+ \frac{a+c}2 (f_{\mu \nu }-\Phi _a\Lambda_{a\mu \nu })=0, \]
 vanishes if $a=-2b=-c\ (=1)$; this case correspond to the vacuum
  equation of GR;\footnote{One can either add a quite
  arbitrary (with some
  exceptions) skew-symmetric equation, or allow
  local Lorentz rotations of the frame field.} one can compare
 the Lagrangian expression, ${\cal L}$,
 and the scalar curvature (\ref{sca})
\[ R=-2\Phi_{\mu ;\mu }-\frac12\Lambda_{abc}\Lambda_{abc}
+\frac{1}{12}S_{abc}S_{abc}+\Phi_a\Phi_a\,; \]
 the  only difference is the term
  $\Phi _{\mu ;\mu }$ which is of no importance for the action.

If $a+c=0$, $a+2b\neq 0$, the equation ${\bf B}_{[\mu\nu ];\nu}=0$
looses its highest-order terms but not the others; that is, it
becomes irregular (in the first jets). If the other case,
$a+c\neq0$, $a+2b= 0$, then an irregular equation is  ${\bf
B}_{[\mu\nu ;\lambda ]}=0$.\footnote{* For more details see
\href{http://www.arXiv.org/gr-qc/0610076}{arXiv:gr-qc/0610076}.}

The third variant (I$_{221}$) of compatible equations is possible
(i.e., the necessary identity exists) only for $D=4$; following
\cite{ei4}, let's write separately the symmetric and antisymmetric
parts of this class of equations:
\begin{equation}
\label{sys3} {\bf C}_{\mu \nu } \ \ \ \ \left\{
\begin{array}{rcl}
{\bf C}_{(\mu \nu )}& {=}&
2G_{\mu\nu}+2(1-\sigma)(\Phi_{(\mu;\nu)}
-g_{\mu\nu}\Phi_{\tau;\tau})+\\
& &\mbox{} +(\sigma-1)^2(\Phi_\mu\Phi_\nu+
g_{\mu\nu}\Phi^2/2)=0\\
{\bf C}_{[\mu \nu ]}&{=} &
 S_{\mu\nu\lambda;\lambda}+ \tau  f_{\mu \nu }=0
\end{array} \right.
\end{equation}
Differentiating both parts and taking into account symmetries of
tensors and the identity $G_{\mu \nu ;\nu }\equiv 0$, we derive
the next equations:
\[ {\bf C}_{(\mu \nu );\nu }\equiv (1-\sigma )f_{\mu \nu ;\nu }
-\frac{D-4}{D-1} \left(\frac{1-\sigma }2 R \Phi _{\mu }+ \frac32
(1-\sigma )^{2} \Phi ^{2}\Phi _{\mu }\right) + ({\bf C}\Phi ) \ ,
\]
\[{\bf C}_{[\mu\nu];\nu }\equiv
 \tau  f_{\mu\nu;\nu}=0~; \]
 that is, the identity  (and compatibility of this system)
  exists if  $D=4$. Any
 variation of parameters in
 (\ref{sys3}) does not allow compatibility of this system at
 another $D$.

These equations are not so interesting because the current of the
Maxwell equation has vanished. The equations preserve
compatibility at adding the following equation:
\[f_{\mu\nu}=0~,\ \ \mbox{ or \ \ }\Phi_\mu=\Psi_{,\mu}~; \]
it is interesting to note, however, that the replacement
 $\Phi \to \Psi' $ reduces the symmetric equation (\ref{sys3})
 into the form of scalar-tensor (vacuum) equation of Branse--Dicke
 theory \cite{ligh,wein}.

At last, the last and most interesting (and nontrivial) class of
equations, the one-parameter class II$_{221221}$ of \cite{ei4},
can be written as follows:
\begin{equation} \label{sys4}
{\bf D}_{a\mu}=L_{a\mu\nu;\nu}-(1-2\sigma)
(f_{a\mu}+L_{a\mu\nu}\Phi_{\nu})=0\, ,
\end{equation}
\[\mbox{where \  }
L_{a\mu \nu }=L_{a[\mu \nu ]}=\Lambda_{a\mu \nu }-S_{a\mu \nu }-
2\sigma h_{a[\mu }\Phi_{\nu] }\,,
 \mbox{\, and \ } {\bf D}_{a\mu}=
 ({\bf G}_{a\mu}-{\bf H}_{a\mu})/2 \,;\]
that is, the equation (\ref{sys4}) contains in equal proportions
both parts, (\ref{systa}) and (\ref{systb}) (symmetrical and
antisymmetrical).

These equations are not supplied in \cite{ei4} in an explicit
form, however they can be recovered using the Table at the end of
this paper.  (But for all that, in the equation (1), one should
replace $a_{1}$ and $a_{2}$: suffice it to compare the Eq.~(1) and
Eq.~(6) of \cite{ei4}. The parameter $q$ of this paper, which
defines the ratio of symmetrical and antisymmetrical parts,
relates to $\sigma $ in this way: $\sigma =q/(2q-1)$.)

That to prove that the  identity necessary for compatibility of
 the system (\ref{sys4}) exists,
 let's proceed in the following way:\\
1. \ Pick out the antisymmetric part of (\ref{sys4}),
\[
{\bf H}_{\mu\nu }=S_{\mu\nu\lambda ;\lambda }+ (1-3\sigma )
(f_{\mu \nu }-S_{\mu \nu  a}\Phi _{a})=0\ ; \]
 its differentiation
gives a Maxwell equation,
\begin{equation}
\label{mm4} f_{\mu \nu ;\nu }=(S_{\mu \nu  \lambda }\Phi _{\lambda
})_{;\nu } \,, \mbox{\, or \ \ \ }  f_{\mu \nu ;\nu }= \mbox{}
-\frac 12S_{\mu \nu  \lambda }f_{\nu \lambda }+ (1-3\sigma )f_{\mu
\lambda }\Phi _{\lambda } \,.
\end{equation}
2. \ On the other hand, differentiation of the whole system
(\ref{sys4}) leads to another Maxwell equation, which, however,
can be reduced to the first one [given the equations (\ref{sys4}),
i.e., `on the equations']):
\[ {\bf D}_{a\mu;\mu }=\mbox{}- (1-2\sigma )
(f_{a\mu ;\mu }-\frac12 L_{a\mu \nu } \Phi _{\nu })=0\ ,\]
\[ f_{\varepsilon \mu ;\mu }=(h_{a\varepsilon }f_{a\mu })_{;\mu }=
h_{a\varepsilon }f_{a\mu ;\mu }-\frac12 \Lambda _{\varepsilon
ab}f_{ab} =\cdots= -\frac12 S_{\varepsilon ab}f_{ab}+(1-3\sigma
)f_{\varepsilon a}\Phi _a \,.\]
 This  coincidence can be written as an identity
(here $\tau =1-3\sigma $):
\[ \tau {\bf G}_{a\mu;\mu }+(1-\sigma)
{\bf H}_{a\mu;\mu }\equiv(1-2\sigma) \{\Lambda_{abc}{\bf
H}_{bc}+\tau({\bf G}_{ab}+{\bf H}_{ab})\Phi_{b}\}\,.
\]

These equations (\ref{sys4}) are remarkable in many respects. The
electromagnetic current, in contrast to the previous class of
equations, does not vanish, and it is linear in tensor
 $f_{\mu\nu}$ [see (\ref{mm4})]. Therefore, there exist solutions
 with zero electromagnetic field; that is, the system preserve
 compatibility at addition of equation $f_{\mu \nu} =0$.

\section*{1.4 \ Chances for unification of gravitation
 and\\ \hspace*{11mm}
  electromagnetism, and the {\em optimum equation}}
  We are going to show that there exists a unique system of
 equations which allows to match (up to a factor)
 tensor $f_{\mu\nu }$ with electromagnetic field $F_{\mu \nu }$.
 During derivation of these `optimum' equation,
 we will usually use Latin indexes, sometimes switching to Greek
 or mixed indexes, if this makes an expression more clear [one can
 compare, for example, the two forms of identity:
   (\ref{iden}) and (\ref{iden}$'$)].
 Latin indexes are convenient by the simple rule to reshuffle
 differentiation indexes [see (\ref{abc})]:
\begin{equation}
\label{abc1} \Psi_{,a,b}=\Psi_{,b,a} +\Psi_{,c}\Lambda_{cba};
\end{equation}
here $\Psi $ $-$ is a $D$-scalar, i.e., it has only Latin indexes.
 Greek indexes are useful in equations similar to  Maxwell equation
(with divergence of skew-symmetric tensor), where symmetrical
 connection is especially helpful.

Let us start with general equations
  [see  (\ref{systa}], (\ref{systb}))
\begin{equation}
\label{sya} {\bf G}_{ab} = \Lambda _{abc,c}+
 \Lambda _{bac,c} + \sigma (\Phi
_{a,b}+ \Phi _{b,a}-
 2\eta _{ab}\Phi _{c,c})+Q_{ab}(\Lambda ^{2})=0\ ,
\end{equation}
\begin{equation}
\label{syb} {\bf H}_{ab} = S_{abc,c}+\tau
  f_{ab}+W_{ab}(\Lambda ^{2}) =0\ .
\end{equation}
The first simple question is about localized solutions with
a Coulomb-like asymptotics of tensor
 $f_{\mu \nu }$ (i.e.,  spherical symmetry,
 and linearity -- for superposition).
 Consider Equation (\ref{syb}) in linear approximation
  (weak fields,  $h''\gg h'^{2}$):
\[ S_{\mu \nu \lambda ,\lambda } + \tau  f_{\mu \nu } = 0\,.\]
If $\tau\neq 0$, then $f_{\mu \nu }$ is a divergence of
a totally skew-symmetric rank tree tensor; this fact is hard to
agree with existence of Coulomb-like (or Yukawa-like)
 asymptotics. Indeed, assume that
there is an asymptotic spherical symmetry along
 the tree dimensions (or, $SO_3$ symmetry), and derivatives
 along extra dimensions (if they exist)
are negligible. Then we have (here $i,j,k = 1,2,3;\ \
r^2=x^ix^{i}$):
 \[   S_{0ij} = \varepsilon _{ijk} x^k \alpha (r)\,;\ \ \tau
f_{0i} =-S_{0ij,j}\equiv 0\,.\]

So, let us choose  $\tau  = 0$. In this case,
the highest-order derivatives
($h'''$) do vanish in the equation ${\bf H}_{ab,b} = 0 $; so,
 either a new and irregular second order equation arises
 or this equation turns into identity
 (i.e., other terms vanish too).
 Sure, we need the second choice, and there is the only
  possible equation to reach this:
\[{\bf H}_{\mu \nu } = S_{\mu \nu \lambda ;\lambda } = 0
\mbox{ \ \ or }\]
\begin{equation}
\label{sybb} {\bf H}_{ab} = S_{abc,c}+
 \Lambda _{acd} \Lambda _{cdb} - \Lambda
_{bcd} \Lambda _{cda} + S_{abc} \Phi _c = 0\ .
\end{equation}
The necessary (for compatibility) identity has a very simple form:
\[ {\bf H}_{\mu \nu ;\nu } =
 S_{\mu \nu \lambda ;\lambda ;\nu }\equiv 0,
\mbox{ \ or \ \ } {\bf H}_{ab,b}
 -\frac12 \Lambda _{abc} {\bf H}_{bc} +{\bf
H}_{ab}\Phi _b\equiv 0. \]

 The symmetrical part can be also defined
 from the requirement of compatibility.
 Equation ${\bf G}_{ab,b} = 0$
 can be reduced to the following form:
\begin{equation}
\label{max} {\bf G}_{ab,b} =
 (\sigma -1)( f_{ab,b} -\frac 12 \Lambda_{abc}
f_{bc} + f_{ab} \Phi _b - J_a) = 0\,,
\end{equation}
or to the usual form of Maxwell equation:
\[ f_{\mu \nu ;\nu } -
 J_\mu (\Lambda \Lambda ',\Lambda ^{3})=0\,.\]
The system (\ref{sya}), (\ref{syb})
 will be  compatible, if  equation
\begin{equation}
\label{side}
 J_{\mu ;\mu }= 0 \ \mbox{ \ or \ \ }
J_{a,a} + \Phi _{a} J_a = 0
\end{equation}
turns into identity -- of course, with equations
 (\ref{sya}), (\ref{sybb})  taken into account.

The compatibility theory  \cite{pomm} explains
 this case is such a way
 (see \S 1.2): symbol $G_2$ of system (\ref{sya}), (\ref{syb})
 is not involutive when
 $\tau  = 0$, $\sigma  \neq 1$; however, its prolongation,
 symbol $G_3$, is involutive, so the second identity (\ref{side})
 is necessary (and sufficient) for compatibility of the system.

  In the case $\sigma =1$, equation
${\bf G}_{ab,b}=0$ should turns into identity, but this case
 is of little interest (it is mentioned in \cite{ei4} too):
 only equation of General Relativity (GR) is possible,
\[-\frac12{\bf G}_{\mu \nu } = R_{\mu \nu }
-\frac 12 g_{\mu \nu }R = 0\ ,\] and it is clear that
 $f_{\mu \nu
}$ (or $\Phi '$) does not enter this equation, and does not
influence the metric.
 Moreover, Maxwell equation (\ref{max}) is not defined, i.e.,
  it can be added quite arbitrarily:
\[f_{\mu \nu ;\nu } =W_{[\mu \nu ]}(\Lambda ^{2})_{;\nu }
\ (=J_\mu ; \ J_{\mu ;\mu }=0)\  .\]

So, well, we are going to find a symmetric equation (\ref{sya}),
 which would lead to an identity
 (\ref{side}). Omitting detailed derivations, let's describe
  the way to solve this problem.
Using definitions  (\ref{sdef}),
(\ref{fmn}$'$), equations (\ref{sya}), (\ref{sybb}), `short'
identity (\ref{idef}), but not the `full' identity (\ref{iden}),
the expression for $J_a$ from (\ref{max})
 can be reduced to the following form:
\begin{equation}
\label{ja} (1-\sigma )J_a =
 C_{ab}(\Lambda ^{2})_{,b} + D_{a}(\Lambda ^{3})
=
\end{equation}
\[\mbox{}=[Q_{ab}+\Lambda _{acd}\Lambda _{bcd}-
\frac12 S_{acd}\Lambda _{bcd}+ \frac{1+\sigma}2
 S_{abc}\Phi _c +(\Phi \Lambda
)]_{,b}+\frac{1+\sigma}2 Q_{ab}\Phi _b
 +(\Phi \Lambda ^{2})\ . \]
Here we have omit  other terms containing vector $\Phi$.

Now we should take into account the `full identity' (\ref{iden}).
 Multiplying (\ref{iden}) by $S_{bcd}$,
$S_{abc}$, and $\Lambda _{abc}$, one can  obtain, respectively
 [also reducing to the form of
(\ref{ja})]:
\begin{eqnarray}
\label{ja1} E^1_a&{=}&(\Lambda _{aij}S_{ijb})_{,b} +
 \Lambda _{aij}S_{ijb}\Phi
_b=0 \ \ (\mbox{or \ }(L_{a\mu \nu }
 S_{\mu \nu \lambda })_{;\lambda}=0) ;
\\
\label{ja2} E^2_a&{=}&(\frac16 \eta _{ab}S_{ijk}S_{ijk}{-}
S_{ija}S_{ijb})_{,b}{+} \Lambda _{ija}S_{jkl}
 (S_{ikl}{-}2\Lambda _{ikl})
{+}\cdots=0 ;
\\          \label{ja3}
E^3_a&{=}&[\frac12 \eta _{ab} \Lambda _{ijk} \Lambda _{ijk}
 - 2\Lambda _{ija}
\Lambda _{ijb} + (\Phi \Lambda ^{2})]_{,b} -
\Lambda _{ija}Q_{ij} + \nonumber \\
\lefteqn{\mbox{}+ \Lambda _{ija} \Lambda _{kli}
 (\Lambda _{jkl}+2\Lambda
_{klj}) - \Lambda _{ija}\Lambda _{ikl}
 (\Lambda _{jkl}+\Lambda _{klj}) +\cdots
=0.}
\end{eqnarray}
In (\ref{ja2}), (\ref{ja3}) we omit terms with  $\Phi $ and
cancel the terms proportional to
(\ref{ja1}).

Note in passing that there are no combination of equations
(\ref{ja1})--(\ref{ja3}), a kind of
 $C_{ab,b} +D_{a}=0$, with a skew-symmetric tensor $C_{ab}$.
 This prove the uniqueness of equation (\ref{sybb}).

 Adding a combination $aE^1_a + bE^2_a + cE^3_a$ to the right hand
  part of (\ref{ja}), we receive
\begin{equation}
\label{jaa} (1-\sigma )J_a =
 C^*_{ab}(\Lambda ^{2})_{,b} + D^*_{a}(\Lambda
^{3})\,.
\end{equation}
The requirement $C^*_{(ab)}=0$ defines (symmetric) tensor $Q_{ab}$;
 after that, equation (\ref{jaa}) can be reduced to
the following form:
\[ (1-\sigma ) J_\mu  = A_{\mu \nu ;\nu }+B_\mu\ ,
\mbox{ \ \ \ or }\]
\begin{equation}
\label{jaaa} (1-\sigma )J_a = A_{ab,b}
 -\frac12 \Lambda _{abc}A_{bc} +
A_{ab}\Phi _b + B_a \ ,
\end{equation}
\[\mbox{where \ \ \ } A_{ab}=C^*_{[ab]}\ ,\ \ \ B^*_a
= D^*_a +\frac12\Lambda _{abc}A_{bc}-A_{ab}\Phi _b\ . \]
The necessary identity exists [see (\ref{side}), (\ref{jaaa})],
 and equation $B_{\mu ;\mu }=0$
turns into identity only if  $B_\mu =0$.

One can easily prove that the `senior' terms (which do not contain
 $\Phi $)  in $B_a$ can be cancelled in the next two cases
 [see  (\ref{ja})--(\ref{jaaa})]:
\[ 1)\ \ \ a = b =-1/2\ ,\ \  c = 1\ ;\
 \ \ \ \ 2)\ \ \ b = c = 0\ .\]
Next, one should cancelled other terms of  $B_a$  (`junior' terms,
 which contain $\Phi $).

The first case leads  either to  GR equation
 ($\sigma =1$, $D$ is arbitrary)
which was discussed earlier (Maxwell equation can be added quite
 arbitrarily), or to one-parameter class of equation,
  the special case of the class
(\ref{sys3}) with $\tau =0$, where $J_\mu =0$ and $D=4$,
($\sigma $ can be arbitrary). Both variants are of no interest.

The second case gives: $\ \ \ a=-1/2\ ,\  \ \sigma  =1/3\,;$
\begin{equation}
\label{mm}
 f_{\mu \nu ;\nu } =J_\mu \ ,\ \ \ J_\mu  = (S_{\mu \nu \lambda }
 \Phi _\lambda )_{;\nu } = -\frac12
  S_{\mu \nu \lambda }f_{\nu \lambda }\ ;
\end{equation}
\[ Q_{ab} =\Lambda _{acd}\Lambda _{cdb}+
 \Lambda _{abc}\Phi _c+\frac29
(\Phi _a\Phi _b - \eta _{ab}\Phi _c\Phi _c) + (ab)\ . \]

It is convenient to write down the end results as follows:
\begin{equation}
\label{equ} {\bf E}_{a\mu }=
 \frac12({\bf G}_{a\mu }-{\bf H}_{a\mu }) = L_{a\mu
\nu ;\nu } -\frac13 f_{a\mu } -\frac13
 L_{a\mu \nu } \Phi _\nu  = 0\ ,
\end{equation}
 where $L_{abc}=L_{a[cb]}=\Lambda _{abc}-
 S_{abc}-\frac13 \eta _{ab}\Phi _c
+\frac13 \eta _{ac}\Phi _b\ $.
This equation (let's call it
{\it optimum\/} equation) is a particular case of the class
{\bf D} (\ref{sys4}).

The trace equation
 ${\bf G}_{aa}= 0$ [see (\ref{sya}),
(\ref{equ})] becomes irregular (in the first jets) if $D= 4$:
\[ {\bf G}_{aa} = \frac23 (4 - D) \Phi _{a,a}+Q_{aa} = 0
\ \ (Q_{aa}\not\equiv 0). \]
 This means that some additional space
 dimension(s) is(are) necessary.
  Later, in the next chapter,
we will show that the most appropriate choice for equation
(\ref{equ}) is dimension $D=5$ (because of the absence, or rarity,
of singularities of solutions).

An arbitrary system (\ref{sya}), (\ref{syb})
 can have a forbidden number of dimensions: $D_{0}=1+1/\sigma$,
when the trace equation (if $D_{0}$ is an integer  $ \ge
2$):
\[
{\bf G}_{\mu \mu }= 2\sigma (D_{0}-D)\Phi _{\mu ;\mu }+Q_{\mu \mu
}=0 \ \ (\mbox{usually \ } Q_{\mu \mu }\not\equiv 0). \]

Linearized equations
(\ref{sya}), (\ref{syb}) for weak fields
\[h_{a\mu }=\eta _{a\mu }+ \epsilon _{a\mu }
\ \ (\epsilon _{a\mu }\ll  1)\]
 are invariant with respect to infinitesimal conform transformations
$\epsilon^{*}_{a\mu }=\epsilon _{a\mu }+ \lambda \eta _{a\mu }$
if  $D=D_{0}$ (the trace equation is lacking
 but one cad add `by hands'
$\Phi _{\mu ,\mu}=0$);
at that, vector $\Phi _{\mu }$ undergoes
 the following gradient transformation:
\[\Phi ^{*}_{\mu }=\Phi _{\mu } + (D-1)\lambda _{,\mu }.\]
Therefore, the
{\it optimum\/} equations (\ref{equ})  with
$D>D_{0}=4$ [$\sigma  =1/3$; having in mind some reduction
 of the extra dimension(s)]
are remarkable also by this `masking'  of the gauge symmetry at
the usual spacetime dimensionality and in the weak
 field approximation.

The Table in \cite{ei4} contains several particular variants
 which are commented (in the column `Notes') as follows:
  ``the symmetrical equation is lacking'', and further analysis
   does not follow.  In some  cases this note is not true.
   For example, the system of equations relating
to variation of the next action,
\[ \int h S^{2}\, dx \ \ \ ( S^{2}=S_{abc}S_{abc}\ ,
\ \ \ h=\det h^{a}{}_{m} = \sqrt{-g} )\,,           \] has the
second-order derivatives, which participate only in the
antisymmetric part: $S_{\mu \nu \lambda ;\lambda }$. However, as
it is not difficult to understand
 [see also (\ref{sys2})], this system contains also
  the term $g_{\mu \nu } S^{2}$. Hence, the
symmetrical equation (part) does exist, but it is irregular.
 The same is valid in the case of equation
$S_{a\mu \nu ;\nu } =0$  [which `abuts on'
 the class of equations (\ref{sys1})].

\section*{1.5 \ Updated (overdetermined) equations}
In this paragraph, on the way to find a
 physical interpretation of the
 {\it optimum\/} equations (\ref{equ}) and covariants of the theory
 (e.g.,  juxtaposing $f_{\mu \nu }$ with
electromagnetic field), we consider possibility to add to system
(\ref{equ}) some additional equations; this corresponds
 to particular solutions when, say, one or another
 covariant vanishes.

System (\ref{equ}) [as well as the equations of class ${\bf D}$
(\ref{sys4})] remains compatible at adding the equation
 $f_{\mu\nu }= 0$, because the arising irregular equation
$J_\mu = 0$ turns into identity, see (\ref{mm})
 [see also (\ref{mm4})]. This means that solutions with zero
  electromagnetic field are possible.

Is it possible to have solutions with zero current,
 $J_\mu =0$, but with
 $f_{\mu \nu}\neq0$\,? As $J \sim S f$ [see (\ref{mm})], one could
 consider possibility to add
 $S=0$. In generally, this equation, $ S_{\mu \nu \lambda }=0$,
can not be added because, in combination with
 identity (\ref{ides}), this gives rise to an irregular equation:
\[ S_{[\mu \nu \lambda ;\epsilon ]}\equiv
3/2 \Lambda _{a[\mu \nu }\Lambda ^{a}{}_{\lambda \epsilon ]}
 \to \  \Lambda
_{a[\mu \nu }\Lambda ^{a}{}_{\lambda \epsilon ]}= 0 \,. \]
 However, this irregular equation turns into identity in
  the presence of
`high' symmetry (say, spherical symmetry), when it is impossible
 (no means) to build a totally skew-symmetric forth order tensor.

In the case $D=3$, skew-symmetric equations (\ref{sybb})
 can be integrated; so, the requirement of field triviality
  at space infinity leads to the equality $S{=}0$:
 \[ (hS^{\mu\nu\lambda})_{,\lambda}=0;
 \ (hS^{012})_{,2}=0,  \ (hS^{120})_{,0}=0,\ldots,
  \to \ hS^{012}={\rm const}=0. \]
So, in the case $D=3$, all localized solutions have zero current,
and electromagnetic field seems to be more `independent'
 (with less interactions).

In the case $D=5$ (this choice of spacetime dimension will
 be substantiated in the next chapter devoted
 to singularities of solutions)
one can introduce a dual tensor\ $\varsigma _{\mu \nu
}=\varepsilon _{\mu \nu \alpha \beta \gamma }S_{\alpha \beta
\gamma }\,,$ which obeys
[as it follows from (\ref{ides}), (\ref{sybb})]
 the following Maxwell-like equations:
\begin{equation}
\label{dm} \varsigma_{[\mu\nu;\lambda]}=0\,; \ \
\varsigma_{\mu \nu ;\nu}=\Upsilon_\mu\,, \ \
\Upsilon_\mu =-\frac32 \varepsilon_{\mu abcd}
\Lambda_{pab}\Lambda_{pcd}\,.
\end{equation}

Given some symmetry (of a solution), one can find the number
 of essential, `important' vectors. Say, in the case of axial
 symmetry (in `usual' space; e.g., a rotating star or planet),
 there are three space-like vectors  (axial, radial, and along
 the extra dimension) and one time-like; therefore,
 the space-like component of
pseudo-current $\Upsilon _\mu $ [in contrast  to
the time-like one, see (\ref{dm})] can be non-zero. That is,
 the components $S_{0ij} \ \ (i, j=1,2,3)$ can `start' with
 a dipole term, while components $S_{ijk}$
 -- only with a quadrupole one.

The next additional equation,
\[ \Phi _{\mu }=0 \mbox{ \ or \ }
 \Phi _{a}=C_{a}={\rm const}, \]
can not be added to the system, (\ref{equ}) [or (\ref{sys4}],
because the trace part gives an irregular equations
 ($\Phi _{\mu }=0$ only for the trivial solution):
\[   {\bf D}_{aa}=\Lambda _{acd}\Lambda _{cda}
+ [3\sigma  + 2\sigma (1-D)]C_{a}C_{a} = 0 \
 \mbox{ [for (\ref{equ}) \ }
\sigma =1/3].   \]

At last, one should note the absence of solutions
 `without gravity field',
that is, of zero curvature. If one combine (\ref{equ}) with
equation
\[ R_{abcd}=0 \ \mbox{ (or \ } R_{ab}=0, \ \mbox{ or \ } R=0)\ ,\]
then, again, the combination of the trace equation and  $R=0$
 [see (\ref{sca})] leads to an irregular
 equation.\footnote{* The same is valid for the Weyl tensor,
 but the proof needs some handling of Bianchi identity.}

\section*{1.6 \ Energy-momentum tensor and conservation laws}

Of course, this paragraph pays special attention to the optimum
equation,   (\ref{equ}).
 Let us consider possible ways to define (or to find)
  an energy-momentum tensor.
Let $M$ is a covariant (of a single dimension --
 with respect to the scale transformations) composed from
 the metric
(or Riemann curvature). Then its variational derivative by the
 metric gives a symmetric  tensor  $D_{\mu \nu }$
  and an identity [see \cite{lan}]:
\begin{equation}
\label{ide} \delta(hM)/\delta g_{\mu \nu }= h D^{\mu \nu }\ \
 \ (h=\sqrt{-g}),
\ \ \ D_{\mu \nu ;\nu }\equiv0\,.
\end{equation}
Now, using the field equations, (\ref{equ}), one can try to
exclude  in $D_{\mu\nu }$ the `linear' terms
 ($\sim \epsilon$): \
 $ D_{\mu \nu }= T_{\mu \nu }(\epsilon ^{2}) $ \
 (so, `on
equations', there will be $T_{\mu \nu ;\nu }=0$); at last it is
necessary to check for weak field,
\[ h_{a\mu }=\eta _{am}+\epsilon _{a\mu }\,,
 \ \ \epsilon _{a\mu }\ll 1
\mbox{ \ (or \ \ } \epsilon \ll1), \] that the `energy' $
{\displaystyle\int} T_{00} (\epsilon^{2})\,dV$ is of constant sign.

 Thus, the simplest case,
\[    M^{(0)}=R \ \ \ (D^{(0)}_{\mu \nu }=
-G_{\mu \nu }= g_{\mu \nu }R/2 - R_{\mu \nu })\,,  \] does not
 result in a success (while at $\sigma =1 $ the linear terms cancel
 in (\ref{sys4}), but the  rest terms are not of constant-sign).

Therefore, one should   proceed to consider  quadratic invariants,
 $M\sim R^{2}$. Using the Bianchi identity, one can easily find
  all three independent tensors which lead to an identity:
\begin{equation}
\label{d1} D^{(1)}_{\mu \nu }=G_{\mu \nu ;\lambda ;\lambda }+
 G_{\epsilon \tau
}(2R_{\epsilon \mu \tau \nu } -
 \frac12g_{\mu \nu }R_{\epsilon \tau })\ \ (=T_{\mu
\nu } (\Lambda '^{2},\cdots));
\end{equation}
\begin{equation}
\label{d2} D^{(2)}_{\mu \nu }=R_{;\mu ;\nu }-
 g_{\mu \nu }R_{;\lambda ;\lambda
}- R_{\mu \nu } R+ \frac14g_{\mu \nu }R^{2};
\end{equation}
\begin{equation}
\label{d3} D^{\mu \nu }_{(3)}=(-g)^{-1}[\mu \nu ,\alpha \beta ,
 \gamma \delta
,\epsilon \tau ,\rho \varphi ]R_{\alpha \gamma
 \beta \delta } R_{\epsilon \rho
\tau \varphi }\ \ \ (D\geq5).
\end{equation}
 The expression with square brackets in (\ref{d3}) denotes
 a co-rank-5 minor of metric,
\[{\partial{}^{5} (-g)/} {\partial g_{\mu\nu }
 \partial g_{\alpha\beta}
 \partial g_{\gamma \delta} \partial g_{\epsilon\tau}
  \partial g_{\rho \varphi} }\,,\]
 which vanishes identically if $D\leq4$ (a `tricky' identity).

Since  $D^{(1)}_{\mu \mu }=(2-D/2)G_{\epsilon \tau }
 R_{\epsilon \tau } +
C^{1}_{\mu ;\mu }$, it is clear that
 $M^{(1)}= -G_{\epsilon \tau }R_{\epsilon \tau}$.

Taking into account ${\bf E}_{\mu \mu }$ and (\ref{sca}) one
 obtains (approximate equalities are exact up to terms
   $\sim \epsilon ^{2}$):
\[ R=r(\Lambda ^{2}),\ \ \ D^{(2)}_{\mu \nu }
\simeq(r(\eta _{\mu \tau }\eta _{\epsilon \nu }-
 \eta _{\mu \nu }\eta
_{\epsilon \tau }))_{,\epsilon \tau } =
 A^{(2)}_{\mu \epsilon \nu \tau
}(\epsilon ^{2})_{,\epsilon \tau }\ . \]
 Since $ D^{(3)}_{\mu \nu ,\nu
}\cong0$
 and already \ $ D^{(3)}_{\mu\nu}\sim \epsilon ^{2}$,
it is clear that the third case,
$D^{(3)}_{\mu \nu }$, also comes to
 the trivial form: $\ D^{(3)}_{\mu \nu }\simeq A^{(3)}_{\mu
\epsilon \nu \tau }(\epsilon ^{2}) _{,\epsilon \tau }\ $
 where tensor $A_{\mu \epsilon \nu
\tau }$ has the same symmetries as the Riemann curvature tensor,
 i.e., the same Young diagram; such terms
  add nothing to energy-momentum
  and angular momentum.

One can cancel `linear' terms in
 $D^{(1)}_{\mu \nu }$ and, then, using linearized field equations,
  extract `trivial terms' (second derivatives)
in such a way that to reduce  $ T_{\mu \nu }$ to $\Phi'^{2}$-terms;
 as a result, $T^{\mu \nu }$ from (\ref{d1}) can
 be written as follows:
\begin{equation}
\label{tmn} T_{\mu \nu }\simeq (\eta _{\mu \nu }f_{\epsilon \tau }
 f_{\epsilon
\tau }- 4f_{\mu \epsilon }f_{\nu \epsilon })/18 +
 A_{\mu \epsilon \nu \tau
}(\epsilon ^{2})_{,\epsilon \tau }\ .
\end{equation}
A simpler way to prove this result is through recasting
 the variational derivative
\[ \delta (hM^{(1)})/\delta h^{a}{}_{\mu }=
2hT_{a}{}^{\mu }\ \ \ (M^{(1)}=
 -G_{\epsilon \tau }R_{\epsilon \tau }), \]
or more exactly, the `Lagrangian' itself. One can remove in
 $M^{(1)}$ terms\\
1)  like  $C_{\mu ;\mu }$; their contribution in density
 $hM$ is  trivial, just
$(hC^{\mu})_{,\mu}$;\\
2)  cubic and quadratic  $\Lambda^{2}\Lambda ', \Lambda ^{4}$
 (invariant!), because their contribution
$(\sim \epsilon^{2})$ is trivial:
$\Delta (hT_{a}{}^{\mu })\sim B_{a}{}^{[\mu \nu ]}
{}_{,\nu }\,$ \
 [$\Delta T_{a\mu }$ has an identity similar to  (\ref{ide})];\\
3)  quadratic in equations  (i.e., like ${\bf E}^{2}$) because
variational derivation is linear with respect to differentiation
 by field.

In this sense there exists an equivalence:
\[ M^{(1)}= -G_{\epsilon \tau }R_{\epsilon \tau }
\approx 1/9 f_{\epsilon \tau }f^{\epsilon \tau } = M^{*}.\]
Variation (of $hM^{*}$)  by  $\Phi _{\epsilon ,\tau }$
 gives terms which
reduce [through application of the `Max\-well equation'
 (\ref{mm})]  to trivial terms
 $B_{a[\mu \nu] }(\epsilon ^{2})_{,\nu }$, while variation
by metric gives the electromagnetic contribution (\ref{tmn}).
 One can consider this as the existence of `approximate'
  (exact for the linearized equations)
 {\it weak\/} Lagrangian
(the term in the sense used by Ibragimov \cite{ibra}),
 which admits both electromagnetic and
gravitational
 form\footnote{* In fact, there exists an exact (!) although
 trivial Lagrangian (quadratic on the
field equations: ${\bf E}_{(ab)}^2 +\kappa{\bf E}_{aa}^2$)
 which looks as a `modified gravity':
$-M^{(1)} +M^* +C_{\mu;\mu}+(\Lambda'\Lambda^2, \Lambda^3)$; this
fact is of great importance for  constructing a `derived'
Lagrangian 4$D$-phenomenology
 of topological quanta.}
 [unlike approximate Lagrangian
$S_{\mu\nu\lambda}^{2}$ $-$ for Eq.~(\ref{equ})].

Following this way, one can show that further invariants,
   $M\sim R'^{2}, R''_{2}$
et cet.,  are equivalent to zero (that is, other approximate
 `weak Lagrangians' are
absent). So, there exist the energy-momentum tensor,
 $T_{\mu
\nu }\ \ (T_{\mu \nu ;\nu }= 0)$,
as well as approximate conservation law
 $T^{\mu \nu }{}_{,\nu }\simeq 0$.
Note the special importance of  component (tensor)
  $f_{\mu \nu }$: weak fields with $f_{\mu \nu }=0$
do not carry $D$-momentum and angular momentum.
 Perhaps, this
circumstance allows to reach
 $g_{\mu \nu }+f_{\mu \nu}$-unification
in a single geometrical structure
 (Lorentz structure of AP):
although second derivatives
  $h''$ corresponding to curvature
 ($R_{abcd}$) and electromagnetic field ($f_{ab}$) should
  be of the same scale
  (or even $f < R$), only $f$-component
transfers  $D$-momentum.

As was mentioned earlier (see
  \S1.2), if symbol $G_q$ is involutive, identities in the next
   orders ($>q$) hold straightforwardly;
that is, correspond to trivial conservation laws.
So,  trying to isolate the highest-order derivatives
in the LHS of (\ref{d1}),
\[ (-g)D_{(1)}^{\mu \nu }=\left( [(-g)(g^{\mu \nu }
 g^{\epsilon \tau }
-g^{\mu \tau }g^{\nu \epsilon })]_{,\alpha \beta }g^{\alpha \beta }
\right)_{,\epsilon \tau } + t^{\mu \nu }(g'g''',g''^{2})
 =(-g)T^{\mu \nu }\ ,\]
in an attempt to receive pseudo-tensor of energy-momentum
(see \cite{lan}), one can easily
check that this pseudo-tensor is trivial. This means, though, that
trivial terms in energy-momentum tensor (\ref{tmn}) can be
 chosen in such a way that
$T_{\mu \nu }$ [instead of (\ref{tmn})] will depend only on metric,
\[T^{\mu \nu }\simeq t^{\mu \nu }(g'g''',g''^{2})
+A^*_{\mu \epsilon \nu \tau ,\epsilon \tau }\,; \]
 this feature  stresses again the fact of unification of
 $g_{\mu \nu }$ and $f_{\mu \nu }$ in a single structure.

The low-order identities also give no nontrivial
 conservation laws for
eq.\,(\ref{equ}). Thus, skew-symmetric  equation (\ref{sybb})
itself has a form of `linear' conservation law:
  $h{\bf H}^{\mu \nu }=(hS^{\mu \nu \lambda
})_{,\lambda }$; however, the conserving integrals should vanish for
smooth initial data (of Cauchy problem)
localized on spatial infinity  (that is,  Fourier transforms
 are smooth functions too). One can rewrite the last
  equation in a different form:
\[{\bf H}_{a\mu }=0; \
S_{a\mu \nu ;\nu }= -\frac12 \Lambda_{aij}S_{ij\mu }
 \ (=I_{a\mu }); \] however, the  current arising here,
 $I_{a\mu ;\mu }=0$, is trivial.

Also trivial is the  electromagnetic current of Maxwell equation
 (\ref{mm}):
\[ hJ^{\mu }=(hS^{\mu \nu \lambda }\Phi _\lambda )_{,\nu }\ . \]
 On the one hand, this is a good feature as guaranteeing the
global electro-neutrality. And in general, the absence of
conservation laws for  system (\ref{equ})
means that there is no problem how  to choose conserving values
 (charges, a kind of `fundamental constants').
 On the other hand, that to make possible the
 existence of solutions
  with Coulomb-like asymptotic,
one should put forward some unusual, at the first sight,
 assumptions  about the extra dimension
 and mechanism of its reduction.
That is, one should assume that a typical scale
along this coordinate, $L$, is large enough,
 so that elementary charged particle-like
field configuration\footnote{The third chapter will deal with
 such non-linear configuration having a topological (quasi)charge.}
 look like a dipole (or a quadrupole) oriented
along the extra dimension:
 a charge of one sign concentrated near a
 `central region' is totally
 compensated by the opposite charge of (one of) the `fringe
regions'. That is, at large enough distances,  $r>L$,
  the Coulomb law should
 change into another multipole,
another power law. The experiments for testing Coulomb law
 \cite{oku} (including observations of
 the Jupiter's magnetic field,
\cite{qu1}) give an upper bound $L>3\cdot 10^{10}$\,cm.
 One should stress that this effect of the `huge extra dimension'
  is not equivalent to the existence of non-zero  photon mass:
 the latter leads to Yukawa potential which decreases faster
 than any power law.
Therefore, the existence of far-extended  ($\sim 10^{20}$\,cm)
 galactic magnetic fields
\cite{qu2}, perhaps, does not constrain $L$ (but, sure,
 constraint the photon mass)\footnote{I appreciate
P.V. Vorobjov for useful discussion of this issue.}.

This strange assumption about the huge extra dimension
can be a simple result (just the relativistic effect)
of an elementary cosmological scenario:
spherically-symmetric  ultra-relativistic
expansion (a kind of single wave running along the radius).
 The detailed discussion of a cosmological solution and its
  parameters (which could serve as `fundamental constants')
is delayed until the next chapters (see \S2.5, \S3.4).
At the same time, let's mention that symmetries of solutions
 is of great importance for their stability
 (and what could be simpler than the spherical symmetry\,!).
 Moreover, the presence of symmetries, i.e., Killing vectors
  (exact or almost exact), combined with the existence of
   energy-momentum tensor, leads to
 (almost exact)
conservation laws (of $D$-momentum): at the later stage of
expansion, when the spherical wave is  indistinguishable from a
plane wave, there is an approximate Killing vector relating to the
translation along the extra dimension (and the time).

\section*{1.7 \ Post-Newtonian effects}
This theory does not allow to cancel
 (by a coordinate transformation)
 in a point all terms $h'$ -- part of them is a tensor
  (of course,  $g'=0$ is possible); moreover, the theory can be
   formulated using an asymmetric connection (this is reflected
 in the name -- AP). Therefore, there is a natural question:
 in what way will a local perturbance (a short-period wave-packet)
  move on a given background solution?

For comparison sake, let us consider one-parameter class of
 equations (\ref{sys4}) (which admit solutions
with $f=0$); its  special case is the  {\sl
optimum\/} equation, (\ref{equ}). The skew-symmetric part
of (\ref{sys4}) and Maxwell equation [see also (\ref{max})]
look as follows:
\begin{equation}
\label{as} {\bf H}_{\mu \nu }=2{\bf E}_{[\mu \nu ]}=
 S_{\mu \nu \lambda ;\lambda
} + (1-3\sigma )(f_{\mu \nu }- S_{\mu \nu \lambda }
 \Phi _{\lambda })=0~;
\end{equation}
\begin{equation}
\label{ml} f_{\mu \nu ;\nu }=J_{\mu }; \ \ J_{\mu }
 =(S_{\mu \nu
\lambda } \Phi _{\lambda })_{;\nu }= -\frac12
 S_{\mu \nu \lambda }f_{\nu \lambda } + (1-3\sigma
)f_{\mu \nu }\Phi _{\nu }\,.
\end{equation}

Let's take a background solution  $h_{a\mu }(x)$ such that
$f_{\mu \nu }=0$ and $ S_{\mu \nu \lambda }=0$: this is possible
(and even necessary) in presence of high symmetry, e.g., spherical
symmetry
 ($\Phi _{\mu }=0$ only for trivial solutions, see \S1.5).

If there is given a field perturbation
$$\delta h_{a\mu }: h_{a\mu }\rightarrow
 h_{a\mu }+\delta h_{a\mu },$$
than Eq-n  (\ref{ml})
 leads to an equation for $\delta f_{\mu \nu }\  \
(\delta f^{\mu \nu }=g^{\mu \epsilon }g^{\nu \tau}
 \delta f_{\epsilon \tau}$, \
because the background part has \ $f=0)$:
\[
(\delta f^{\mu \nu })_{;\nu } =
 (1-3\sigma )\Phi _{\nu } \delta f^{\mu \nu } \
\ ((\delta f_{[\mu \nu })_{;\lambda ]}\equiv 0);
\]
that is, if $\sigma =1/3$  [and this is the case of {\it
optimum\/} equation (\ref{equ})], there is a perfect illusion that
an `independent' field $\delta f_{\mu \nu }$ is immersed into
 a background Riemann space and moves in the usual manner --
along a Riemannian  geodesic (as a matter of fact,
 all is united in
 single field $h^{a}{}_{\mu }$, and there is  more `stuff'
  there, besides metric, for example,
$\Phi_{\mu}$ and $S_{\mu\nu\lambda}$ ).
 This feature is of great importance because
just tensor $f_{\mu \nu }$ transfers energy;  one can relates the
`equivalence' (principle) with the universal
 (electromagnetic) structure of $T_{\mu\nu }$.
 If $\sigma \neq 1/3$, the motion of $\delta f_{\mu\nu}$
depends not only on metric, as well as the motion of component
 $\delta S_{\mu\nu\lambda}$ -- for any  $\sigma$ [see
(\ref{ides})].

Stars and planets are of almost spherical form, but their rotation
breaks spherical symmetry and should induce a field
 $\Delta S_{\mu \nu \lambda}$. In turn,
$\Delta S_{\mu\nu\lambda}$ can give rise to a non-zero
 field $\Delta
f_{\mu \nu }$, see (\ref{ml}) (seemingly of any sing!). This can
give a contribution to the magnetic fields of rotating
 stars and planets  (there are different opinions about
 the origin
  of these fields
\cite{ei1,dol,run}).

It is naturally to expect that, say, in the Shapiro's experiment,
field $\Delta S_{\mu\nu\lambda}$ has an influence only on the
polarization of
 $\delta f_{\mu\nu}$-packet, because tensor
  $\Delta S_{\mu \nu \lambda}$ is unable to enter the
  eikonal equation
$$\psi _{;\mu }\psi ^{;\mu }=0\ \ (\delta  f\sim a_{0}e^{i\psi }).$$
Thus,  if $\sigma =1/3$ (i.e., the {\it optimum\/} equations), an
electromagnetic wave-packet moves along a Riemannian geodesic, and
the classical (independent on a polarization or spin)
post-Newtonian corrections are the
 same as in the case of GR  (if the Newton law is possible).
 In other cases, $\sigma \neq1/3$, the eikonal equation
contains additional term:  $\Phi ^{\mu }\psi _{;\mu }$, see
(\ref{ml}).

Let's consider a plane wave $\delta f_{\mu \nu }= a_{\mu \nu }e^{i
k_\lambda x^{\lambda }}$ on a background with $S$-field with
non-vanishing component
\[ S_{0ij}=\varepsilon _{ijk} \alpha _{,k}, \ \ \alpha_{i,i}=0. \]
 Here $i,j=1,2,3$; equations  (\ref{ides}), (\ref{sybb}) are taken
  into account. Component
$S_{ijk}$ can `start' with a quadrupole term so it can be neglected.
 The simplest solution for $\alpha $ is the dipole solution
 $\alpha =\Omega _{i}x^{i}/r^{3}$, where $\Omega _{i}$ is
a (pseudo)vector similar, i.e. proportional to the rotation
vector.

In the first approximation,  $S$-field contribution does not
 depend on the small non-Euclidean metric deviation,
 so, assuming that
metric is trivial, $S$-field is slowly varying,
and taking
\[ k_{\mu }=(-\omega,0,0,k) , \ \ a_{0i}=e_{i}, \ \ a_{ij}=
\varepsilon _{ijk}h_{k},\] one can obtain the next equations
 for the plane wave amplitudes [see (\ref{mm}), (\ref{ml})]:
\[   h_i= \frac k \omega \varepsilon _{3ij}e_j,
\ \ \ \omega e_i+k\varepsilon _{3ij}h_j+
 i\varepsilon _{ijk}e_j \alpha
_{,k}=0.\]
Excluding $h_i$ and neglecting (in the first approximation)
 amplitude $e_3$
[$e_3\sim O(\alpha )$], one can obtain
\[ \left(
\begin{array}{cc}
\omega^2-k^2 & -i\omega \alpha _{,3}\\
i\omega \alpha _{,3}& \omega^{2}-k^2
\end{array}       \right)  \left(
\begin{array}{c}
e_1 \\ e_2
\end{array}                \right) =0 \,; \]
this results in a dispersion relation and its solution:
\[ (\omega^{2}-k^2)^2 =\omega^2\alpha _{,3}^2 \ ,
\ \ \ k_{\pm}=\omega \pm \frac12 \alpha _{,3} \ . \]
It is easy to check that the eigenvectors correspond to left
 and right polarizations.
 The difference of
 $k_+$ and $k_-$ means that the plane of polarization of a
plane-polarized wave will rotate,
 \ $d \varphi = \alpha _{,i}k_i/k\,dl $, the full rotation
  angle depends only on the initial and end point:
 \ $\Delta \varphi =\alpha _{f} -\alpha _{i}$.

The exact phenomenology of  massive matter
 (and its interaction with $S_{\mu \nu
 \lambda }$) still is not clear in framework of AP (the extra
 dimension is of true importance), so, unfortunately, it is
  quite difficult to give some qualitative estimations.

\chapter*{Chapter 2. \ Singularities of solutions}
\addtocounter{chapter}{1} \setcounter{equation}{0}

Imagine that there is available a super computer, and a nearby
 student (who has no enthusiasm
about supersymmetry) would like to calculate solutions to
 an `appropriate' system of field equations.
Appropriate to him  are those equations which are:

\begin{itemize} \itemsep -2.mm
\item first, of hyperbolic type
(i.e, Lorentz invariance should exist
in the weak field approximation;
yes,  and the dimensionality of space, $D$, should be not less than
 $1+3$);
\item second, nonlinear (in many cases superposition is
 experimentally lacking; linearity is too simple);
 \item and, in conclusion, that equations should be self-consistent;
that is, solutions to Cauchy problem should have no limits in time
(relating to appearance of singularities or many-valuedness of
 solution), excepting, perhaps, only very rare, of zero measure,
 solutions (or variants of initial data for Cauchy problem).
\end{itemize}              \vspace{-2.5mm}
Arguing for the last requirement, that fella is ready to use
 an analogy from theology:
 `Lord is not able to restrict His power, respectability and
competence'\footnote{By the way, the custom to {\it throw dice}
 means evident absence of competence
(Lord is `not Copenhagen').

 {*} In informal Russian, to be not
Copenhagen means to be incompetent.}

As a rule, solutions of nonlinear equations (PDEs) can
become many-valued  \cite{vino}. The other name of this
 phenomenon is the {\it gradient catastrophe}, which is more
  familiar in the area of gas dynamics equations \cite{rozh}.
In the case of generally covariant equations (if it is possible
 to rewrite equations in such a way)
 one can obviate  bifurcation through a many-valued coordinate
  transformation, but instead of
`points of a catastrophe', there arise singularities of
 co-metric (and/or co-frame) field, i.e.,
co-singularities, where $g_{\mu\nu}$ (and $h^a{}_\mu$) is
 a degenerate matrix.

The other sort of singularities concerns contra-singularities,
 with degeneration of
 (contra-variant) frame $h_a{}^\mu{}$ ($g^{\mu \nu
}$), or, more generally, frame density of some weight.

This chapter deals with formal solutions (solution jets)
which grow `from a singularity' or infinitely close to that;
 the necessary conditions of existence
 of `singular jets' are to be discussed.
 The main purpose here is to find
 `singularity-free' AP equations which solutions
of {\it general position\/} have no arising singularities.

\section*{2.1 \ Gradient catastrophe and co-singularities}
The first problem which could damage a non-linear (PDE) equation
 is the
problem of gradient catastrophe, i.e., possibility of many-valued
 solutions,
 development of many-valuedness. Even the simple equation
 (a model equation of gas dynamics\,\cite{vino})
\begin{equation}
\label{gd}
\frac{\partial u }{ \partial x^0}=u\frac{\partial u}{\partial x^1}
\end{equation}
  \\[1.mm]
\parbox[b]{8.1cm}{
has solutions $u(x^0,x^1)$ which formally
 become many-valued (and
the first de\-ri\-vatives
 diverge in some points), see fig.\ 1:
$u(x^0,x^1)$ is a section of trivial fiber bundle
 $ (u;x^0,x^1)$; its projection on the base  $(x^0,x^1)$
has a cusp singularity (cusp catastrophe) \cite{arno}.
}\,\hspace{\parindent}\parbox[b]{6.1cm}{\vspace{-1.6cm}
 \includegraphics[bb=0 0 160 117,scale=1]{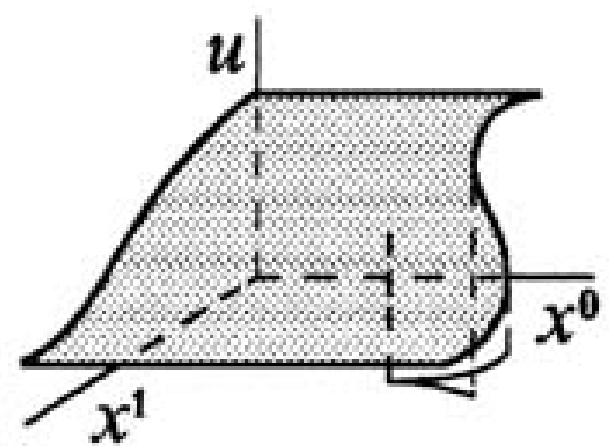}
 \vspace{-.25cm}
 {\normalsize Fig.\ 1. \
Development of many-\\[-1mm]
valuedness in solution $u(x,t)$.}}

Many-valuedness and infinite derivatives are physically
 not appropriate.
As a way out, additional information
 (free path, Hugoniot conditions, that
are not contained in gas dynamics equations) takes  effect,
whereas equations themselves acquire a  status of
 `phenomenology with
limited responsibility'.

The vacuum GR equation is seemingly  something greater
 than a mere phenomenology.
However, at explicit conditions, GR solutions inevitably
 becomes singular
 [3,7--10] (some spacetime points are singular).
This also gives rise a question on how these singularities
 relate to the gradient catastrophe.

Let there is a many-valued solution (local one, or a germ)
$g(x)$, and let's choose new coordinates `along the solution
 section' (Fig.~1), i.e., in such a way that
$g^*(y)$ is a  single-valued mapping. But for all that,
 $x(y)$ is a singular mapping: the metric in the new coordinate
system
\[ g^*_{\mu \nu}(y)=g_{\epsilon \tau}\,
\frac{\partial x^\epsilon}{\partial y^\mu}\,
 \frac{\partial x^\tau}{\partial y^\nu}
\]
is degenerate in those points with a fold where matrix
$\partial x/\partial y$ is degenerate.
Singularities of this kind concerns  elements of covariant
matrix $g_{\mu \nu}$ as independent variables (while this
matrix degenerates), so one can call them `co-singularities'.

Such a solution,
 $$g^*(y)\,, \, {\rm rank}\, g^*_{\mu\nu}(y_0)<D\,,$$
is possible if, firstly, the symbol of equations remains
 {\it
involutive\/} at degeneration of $g_{\mu \nu}$
(that is, equations themselves and their prolongations should
 keep the highest-order terms, highest derivatives
 \cite{vino,pomm,z5}). Secondly, one has to know
whether  the other terms (non-principal terms) diverge or stay
finite at co-metric degeneration.
The first case will be referred as `strong' singularities,
 while the second case -- as `weak' ones, which do not break
  solution off.

Following this point of view, let's rewrite equation
 (\ref{gd}) in a generally covariant form:
\begin{equation}
\label{gdc} \left\{
\begin{array}{rcl}
u_{,\mu}c^a(u)h_a{}^\mu h&{=}& 0
\ \ (c^a=(1,-u),\ h=\det h^a{}_\mu),\\
2h^a{}_{[\mu,\nu]} &{=}&0 \ \ (=h^a{}_{\mu,\nu}-h^a{}_{\nu,\mu}).
\end{array} \right.
\end{equation}
Here
 $h^a{}_\mu \ (h^a{}_\mu h_b{}^\mu=\delta^a_b)$
is a 2-frame field.

The condition of spacetime triviality
 [i.e., flatness: see the second equation in
  (\ref{gdc})] still is regular for degenerate
 matrices $h^a{}_\mu$.
Co-rank one minors of matrix
$h^a{}_\mu$,
$$ hh_a{}^\mu=\frac{\partial
h}{\partial h^a{}_\mu}\,, $$
 keep finiteness,
and the regularity of  (\ref{gdc}) preserves when
 ${\rm rank}\,h^a{}_\mu =1$,
 excepting the case $c^ah\,h_a{}^\mu =0$.
Therefore, system (\ref{gdc}) has formal solutions with a
degenerate matrix
 $h^a{}_\mu(x_0)$ as an initial term of series.
Such a singularity in  point
  $x_0$ is physical, because invariants
 $u_{,\mu }h_a{}^\mu $ come into infinity.
Of course, the possibility of formal solutions with
co-singularity, being a necessary criterion, does not yet
guarantee the existence of a global solution complying with a
certain boundary condition.

Co-singularities exist also in GR solutions, although look
 some more complex.
Using rank $D-3$ minors of metric
  $g_{\mu \nu }$
 \[ [\mu \nu ,\varepsilon
\tau ,\alpha \beta ] \equiv \frac
 {\partial^3 (-g)}{\partial g_{\mu \nu }
\partial g_{\varepsilon \tau }\partial g_{\alpha \beta }}, \]
and taking into consideration their symmetry features,
one can rewrite the (vacuum) equations of General Relativity
as follows:
\begin{equation}
\label{o1} g(g^{\mu \nu }R-2R^{\mu \nu })= [\mu \nu ,\varepsilon
 \tau , \alpha
\beta ](g_{\alpha \beta ,\varepsilon \tau }+
 g^{\rho \phi }\Gamma _{\rho
,\varepsilon \tau } \Gamma _{\phi ,\alpha \beta })=0\,.
\end{equation}
The symbol of this equation (highest-order derivatives) is regular
 and involutive if
${\rm rank}\, g_{\mu\nu}\geq D-1$, but the minor terms diverge at
degeneration of metric
 $g_{\mu \nu }$.
Therefore, a point of co-singularity itself, seemingly, is
 unachievable, but can be approached infinitely near.
That is, co-singularities in GR are `strong' in contrast to
`weak' co-singularities of system (\ref{gdc}).

A general GR solution in vicinity of singularity is produced in
 \cite{blh}; it has complex, oscillating behavior,
due to divergence of the minor terms. One should emphasize that the
 presence of symmetry
can permit additional types of singularities (e.g., co-singularities
 of co-rank greater than unity).

In the next paragraph, we will find the AP equations which do not
admit
 co-singu\-lari\-ties, because their
symbol does not keep involutive at degeneration of co-frame
 $h^a{}_\mu$.

Absolute Parallelism, just as General Relativity, is remarkable
 for the following important
feature:
 without making differentiation of field components
 (frame field $h^a{}_\mu$), i.e., using only the zeroth jets,
it is impossible to make an invariant, a scalar
 field;\footnote{This is not the case
for bi-metric theories or a theory of non-symmetric metric.}
 it is not possible to distinguish one (non-singular) point from
  another without coming to their
neighborhoods. The symbol of AP equations is defined
 just by  zeroth
jets, so the all
 variety of singularities in AP
is concerned with the single parameter, their rank:
 ${\rm rank\,}h^a{}_\mu$
  (nevertheless, the other parameter,
 if different from the
first one, is metric (co-)rank
 ${\rm rank\,}g_{\mu\nu}$).

\section*{2.2  \ Field equations\\ \hspace*{11mm} free from
co-singularities of solutions}

Let's try to rewrite AP equations
  (\ref{sya}), (\ref{syb})
in such a way that the coefficients at the second derivatives
 of co-frame field
$h^a{}_{\mu,\nu \lambda }$ (which define the symbol)
would be expressed through  minors (of matrix
$h^a{}_\mu$) of co-rank as large as possible.

We use the following notation for
 $k$-minors (i.e., minors of rank  $D-k$)
 of matrix $h^{a}{}_{\mu }$ and
$g_{\mu \nu }$, respectively:
$$ \pmatrix{\mu_{1}\;\cdots\; \mu_{k}
\vspace{-3.5mm}\cr a_{1}\;\cdots\;a_{k}}
 =\frac{\partial^k h}{\partial
h^{a_1}{}_{\mu_1}\cdots
\partial h^{a_{k}}{}_{\mu _{k}}}
= k!\, h h^{\mu _{1}}_{[a_{1}}\cdots
 h^{\mu _{k}}_{a_{k}]} \ ; $$
\[
[\vspace{1.5mm}\mu _{1}\nu _{1},\ldots ,\mu _{k}\nu _{k}
 \vspace{-1.5mm}] =
\frac{\partial^{k}(-g)}{\partial g_{\mu _{1}\nu _{1}}
\cdots \partial g_{\mu _{k}\nu _{k}}}= {1\over k!}
 \pmatrix{\mu_{1}
\cdots \mu_{k} \vspace{-3.5mm}\cr a_{1}\cdots a_{k}}
 \pmatrix{\nu_1\cdots \nu_k
\vspace{-3.5mm}\cr a_1\cdots a_k}.
\]
These minors, just as the determinant, are `polylinear'
expressions with
 respect to the elements of
 $h^{a}{}_{\mu }$ (or $g_{\mu \nu }$),
 and they are skew-symmetric
expressions in both the `row indices' and the `column indices'
 (taken separately, of course).

It turns out, that (almost) any system of  AP equations,
(\ref{sya}), (\ref{syb}), can be rewritten in such a way that
 the coefficients at the second-order derivatives,
 $h_{a\mu ,\nu \lambda }$, are quadratic in
 2-minors of matrix $h^{a}{}_{\mu }$. For example,
the case $\sigma =0$, $ \tau =1$, i.e., a system similar to
(\ref{ex}), easily leads to such a form:
\begin{equation}
 \label{ex1}
 h^{2}{\bf E}_{a}{}^{\mu }=
[h_{a\alpha ,\beta \nu }- (\alpha \beta )]
 (-g)g^{\alpha \mu}g^{\beta
\nu}+\cdots
= h_{a\alpha ,\beta \nu }[\vspace{1mm}
 \alpha \mu ,\beta \nu \vspace{-1mm}]
+(h^\prime{}^{2})\ .
\end{equation}
A finite degenerate matrix $h^a{}_\mu$
 of rank $D-1$ can be reduced,
 by means of a `left-right' transformation  (\ref{lrsy}),
  to the diagonal form $h^{a}{}_{\mu }
  ={\rm diag}(1,\ldots ,1,0)$.
Then one can check that the leading terms of system (\ref{ex1})
are regular (and the symbol is involutive) if
 ${\rm rank}\,h^{a}{}_{\mu
}=D-1$; if ${\rm\, rank}\,h^a{}_\mu\le D-2$,
 some equations of
(\ref{ex}) [or system (\ref{sya}), (\ref{syb})]
 loose their leading  terms
 (with the second derivatives).
Co-rank one matrices
$h^a{}_\mu$ split into three cases: $L$-vector
$C_a$, annihilating $h^a{}_\mu$ (i.e., $C_ah^a{}_\mu=0$),
 can be space-,
 time-, or light-like depending on the value of
 $C_aC_b\eta ^{ab}$; only the first case correspond to {\it
arising\/} singularity. In the last case, (co-)metric
$g_{\mu \nu }$ has rank two, and $h^a{}_\mu$ can be
reduced to other `canonical' form
 (non-diagonal); in this case regularity of
equations (\ref{ex1}) is lacking (i.e., these points
of jet space $J_2({\cal E})$ (${\cal E}=Mat(D\cdot D)$)
have to be removed from the `system set'
 ${\cal R}_{2}$).

A systematic approach to analysis of an arbitrary AP system
consists in the following.
Generally, an equation similar to
  (\ref{ex1}) could contain some other combination of symmetric,
 anti-symmetric, and `trace' equation, therefore it is necessary
to consider the leading terms of the most general form:
\begin{equation}
\label{ea} {\bf\tilde{E}}_{a}{}^{\mu }=(a_{1}{\bf G}_{ab}
+a_{2}{\bf H}_{ba} + a_{3}\eta_{ab}
 {\bf G}_{cc})H_{b}{}^{\mu}H^{2p}=0;
\end{equation}
here, for more generality, we use the  substitution
 $h^{a}_{\mu }=H^{p}H^{a}_{\mu }$
(co-frame density of some weight as the set
 of independent variables).

Assigning
 $$H^{a}{}_{\mu }={\rm diag\,}(1,\ldots ,1,\delta )$$
 and cancelling terms
 $\sim 1/\delta ^{4},1/\delta^{3}$ in the leading derivatives
 (such terms arise not in all components of the system,
 so they should be
  cancelled: if one takes them as the `main' terms and
eliminate their divergence through multiplication on determinant
 $h$ to some power, then the equations
  will be loosing their regularity
at degeneration of co-frame $h^a{}_\mu$), one can find
 the free parameters
(the overall coefficient for $a_i$ does not matter):
\[a_{1}=\tau ,\ \  a_{2}=1-\sigma,\ \ a_{3}=0;
\ \ p=0\] (if $\tau =0$, then  $\,p\,$ is indefinite).

 Now one can write
the equations
 (their leading part, the highest derivatives)
in the desired form:
\[ h^{2}{\bf \tilde{E}}_{a}{}^{\mu}= 2h_{b\alpha ,\beta \nu }
\left\{ (\tau +\sigma)\eta_{ab}[\alpha\mu,\beta \nu ]
 -\sigma \pmatrix{\alpha\;
\beta\vspace{-3.5mm}\cr a\; c} \pmatrix{\mu\;
\nu\vspace{-3.5mm}\cr b\;c}- \right. \]
\begin{equation}
\label{ea1} \ \ \ \ \ \ \
 \left. -\sigma\tau \pmatrix{\alpha\;
\beta\vspace{-3.5mm}\cr b\;c}
 \pmatrix{\mu\; \nu\vspace{-3.5mm}\cr
a\;c} \right\} +(h'^2)\ .
\end{equation}
Remarkably, the required combination of equations turns out
 to be the
very same combination which enters the leading part
 (with the highest
derivatives) of the identity, which ensures the system's
compatibility; see, e.g., the identities for systems
 (\ref{sys1}), (\ref{sys2}), (\ref{sys3}), (\ref{sys4}).

The symbol of system
 (\ref{ea1}) is involutive if
\[ {\rm rank\,}h^a{}_\mu,
 {\rm rank\,}g_{\mu \nu }\ge D-1\ , \]
excepting the cases $\sigma =1$ or $\tau =0$, when equation
(\ref{ea}) [and (\ref{ea1})] contains, respectively, only
symmetric or only antisymmetric part of equations;
 these cases requires an additional
analysis.

First, in both these cases, the two  indexes of equation are
evidently equivalent, equal in rights; so, one should consider
equation
  $h^{2}{\bf \tilde{E}}^{\mu\nu }=0$ and to note that
 3-minors appear in the leading terms.
 For the case $\sigma =1$, equation (\ref{ea1}) gives
\begin{equation}
\label{eg1}
 h^{2}{\bf \tilde{E}}^{\mu \nu }=
\frac\tau 2\left\{ \pmatrix{\mu\ \lambda \vspace{-3.5mm} \cr c\ d}
\pmatrix{\nu\ \alpha\ \beta \vspace{-3.5mm}\cr b\ c\ d}
 +\pmatrix{\nu\ \lambda
\vspace{-3.5mm}\cr c\ d} \pmatrix{\mu\ \alpha\ \beta\vspace{-3.5mm}
 \cr b\ c\
d}\right\}h_{b\alpha ,\beta \lambda } + (h^{\prime 2})\ .
\end{equation}
The symmetric equation, ${\bf G}_{(\mu \nu )}=0$ (\ref{sya}),
 in its leading terms,
 coincides at $\sigma =1$ with the GR equation:
one can compare the previous equation, (\ref{eg1}),\ and the
leading terms of GR equation (\ref{o1}).

The antisymmetric part can be written in other form,
 in other indexes, than symmetric
one. 
 This time 1-minors do inevitably arise;
 nevertheless, the  equation
\begin{equation}
\label{eh1} h^{2}{\bf H}_{ab}= h_{c\mu ,\nu \lambda }
\pmatrix{\lambda\vspace{-3.5mm}\cr d} \left\{
\eta_{ac}\pmatrix{\mu\ \nu\vspace{-3.5mm}\cr b\ d}-\eta_{bc}
\pmatrix{\mu\ \nu\vspace{-3.5mm}\cr a\ d}+(1-\tau
)\eta_{cd}\pmatrix{\mu\ \nu\vspace{-3.5mm}\cr a\ b} \right\} +
(h^{\prime 2})=0\,
\end{equation}
(or its symbol), as well as the system  (\ref{eg1}), (\ref{eh1})
as whole, is regular (compatible) if
 ${\rm rank\,}g_{\mu \nu }\ge D-1$.
 Counting equations (in the leading terms),
 one can use the signature
 $\delta _{ab}$ and
 the indexes  $a,\mu =1,\ldots ,D$ as more convenient.
Choosing (here $D$ serves as a fixed index value)
\[h^{a}{}_{\mu}=\delta_{a\mu}-\delta_{Da}
\delta_{D\mu},\ \ \ \pmatrix{\mu\vspace{-3.5mm}\cr a}=
\delta_{Da}\delta_{D\mu } \mbox{ \ et cet.}\,,\]
 one can easily
check that the highest-order derivatives of
 $iD$-component ($i<D$) in (\ref{eh1}) do not vanish,
 and do not coincide
with analogous terms in  (\ref{eg1}).

It remains to consider the most interesting case:
 $\tau =0$, which has
a free parameter $p$.
It is easy to rewrite the antisymmetric part
[see (\ref{ea1}); minors here relates to the matrix
 $H^{a}{}_{\mu }$],
\[H^{4p+2}{\bf H}^{\mu\nu}= \frac12\pmatrix{\alpha\ \beta
\vspace{-3.5mm}\cr b\ c} \pmatrix{\mu\ \nu\
 \lambda\vspace{-3.5mm}\cr a\ b\
c}H_{a\alpha,\beta\lambda }+ (H^{\prime 2})\ .  \]
One can choose
  $p$ in such a way that the lower terms of this equation
   will be also finite
at degeneration of  $H^a{}_\mu$. However, it turns impossible to
write the symmetric equation (or separately the trace equation and
traceless symmetric part) in such a way that to keep the system
regular
  (and the symbol involutive) for degenerate co-frame density,
  ${\rm rank\,}H^a{}_\mu=D-1$:
  the symmetrical part has too many components
(as compared with the skew-symmetric one), while 1-minors vanish
too simultaneously.

Thus, the AP equations with
 $\tau =0$, $\sigma\neq1$ are of special interest
  because their solutions
are free of co-singularities and, hence, gradient catastrophe.
Just this case was specially considered in
 \S 1.4, and only two compatible systems
(with $\tau=0$) were found.
The first variant (system) is the
  {\it optimum\/} equation
(\ref{equ}); denote it $ Eq.{\cal A} $
(equation ${\cal A}$); the spacetime dimension
 $D$ is not fixed here:
\[ Eq.{\cal A} \ \to\ \ (\ref{equ}),
\ \ \ \ \tau =0, \ \sigma =\frac13, \ D\neq 4 \ (D_0=4); \]
the second variant
(equation ${\cal B}$) is a particular case (with $\tau =0$)
of the two-parameter class
 {\bf C} (\ref{sys3});  the spacetime dimension is fixed here while
parameter $\sigma $
($\sigma \neq1$) remains free:
\[ Eq.{\cal B} \ \to\ \ (\ref{sys3}),
\ \ \ \ \tau =0, \  \sigma \neq\frac13, \ D= 4
 \ (D_0=1+\sigma ^{-1}). \]
 The variant  $Eq.{\cal A}$ is more preferable (from the
 experimental point of view)
  because `electrically charged' solutions
are possible; nevertheless it would be desirable to give
 {\em a priori\/} reasons, motivations of this choice, as well as to
`calculate'  the spacetime dimension $D$.

Notice in this connection that the {\em supplemented\/} equations
{\bf C} and {\bf D} (when equation $f_{\mu \nu }=0$ is added)
admit co-singularities at any  $\tau $ (the symbol $G_2$ is
involutive if
 ${\rm rank\,}h^a{}_\mu\ge n-1$),
 that is, $Eq.{\cal
A}$ and $Eq.{\cal B}$  loose their {\it singularity-free\/}
property if $f=0$.

A {\it strict localization\/} condition
 (`island' solutions; as well as
the smoothness condition,
$h(x)\in C^{\infty}$) could be imposed on initial data.

If $\tau\neq 0$,  a system can be rewritten as [see for instance
(\ref{sys2})]
$$K_{a[\mu\nu];\nu}=I_{a\mu}(\Lambda^{2}),$$
  which gives a non-trivial conservation law
$I_{a\mu;\mu}=0$; then, such a strict localization is impossible:
there will a `long', Newton-like asymptotic behavior.

 In the case of
 $Eq.{\cal B}$, the absence of the current in
  `Maxwell equation' ($f_{\mu
\nu ;\nu }{=}0$) gives rise to a guess-work that a general
strictly localized field configuration can evolve to provide in a
finite while a region with $f_{\mu \nu }=0$. This observation
perhaps enables one to discard the variant
 $Eq.{\cal B}$, as well as the case $D=3$
for $Eq.{\cal A}$:  integration of equation
 $(hS^{\mu\nu\lambda}){}_{,\lambda}=0$
gives (the current vanishes)
\[ S_{012 }=0 \ \ \Rightarrow \ J_{\mu }=0\ . \]

In the following paragraphs, another type of singularities,
 co-singularities will be
considered for
 $Eq.{\cal A}$; the requirement of absence of these
  singularities in solutions
of general position can serve as a good reason to make
 the choice of spacetime dimension;
at the moment, in one way or another, the variants
 $D\leq4$ are discarded.


\section*{2.3 \ Contra-singularities}
Contra-singularities of GR equation concern with the choice of
tensor  density
$$G^{\mu \nu
}=\sqrt{-g}g^{\mu \nu },$$
 but not $g^{\mu \nu }$,
as  a new set of dependent variables.
As is well known \cite{lan}, this equation can written as follows
 [see also (\ref{o1})]:
\begin{equation}
\label{o2} (G^{\mu \varepsilon  }G^{\nu \tau }-
 G^{\mu \nu }G^{\varepsilon \tau
})_{,\varepsilon \tau }+ 2(-g)t^{\mu \nu }(G^{\prime 2})=0\,.
\end{equation}
Using the prescriptions of compatibility theory, which were
 discussed in
  \S~1.2, one can prove that the symbol of equation
   (\ref{o2}) is involutive if
 ${\rm rank\,}G^{\mu\nu}\geq 2$.
As it is shown in \cite{lan} (see Eq.~96.9 for $(-g)t^{\mu \nu
})$, the terms quadratic in the first derivatives still contain
low indexes matrix $G_{\mu \nu }$, that is, these (lower) terms
diverge at degeneration of matrix   $G^{\mu \nu }$
(contra-singularities of GR are `strong' too).

In the case of AP equations, there is an analogous
 change of variables:
\begin{equation}
\label{ch} H_a{}^\mu{}=h^{\frac\sigma {\sigma +1}}
 h_a{}^\mu{} \, ,
\ \ h_{a}{}^{\mu } =H^{-p} H_{a}{}^{\mu }\, ;
\end{equation}
\[ \mbox{ here \  }  p =(D_0- D)^{-1},
\ \ D_0=1 + \frac1 \sigma; \ \
 \ H = \det  H^{a}{}_{\mu } \,. \]
Substituting (\ref{ch}) into (\ref{gala}), one obtains
\begin{equation}
\label{lap} \Lambda_{abc}=2H^{-p}\left(
 H_{a\mu }H_{[b}{}^{\nu } H_{c]}{}^{\mu
}{}_{,\nu } +p\eta _{a[b}H_{c]}{}^{\nu }A_{,\nu }\right),
\end{equation}
\[ \Phi _{c}= H^{-p}\left( H_{c}{}^{\nu}{}_{,\nu}
+{p\over \sigma}A_{,\nu }H_{c}{}^{\nu }\right),
 \mbox{ \ here \ } A = \ln  H
\, . \]

As it turns out
\cite{z3,z4,zz3},  an arbitrary AP system can be rewritten
 in such a way that
 the leading part (with the highest-order derivatives)
  would contain only matrix
 $H_{a}{}^{\mu }$ (and its second derivatives) but not
 the inverse matrix,
  $H^{a}{}_{\mu }$ ($H^{a}_{\mu }H^{\nu }_{a}=\delta ^{\nu
}_{\mu}$;
 i.e., the coefficients at the main derivatives \ $H^{\mu
}_{a ,\nu \lambda }$ are quadratic in  $H_{a}{}^{\mu }$):
\[H^{3p}({\bf G}^{\mu }{}_{b}+{\bf H}^{\mu }{}_{b})/2=
H^{2p} H^{\mu }_{a}(\Lambda _{abc,c}+\alpha\Phi_{a,b}+
 \beta\Phi_{b,a}-\sigma
\eta_{ab}\Phi_{c,c}+(\Lambda ^{2}))=\]
\begin{equation}
\label{egh} =(H^{\mu }_{c,\nu \lambda }
 H^{\nu }_{b}-H^{\mu}_{b,\nu\lambda}
H^{\nu }_{c}+\beta H^{\nu }_{b,\nu \lambda}H^{\mu }_{c})
 H^{\lambda }_{c}
+H^{\nu}_{c,\nu\lambda}(\alpha H^{\mu }_{c}
 H^{\lambda }_{b} -\sigma H^{\lambda
}_{c}H^{\mu}_{b}) +
\end{equation}
\[+\Pi^{\mu}{}_{b}(H^{\prime 2})=0\, ; \ \ \mbox{
here \ } \alpha ={\sigma +\tau-1\over 2}, \ \
 \beta =\sigma -\alpha . \]
 It is necessary to emphasize that in this case
 (contra-variant frame density as the independent variables)
 the  `right' combination includes in equal
 proportions both symmetric and
 antisymmetric part of
 equations, and the choice of  coefficient $p$ (the weight
 of the frame density) helps
 to cancel the terms like
$H^a{}_\mu H_a{}^\mu{}_{,\nu \lambda }$.
 This change of variables, $h^a{}_\mu\to
H^a{}_\mu$ (\ref{ch}),  ``repairs''  the system's non-regularity
(in the trace equation) at  $D=D_0$ (the forbidden dimension -- if
$D_0$ is an integer), while it is irreversible in this case.

Quadratic terms $V_{\mu \nu }(\Lambda ^{2})$
 of antisymmetric equation
(\ref{systb})
may include the next three terms:
\[ V_{\mu \nu }=a_1 S_{\mu \nu \lambda }\Phi _{\lambda }+
a_2 \Lambda _{\lambda \mu \nu }\Phi _{\lambda }+
 a_3 (\Lambda _{\mu \epsilon
\tau } \Lambda _{\epsilon \tau \nu }-
 \Lambda _{\nu \epsilon \tau } \Lambda
_{\epsilon \tau \mu }). \]
 Multiplying (\ref{egh}) on $H^{\nu}_{b}$, and
 examining at which restrictions
 the lower terms,
$\ \Pi^{[\mu }{}_{b} H^{\nu] }_{b}\,, $ include only the matrix
$H^{\lambda }_{c}$ and its first derivatives (have a four-linear
form), one can find the following necessary conditions:
\[ \tau=1-3\sigma =-a_{1},
\ \ a_{2}=a_{3}=0. \]
 Among the compatible AP equations considered
 in   \S~1.3, only the one-parameter class
${\bf D}$ (\ref{sys4}) (class ${\bf A}$
(\ref{sys1}) intersects
class ${\bf D}$ at the point $\sigma
=-\tau =1/2$) fulfills  these `conditions of poly-linearity';
and the equations of this class take on a
 3-linear form [in analogy with
Hirota's bilinear equations; see  (\ref{egh})]
\begin{equation}
\label{egh3} \left(N^{\mu }_{ab}H^{\nu }_{b}
 -2H^{\nu }_{[a,b]}H^{\mu
}_{b}\right)_{\!,\nu }+ (1-2\sigma )
 \left(2H^{\mu }_{b}C_{[a,b]}- N^{\mu
}_{ab}C_{b}\right)=0,
\end{equation}
\[ \mbox{where\ \ } C_{a}=H^{\lambda }_{a,\lambda };
\ \ {}_{,a}={}_{,\lambda }H^{\lambda }_{a}; \
 \ N^{\mu }_{ab}=H^{\mu
}_{a,b}-H^{\mu }_{b,a} + \sigma (H^{\mu }_{a}
 C_{b} - \eta _{ab} H^{\mu }_{d}
C_{d}).  \]
 The case $\sigma =1/3$ in (\ref{sys4}) [and
(\ref{egh3})] corresponds to  the {\it optimum\/} equation
discussed in the first chapter.

Equations (\ref{egh3}) keep regularity and compatibility
for the case of degenerate (but finite)
matrix $H_{a}{}^{\mu }$ if
 ${\rm rank\,}H_{a}{}^{\mu }H_a{}^\nu \ge 2$;
 that is, a formal solution,
as a series, can have
 a degenerate matrix $H_{a}{}^{\mu }(x_{0})$
as the initial term.

Hence, the contra-singularities for class
 ${\bf D}$ of AP equations,
(\ref{sys4}), including the optimum equation, (\ref{equ}),
are `weak', whereas in the case of other AP equations
(including the case
$Eq.{\cal B}$ of \S 2.2), contra-singularities are `strong'.

In order to demonstrate that  solutions with arising
 contra-singularities
 (with  ${\rm rank\,}H_a{}^{\mu}=2$) do exist for
equations (\ref{egh3}), suffice it to consider the
spherically-symmetric problem; this will be done in the end of
this chapter.

The difference of tri-linear equations from others
 can be illustrated by
the difference of
the following ordinary differential equations:
\[1)\ \ \varphi^\prime =\varphi;\ \ 2)\ \
 \varphi^\prime=-1/\varphi \]
(the case $\varphi =0$ corresponds to
 ${\rm rank\,}H_a{}^\mu{}<D$; at that,
the leading term remains regular. Solutions of the first equation
exist for all values of independent variable, but it is not the
case for the second equation.

\newpage
\section*{2.4 \ Measure of
solutions with arising contra-singularity
 \\ \hspace*{11mm} and spacetime dimension $D$}

 The fact of existence of formal solutions with a contra-singularity
 (which is undoubted
for the case of `weak' singularities) still can
 be an insufficient reason
for singularity appearance in solutions of {\em general position}.

The set of square matrices
 $M={\rm Mat}(D\cdot
D)=\mathrm{R}^{D^{2}}$ contains subsets
 $M_{i}$ of matrices of co-rank
  $i$ ($M_0=GL(D))$.
The dimensions of these subsets are defined as follows
  (for example, see \cite{arno}):
\[ \dim M_i=D^{2}-i^{2} \ \ \
 \mbox{ (or \ }{\rm codim\,}M_i=i^{2}). \]
 A field $H_a{}^\mu{}(x)$ [a solution of
 3-linear equations (\ref{egh3})],
  which has a singular point $x_0$
  of corank  $i$ in a region
  $U^{D}\subset R^{D}$,
in general will have some set of singular points,
  $\Sigma $,
a {\em surface of singularities}:
\begin{equation}
\label{sng} H(x):\ U^{n}\to M; \ \ \ \Sigma =
 \Sigma _1\cup \cdots \cup \Sigma
_i, \mbox{ where \ } \Sigma _j=H^{-1}(M_j).
\end{equation}
For most mappings (\ref{sng})
 (i.e., for solutions of general position),
this image
$$H(U^{D})\subset M$$
intersects the subsets
(strata) $M_i$ {\it transversally\/} (see \cite{arno});
AP field equation are of second order, so the first derivatives
 $H'$ have no restrictions.
Therefore, one can find the dimensions of spaces
  $\Sigma_i$:
\[ \dim \Sigma _i=D-i^{2} \
\mbox{ (or \ } {\rm codim\,}
 \Sigma _i ={\rm codim\,}M_i); \]
  a negative dimension means that
  that space is empty (that is, such
singularities, of such a large corank,
 do not arise in a general
solution). Naturally, it is sufficient
 to consider singularities
of corank one.
 The zeroth term  ($A_a{}^\mu{}$) of a singular germ,
 by means of a left-right transformation,
 can be diagonalized:
\[ H_a{}^\mu(x)=A_a{}^\mu_{|\Gamma}\, x^\Gamma;
\ \ \ A_a{}^\mu{}={\rm diag\,}(1,\ldots,1,0); \]
 The $L$-vector $C_a$, annihilating this matrix
$A_a{}^\mu{}$, is space-like: this preserves
 the equation's
hyperbolicity [and provides characteristic (or systatic
\cite{pomm}) covectors].

These considerations perhaps are inapplicable in the case of
{\em `strong'\/} singularities,
but we are interested only to deal with
$Eq.{\cal A}$, see (\ref{equ}), (\ref{egh3}).

It seems that on the set of singular germs one should discriminate
a subset of `dangerous' germs --
 with {\em arising\/} singularity,
which comply, in some region   $U^{D}$ (around the starting point
of a germ),
with a couple of conditions:\\
(1) the singularity surface
 $\Sigma_1 $ defined by
\[H^{-1}=0,\mbox{ i.e., \ }
 dH^{-1}=\xi_{\mu}dx^{\mu}=0, \ \ \ \xi_\mu
= H^{-1}H^a{}_\nu H_a{}^\nu{}_{,\mu},\]
 is space-like or at least light-like
  (so one can choose a Cauchy surface infinitely close
 or tangent  to  $\Sigma_1$ in a point;\\
(2) one can choose coordinate in $U^{D}$ such that  \ $g_{00}<0$
or $g_{00}\leq0$ (i.e., $x^{0}$ \ is `time-like' coordinate)
on $\Sigma_1\cap U^{D}$.\\
Then one should evaluate    `a quantity', a
measure of `dangerous' germs on the set of all singular germs.

The strict conditions
$\xi_\mu\xi_\nu g^{\mu\nu}, g_{00}<0$
can be met without decreasing the measure of these germs
(restrictions as inequalities are sufficient)
for any $D$, with the except for the `near-critical' dimensions,
$D=3$ and $D=5 $.

Consider the case $D=3$ for the variant
 $Eq.{\cal A}$, when  $p=(4-D)^{-1}=1$
($\sigma =1/3, D_0=4$).
Substituting
 $p=1$ into (\ref{lap}),
one can see that
 $D$-scalar
$\Lambda_{abc}$
remains finite at degeneration of matrix
 $H_a{}^\mu{}$:
\[
\Lambda_{abc}=2\left(H^{-1} H_{a\mu }H_{[b}{}^{\nu }
 H_{c]}{}^{\mu }{}_{,\nu }
+\eta _{a[b}H_{c]}{}^{\nu }H^{-1}H^d{}_\lambda
 H_d{}^\lambda {}_{,\nu }\right),
\]
besides $H_a{}^\mu{}$, this equation contains its minors,
 $H^{-1}H^a{}_\mu$. Moreover, some of these
 $D$-scalars do vanish
`in singularities', as well as
$D$-scalar in the following orders by differentiation:
\[ \Lambda_{abc,d}=\Lambda_{abc,\mu }H^{-1}H_d{}^\mu{}
\ \ \ (h_a{}^\mu{}=H^{-1}H_a{}^\mu{}). \]
It is clear, therefore, that for $D=3$
all invariants (like $\Lambda_{abc,d}\Lambda_{abc,d}$)
are finite and, moreover, vanish (except for
invariants $\Lambda^{2}$) on
$\Sigma$; this seems to be impossible for
 {\it general\/} solutions with a space-like
$\Sigma$.

This situation (and the peculiarity of $D=3$)
 is to be explained perhaps
by the fact that the surface of singularities, $\Sigma $ (where
frame $h_a{}^\mu{}=H^{-1}H_a{}^\mu{}$ does just vanish),
admits no space-like co-vectors: all co-vectors are light-like.
 Surface $\Sigma $ is always light-like and the weak condition
 $\xi_\mu\xi_\nu g^{\mu\nu}\leq0$ holds true,
  as well as the second one:
 $ g_{00}<0$ ($g_{00}\to -\infty$).

An additional analysis of urgency
 of these singularities (when $D=3$;
or, maybe, their
unphysical nature, i.e., possibility to eliminate them
using a non-diffeomorphic coordinate transformation)
 would be desirable;
however, this variant, $D=3$,
 is already rejected in  \S2.2 because of
other arguments.

The other special case is  dimension $D=5$ when, on the contrary,
co-frame $h^a{}_\mu$ remains finite
 at singularity, as it coincides with
the minor of `working' matrix
 $H_a{}^\mu{}$ [see (\ref{ch})]:
\[ p=(4-D)^{-1}=-1, \ \ \ h^a{}_\mu=
\| H_b{}^{\varepsilon}\|H^a{}_\mu \to
 {\rm diag\,}(0,\ldots,0,1)
.\]
 In this case, the surface $\Sigma $ has no time-like vectors,
so there is no an evolution parameter which would serve as the
time. More precisely, the weak condition,
 $g_{00}=0$, can be satisfied in a single point on
 $\Sigma$ (say, the point of the series development),
but to make it hold true in a vicinity
 $\Sigma_1\cap U^{D}$,
one should impose some restrictions on the series coefficients
(of a singular germ) because light-like vectors
 (on $\Sigma$) forms
a sub-set of co-dimension one (or of measure zero);
this reduces the measure of `dangerous' germs to zero.

Thus,  a contra-singularity's {\it initiation\/} in a general
 solution, its appearance as
a result of solution of the Cauchy
 problem\footnote{It can be posed after
fixing the coordinate arbitrariness,  a gauge fixing, -- on
non-singular solution germs.}, seems questionable for $D=3$, and
impossible for $D=5$.
 Interestingly, the case of spherical
symmetry admits beginnings of co-singularities of co-rank  $D-2$;
that is, for $D=3$ these are usual singularities of co-rank one,
while for $D=5$, are `rare' ones, of co-rank three
 (for that singularities, all the tangent space consists of
light-like vectors, so the suppression mechanism does not work).

The next chapter (aiming at a qualitative,
 `topological' analysis of the set
of AP solutions) will pay the most attention
 to the case of spacetime
dimension $D=5$.

Note in conclusion that the group of coordinate diffeomorphisms
acts ineffectively on  a singular germ.
 The series coefficients of
a coordinate transformation,
 $y^\mu(x^{\nu })=a^\mu_\Gamma\, x^\Gamma \ $
($\Gamma $ is a  multi-index),
 `are conveyed' through a degenerate
matrix $A_a{}^\mu{}$ (the zeroth term of a singular germ). It is
conceivable that contra-singular germs of solutions (for $D=3$ and
$D=5$) have a some extended symmetry group which includes
coordinate transformations differing from
 diffeomorphisms --
 {\it
singular\/} transformations (perhaps depending on
a singular solution germ) which transform the singularity surface
 $\Sigma $ in a peculiar way.

 \newpage
\section*{2.5 \  Spherical symmetry;
examples of solutions\\ \hspace*{11mm}
  with arising contra-singularity }

A spherically symmetrical field
 $h^{a}{}_{\mu }$ can be written
 [fixing the global rotation from
 (\ref{lrsy})] as follows
 (see\ \cite{ei5,z4,zz1}; \ $i,j=1, \ldots , D-1$):
\begin{equation}
\label{spsy} h^{a}{}_{\mu }= \pmatrix{a&bn_{i}  \cr
cn_{i}&en_{i}n_{j}+d\Delta _{ij} }\, ; \ \ \ h_{a}{}^{\mu }=\frac
1 \kappa \pmatrix{e & -cn_{i} \cr -bn_i & a n_{i}n_{j}+\frac
\kappa  d \Delta_{ij}}\, ;
\end{equation}
where $\kappa =ae-bc$,
 $x^{2}=x^{i}x^{i},\ n_{i}=x^{i}/x$ is the
unit vector along the radius;
 $\Delta _{ij}=\delta _{ij}-n_{i}n_{j}$
 tensor orthogonal
 to
 $n_{i}$, $\Delta_{ii}=k=D-2$ ($n_{i,j}=\Delta _{ij}/x$);
 $a,\ldots
,e$ are smooth functions of radius and time
 $t=x^{0}$ ($a,d,e$ are even, while
  $b$ and $c$ are odd functions of radius $x$).

  It is evident, for fields
(\ref{spsy}), that  $S_{\mu \nu \lambda }\equiv 0$
 (there are no ways to build a skew-symmetric tensor);
  therefore, the current of `Maxwell equation'
 (\ref{mm}) [for
$Eq.{\cal A}$ (\ref{equ})] vanishes too.
 So, this equation gives the two equations for the component
  $w=hf^{0i}n_{i}$ ($f_{ij}\equiv 0$),
 which can be easily integrated:
\[ (hf^{\mu \nu })_{,\nu }= 0 ;\ \ (w n_{i})_{,i}
= w^\prime  + w k/x =0 ,\ \  \dot{w}=0 ;\ \ \ w=C x^{-k} .
\]
 We use the `dot'-mark to denote differentiation by time, and
 the 'prime'-mark -- differentiation by radius.
 It is clear that
   $C\neq 0$ (integration constant) means, in fact, that a
$\delta$-source is added to the current;
 we should not `modify' our equations, therefore choose
 $C=0$.
 Then we simply have
  $f_{\mu \nu }= 0$ (this just follows from
  the skew-symmetric part of system ${\bf D}$,
(\ref{sys4}), if $\sigma  \neq 1/3$).

 Taking $S=0$, $f=0$, one can
rewrite equation ${\bf D}$
 as follows:
\begin{equation}
\label{eq1} (h L_a{}^{\mu \nu })_{,\nu }=(1-2\sigma )hL_a{}^{\mu
\nu }\Phi _{\nu }, \mbox{ \ or \ }
 E^{\mu}_{a}= (h \psi ^{2\sigma
-1} L_a{}^{\mu \nu })_{,\nu } =0;
\end{equation}
 here $L_{a\mu \nu }=\Lambda_{a\mu \nu }-2\sigma
  h_{a[\mu }\Phi_{\nu]}$
 and (a scalar is introduced)
  $\Phi _{\mu }=\psi _{,\mu }/\psi$.

 The equations $E^{0}_{\underline{0}}=E^{i}_{\underline{0}}=0$
 can be integrated in the same way as the previous
(the 'Maxwell equations'); the integration constant is zero again
 (to suppress extra $\delta (x)$-term), and we obtain:
\[ L_{\underline{0}0i}=L_{\underline{0}}{}^{0i}=0
\ \ \ (L_{\underline{0}ij} \equiv 0)\, .\]
 The underlined indexes are of
  "Latin origin" ($\Phi _{0}\neq
\Phi _{\underline{0}}).$

 The evidently spherically-symmetric form
   (of frame field)
  (\ref{spsy}) allows the next coordinate changes:
 \begin{equation}
\label{chs} x^{*} = X(x,t) ,\ \  t^{*} = T(x,t) ,\ \ n^{*}_{i} =
n_{i} .
\end{equation}
 Let us choose a coordinate system, that is, fix the remaining
 coordinate freedom, taking  $b=c=0$.
 These restrictions lead to first order equations for
 $T$ and $X$, therefore, a  freedom remains
 to choose two functions of
 one variable (`initial conditions');
 in what follows, we have to choose (exactly) two `functions of
 integration', and this will be done from convenience sake and
 with accounting for the boundary condition:
\[h^{a}{}_{\mu }\to \delta ^{a}_{\mu } \mbox{ \ at \ }
x\rightarrow \infty \ . \]
 Note that the component $d$ transforms independently:
  $\ d^{*}=d
x/x^{*}$, so, say, the condition  $d^{*}=1$
 defines  $X(x,t)$
 completely.\footnote{* \ For the trivial solution(s), when
 $\Lambda_{a\mu\nu}{=}0$,
  by integrating  $dy_{a,\mu}{=}h_{a\mu}$,
  one can define  scalars $y_a$ which serve
   as the privileged, inertial coordinates.
   In the same sense, one can find
  some `scalar radius and time'; hence, all $x, t$
   are of covariant, scalar sense (quasi-inertial).\ \
 \  E.g.: (1)  $xd$ is a `scalar radius'; \ \ (2)
 \   eq-n
  $L_{\underline{0}\mu\nu}=0$, or
 $\psi h_{\underline{0}[\mu,\nu]}-
 \sigma h_{\underline{0}[\mu }
 \psi _{,\mu ]}=0$, gives a `scalar time' $\tau$:
 $d\tau_{,\mu}=\psi^{-\sigma}h_{\underline{0}\mu }$.}

 The components of  $\Lambda _{a\mu \nu }$ and $\Phi _{\mu }$
 read
 [see (\ref{gala}), (\ref{spsy}); $b=c=0$]:
\begin{equation}
\label{la} \Lambda_{\underline{0}0i}=
-a^\prime n_{i},\ \ \Lambda_{\underline{i}j0}=
\dot{e}n_{i}n_{j}+\dot{d}\Delta_{ij}, \ \
\Lambda_{\underline{i}jk}=(d^\prime + \frac{d-e}{x})
2\Delta_{i[j}n_{k]};
\end{equation}
\begin{equation}
\label{ph} \Phi_{0}=
h_{\underline{i}}{}^{j}\Lambda_{\underline{i}j0} = \dot{e}/e +
k\dot{d}/d, \ \ \ \Phi_i n_i=\frac{a'}{a} +
k(\frac{d'}{d}+\frac{d-e}{dx}).
\end{equation}
Making integration of equation  [see (\ref{eq1})] $\Phi _{0}=
\dot{\psi}/\psi$, one obtains $\psi =e d^{k}$.
 Now the integration of eq-n  $L_{\underline{0}0i}=0$,
(\ref{la}) and (\ref{eq1}), is in order ($\Phi =\Psi '$):
\begin{equation}
\label{ina} L_{\underline{0}0i} n_{i}= - a^\prime+\sigma
a\psi^\prime /\psi=0, \ \ \ a= \psi^{\sigma }= e^{\sigma }
d^{k\sigma} .
\end{equation}
Using $\sigma\Phi_{\mu }= a_{,\mu }/a$, write the remaining
 components of tensor
  $L_{a\mu\nu }$:
\[L_{\underline{i}j0}=(\dot{e}- e\dot{a}/a)n_{i}n_{j}+
(\dot{d}- d\dot{a}/a)\Delta _{ij}\ , \]
\[ L_{\underline{i}jk}=(d^\prime + \frac{d-e}{x} - \frac{da'}{a})
(\Delta _{ij}n_{k}- \Delta _{ik}n_j)\ , \] and then
 switch to the components of
 $L_{a}{}^{\mu \nu }h\psi^{2\sigma -1}$
 (see (\ref{eq1})--(\ref{ina});
 $ h=aed^{k}=a\psi$, $a=\psi^{\sigma}$):
\begin{equation}
\label{hl} a^{3}L_{\underline{i}}{}^{j0}=\dot{\alpha }n_{i}n_{j}
+\dot{\beta }\Delta_{ij},\ \  a^{3}L_{\underline{i}}{}^{jk}
=\{\alpha\beta(\alpha-\beta)/x-\alpha^{2}\beta^\prime \}
2\Delta_{i[j}n_{k]}
\end{equation}
Here we introduce the new variables
 $\alpha =a/e=d^{k\sigma }e^{\sigma
-1}$ and $\beta =a/d$. The change $e,d\rightarrow \alpha,\beta$
(irreversible at  $k=1/\sigma -1$) corresponds to (\ref{ch})
(components of metric $g^{\mu \nu }$ are given as well):
\[
H_{a}{}^{\mu}=\pmatrix{1& 0\cr 0 &\alpha
n_{i}n_{j}{+}\beta\Delta_{ij}}{=} h^{\sigma
/(1+\sigma)}h_{a}{}^{\mu }; \ \ g^{\mu\nu}{=}\pmatrix{{-}1/a^{2}&
0\cr 0 & n_{i}n_{j}/e^2{+}\Delta_{ij}/d^2}. \]
 Substituting
(\ref{hl}) into (\ref{eq1}), let us write equation
$E_{\underline{i}}^{j}n_{i}n_{j}=0$:
\begin{equation}
\label{eq2} (a^{3}L_{\underline{i}}{}^{j0}n_{i}n_j)^{\cdot}-
(a^{3}L_{\underline{i}}{}^{jk})n_{i,k}n_{j} = \ddot{\alpha
}+\{\alpha\beta(\alpha-\beta)/x-\alpha^{2}\beta^\prime \} k/x =0.
\end{equation}
Equation $E_{\underline{i}}^{0}=0$ reduces to equation
 $f_{0i}=0$, i.e., is already taken into account.
 One should still remember that
 $\Phi_{i}n_{i}=\psi ^\prime /\psi$;
 this gives a simple `constraint equation',
  $\beta =\beta (x,\alpha, \alpha ')$ [see (\ref{ph})], and
  after its substitution into
  (\ref{eq2}) one can arrive to the resulting equation(s)
($k>0$):
\begin{equation}
\label{eq01} \ddot{\alpha
}=\alpha^{2}\alpha^{\prime\prime}+\alpha \alpha'^{2}/k
+(k+2)\alpha ^{2}\alpha^\prime /x~;~~~~\beta
=\alpha+x\alpha^\prime/k~.
\end{equation}
Equation $E_{\underline{i}}^{j}\Delta_{ij}=0$ is consistent with
this system,  (\ref{eq01}) (a consequence of the identity).

The Cauchy problem in the chosen coordinate system reduces
 to solution of a single second order equation;
 one should define even functions
 $\alpha_{0}(x)>0$ and $\dot{\alpha }_{0}(x)$
 as the initial data.
 These functions can be defined in such a way that
  $\beta_{0}(x)$ approaches zero in a point $x_{1}> 0$,
but $\beta (x,t)>0$ at $t<0$
 ($\dot{\alpha }_{0},\dot{\beta}_0<0$
near $x_{1}$). For example:
\[ \beta_{0}(x)=1-3x^{2k}/(2+x^{6k}),
\ \ \beta _{0}(1)=0;\ \ \ \alpha_{0}(x)=x^{-k} \int
\beta_{0}(x)\,dx^{k} > 0 .
\]
The point $x=x_{1}$, $t=0$ is singular, because $\beta =0$; at
 $t>0$, a region emerges where $\beta$  is negative.

 If a region with $\beta <0$ is allowed in initial data, the data
 can be defined in such a way that $\alpha_{0}(x)$
 approaches (is tangent to) zero in a single point
 ($x_{2}$; at that $\beta _{0}$ changes its sing);
 and a region with $\alpha <0$ emerges at  $t>0$
   (if $\dot{\alpha }_{0}<0$ near $x_{2}$).

 That is, all variants of signs of
  $\alpha$  and $\beta$ are possible, as well as the cases
   $r=D-1$ ($\alpha =0$,
$\beta \neq 0$), $r=2$ ($\alpha \neq 0$, $\beta =0$) and even
$r=1$ ($\alpha =\beta =0$),
 although the last case is impossible in a general solution;
 if one takes $$
n_{i}=(1,0,\ldots,0),\mbox{ \ then \ } H_{a}{}^{\mu }=
 {\rm diag\,}(1,\alpha,\beta,\ldots,\beta ); \ \
r={\rm rank\,}H_{a}{}^{\mu}.$$

 Points with $\alpha =0$ or $\beta =0$ are singular
 at $D>D_{0}$ because
$D$-scalars like $\Phi_{\underline{0}}$ ($\dot{\alpha }$,
$\dot{\beta }\neq 0$) turn into infinity:
\[ \Phi_{\underline{0}}=-p/\sigma (\dot{\alpha }/\alpha+
k\dot{\beta }/\beta )\alpha^{p}\beta^{kp}; \ \ \
p=(D_{0}-D)^{-1},\ \ \ D_{0}=1+\sigma ^{-1}. \]
 In the case $\sigma=1/3$ ($D_{0}=4$),
 a region with $\alpha <0$ has no real-valued solutions for
 the frame $h_{a\mu }$ if $D$ is even:
  $$a=(\alpha \beta
 ^{k})^{-p}\, \ (e=a/\alpha, \ d=a/\beta).$$
 It is interesting that this `loose of reality'
(when $\alpha <0$ or $\beta <0$) takes place for any $D$, if
$\sigma$ is irrational!

 The question about possibility of other types of singularities in
 a {\it general\/} spheri\-cally symmetric solution
  is not easy.
 One could consider the full system with five dependable
 variables, which is 2D-covariant, with respect to
  (\ref{chs}), and to check whether the symbol remains involutive
  (and the Cauchy problem remains well-posed)
 or not -- at different ways of degeneration of the frame
 matrix (\ref{spsy}).
\newpage

\section*{2.6 Another co-ordinate
choice; Cosmological solution }

Another conditions, $b=0$, $e=d$, also allow to make similar
integrations  \cite{zz1} and to obtain a system of quasi-linear
first-order equations:
\begin{equation}
\label{eq02} A^{\cdot}=AB^\prime -BA^\prime +kAB/x\ , \ \ B^{\cdot
}=AA^\prime -BB^\prime -(k-1)B^{2}/x~,
\end{equation}
where $H_{a}{}^{\mu }=\pmatrix{1 &Bn_{i}\cr 0 & A\delta _{ij}}
A^{-\sigma /(1+\sigma )};
 \ ~~A=a/e=e^{k\sigma /(1-\sigma )-1},\ ~~B=-c/e$.\\
Here one should define as initial data the even positive function
 $A_{0}(x)$ ($A_{0}\to 1$ at $x\to\infty$) and
 uneven function $B_{0}(x)$. Equations (\ref{eq02}) have no
stationary solutions (if an additional $\delta$-term is not
allowed), excepting the trivial one: $A=1$, $B=0$ [or $\alpha
=\beta =1$ for (\ref{eq01})].
 No singularity
 ($1/A,$ $1/B$ or $A \to 0$)
 can emerge in solutions of system
  (\ref{eq02}) during a finite time.

 The first equation of (\ref{eq02}) reduces to a conservation law:
 $$(x^{k}/A)^{\cdot }=-(x^{k}B/A)^\prime\, ,$$
 so one can introduce a Lagrange variable, $y(x,t)$ (see
\cite{rozh}; $x_{y}=\partial x/\partial y):$
\[ dy=x^{k}/A\, dx - x^{k}B/A\, dt;
\ \ dx=Ax^{-k}\,dy +B\, dt,\ \  B=\dot{x},\ \ A =x^{k}x_{y}.\]
 Substituting $A$, $B$, $B^{\cdot }(x,t)=
\ddot{x}-\dot{x}_{y}\dot{x}/x_{y}$, $B^\prime=\dot{x}_{y}/x_{y}$,
$A^\prime$  into the second equation of  (\ref{eq02}), we obtain
after the change of variable, $z=x^{k}$: $\ \ddot{z}= z^{2}z_{yy}+
z z^{2}_{y}/k\ $ [see (\ref{eq01})];
 at $k=1$, there exists a
Lagrangian,  ${\cal L}=\dot{z}^{2}- z^{2}z^{2}_{y}$; or for
(\ref{eq01}): ${\cal L}=(\dot{\alpha }^{2}- \alpha ^{2}\alpha
^{\prime 2})x^{3}$.

 Switching to  invariants, $u=A+B$, $v=A-B$, one can easily show
 that the system
 (\ref{eq02}) is of weekly-nonlinear type
($u+v=2A>0$)
\begin{equation}
\label{eq04} \left\{ \begin{array}{rcl}
\dot{u}&{=}&vu'+ \frac k {4x} (u^2-v^2)-\frac{k-1}{4x}(u-v)^2 \\
\dot{v}&{=}&-uv'+ \frac k {4x} (u^2-v^2)+\frac{k-1}{4x}(u-v)^2
\end{array}
\right. ,
\end{equation}
 i.e., its solutions do not suffer of the gradient
  catastrophe \cite{rozh}.
 The `cut' system (\ref{eq04}) -- without terms $\sim 1/x$ --
has infinitely many conservation laws and is fully integrable, for
example, using the hodograph transformation, or through transition
to the Lagrange coordinate. A solution (island-like) takes through
a finite time a form of two single waves moving along  $x$ -- one
to the right, the other to the left, with the unit velocity, which
keep their form; due to the symmetry, $v(x)=u(-x)$, it is enough
to consider only the  $v$-wave (where $u=1$), which moves to the
right.

 The lowest terms, $\sim 1/x$, vary, of course, the form of wave
 ($v$-wave acquires a contribution of  $u$-component) and
 seemingly slow down $v$-wave (its top part) -- otherwise the
 gradient catastrophe could  emerge.
  It seems that at the large stage of expansion, the influence of
  the lowest terms is quite moderate; so the solution can be
  described by  a few parameters:
 \ $R\sim t$ is the radius of expansion; \
 $\lambda $ and $a$ the width and the amplitude of the `wave-hump';
    \ $\gamma
 \gg 1$ relativistic factor; $\lambda $ and $a$  change very
 slowly (in comparison with  $R$ and, maybe, $\gamma$) with the time
 and can serve as  `fundamental constants'.
 Some estimates are possible for the amplitude of `extrinsic'
  $u-$component (near the $v-$hump) and relativistic factor:
\[\Delta u=u-1 \sim -a\lambda/R;
\ \ \gamma\simeq1/\sqrt{1-u^{2}}\sim \sqrt{R/(a\lambda)}\sim
\sqrt{t}.\]
 In the comoving coordinate system (moving with the  `hump'),
 the typical size $L\sim \sqrt{t}$ can be large enough
($>10^{10}$cm).

 It seems that a spherically-symmetric solution (large symmetry
 may promote stability of solution)
 can serve as a base of interesting cosmological model --
 ultra-relativistically expanding  wave-guide --
   and solve the problem of reduction of the extra dimension
   (`relativistic', space dimension).

 A more realistic model should contain one more cosmological
 (global, or total) object --
 weak stochastic waves (relic noise, `zero point vibration'),
 which fill the wave-guide and have evolved to a quite symmetrical,
 `thermalized' state (with Lorentz-covariance along the tangent
 coordinates, i.e., usual space), at frequencies smaller than the
 the noise spectrum edge,  $\omega _0$;
 one may think that  $1/\omega _0\sim \lambda_0< \lambda $.
 The already introduced parameters can be accompanied by
 amplitude parameters of the noise (perhaps, one should distinct
 the amplitude of electromagnetic component
  $a_{f}$ and  the `total' amplitude $a_{0}$,
$a_{f}<<a_{0}$).

 The cosmological (wave-guide) solution is characterised by
  $S_{abc}=0$, $f_{ab}=0$
  (and $J\propto Sf=0$; noise contribution, averaged at a scale
   $\sim 1/\omega _0$, vanishes too);
 therefore, given a `perturbation', moving along the wave-guide,
 the electromagnetic current is quadratic on the perturbation
    (the `wave-guide--thermostat' gives no linear contribution,
    and, in this sense, it is unobservable).

 If amplitude $a_{0}$ is large (enough),
 it seems impossible to pick out something,
  some specific perturbations
  (due to fast thermalization in conditions of already
  emergent `heat death'). However, as it will be shown in
  the next chapter, AP grants solutions, localised frame field
  configurations, which carry integer information:
  topological charges and quasi-charges.

\chapter*{Chapter 3. \ Topological quasi-solitons
 \\ \hspace*{47mm} (quanta)}
\addtocounter{chapter}{1} \setcounter{equation}{0}

 This chapter is intent to discuss global properties of localised
 configurations of the frame field $h(x)$ (that is, asymptotically
 plane solutions), the structure of the set of solutions
   ${\cal
C}(h^a{}_\mu(x))$ (the set of localised solutions, which `field
carrier' is topologically trivial),
 as well as the structure of sub-sets of solutions which have one or
 another symmetry.

 Amongst the all solutions, the {\em stable} solutions are of special
  importance (there are a few, slightly different, definitions of
  stability:  Lyapunov stability, Lagrange stability, et cet.\
   \cite{ryba});  and `our experience' demonstrates that
   stable solutions usually have high symmetry.

 As it will be shown below,
 given localised initial data of the Cauchy problem on a
 space-like undersurface,
 one can define an invariant,
 integer-valued quantity -- topological charge -- which remains
 invariable.\footnote{ If only no singularity (which were straggled
 against in the previous chapter) will arise in a solution.}

 Therefore, the set of solutions turns out to be divided into
 disconnected (through a smooth deformation)
 components; each component corresponds to a specific value of
 topological charge.

 Also the question arises, how to choose a component, a homotopy
 class (for a suitable solution).
 It is interesting, that there are only two cases of space-time
 dimension, $D=3$ and $D=7$, when the solution space consists of
 only one component  (the case $D=7$ was considered,
 *quite superficially*, in \cite{z1}),
 but perhaps, the case of multi-component space of solutions has
 its own advantages; this question, together with the problem of
 choice of the space topology, will be broached in the last
 paragraph. Meanwhile, let us dwell on the case of trivial
 topology (of space) as the most simple one.

 For field configurations which have some symmetries (discreet or
 continuous, when Killing's vector(s) exists),
  one can define the notion of
 topological quasi-charges, which conserve while the symmetries are
 conserving; it is tempting to try to interpret (quasi-)charge
  as the presence
 of `particle', or quasi-soliton (quantum).

 It is necessary, first, to make classification
 of possible symmetries
 (composing them from simple components like `inversions', i.e.,
 reversal of $p$ space dimensions, or `rotations', including
 $q$ space dimensions,
 $p,q\leq D-1$); next, to establish possible `equivalence' of some
 symmetries (i.e., the identity of corresponding
 discretizations on
  separated
 components).
 Then one should choose or guess the symmetry
 (it should be not feeble)
  of large-scale,
 cosmological solution (it should be stable in a sense), which
 would define a {\em waveguide} in a sense, where small-scale
 waves, perturbations could exist in some stable manner.

At last, one should list all symmetries which conform with the
symmetry of that {\it cosmological waveguide}, and perform, on
this basis,  classification of  {\it quasi-soliton}
`perturbations' admissible for that cosmological background.

It turns out that in a field theory -- such simple one as the
absolute parallelism -- a rich `particle' (quasi-soliton)
combinatorics can emerge, which includes both bosons, and
fermions; moreover, it can resemble the observing particle
combinatorics (if one believe that the leptons (and the quarks),
the vector bosons (and gluons) are all  elementary).

\newpage

\section*{3.1 \ Topological charges of localized
field configurations}
 Let assume that the Riemannian space defined by the metric
  $g_{\mu \nu }$ is topologically equivalent to the plain
  space:
   no singularities, no worm holes, and so on.
  Let
 \[x^{\mu }\in R^D\;\; (\mu =0,1,\ldots,m\,; \;\; m=D-1) \]
 is a coordinate system, and the (hyper)surface $R^m=\{x;x^0=0\}$
  is space-like.
  It is necessary to note that not any field configuration
  $h(x)$ can confer such surfaces, although the
  `physically-sensible' field configurations, solutions of
  equations of absolute parallelism, should, seemingly, have
  (space-like) Cauchy surfaces.

 A field $h(x)$ is {\it localized\/} in $U^m\in R^m$ if, after
 some global  {\it left\/} transformation from (\ref{lrsy}),
  it has
 the trivial asymptotics:
\begin{equation}
\label{hloc}
 h^a{}_{\mu}=\delta ^a_{\mu}+O(\varepsilon )
  \mbox{ \ on \ }
  R^m\backslash U^m,
\end{equation}
where $\varepsilon $ is small enough, for example  $\varepsilon
=10^{-5}$. At that, the metric (\ref{metr}) is also localized in
the region $U^m$ in the sense of (\ref{hloc}).

 Making a deformation of metric $g(x)$ (on $R^m$) to the trivial,
pseudo-Euclidean form, and accompanying this with a continuous
deformation of the frame field, which would conserve the relation
 (\ref{metr}), one can transform $h(x)$ to a field of Lorentz
 rotations, $\sigma^a{}_{\alpha }(x)$
\begin{equation}
\label{dfm}
\begin{array}{cccc}
g_{\mu\nu}(x) & {\to} & \eta_{\mu\nu}& \;\;\; x\in R^{m},\\
{\Uparrow} & &{\Uparrow} & \\
h^{a}{}_{\mu}(x)& {\to} & \sigma^{a}{}_{\mu}& \;\;\; \ \sigma
^{a}{}_{\mu} \in SO(1,m).
\end{array}
\end{equation}

 Such a deformation can be perform by the following way:
 \begin{equation}
\label{lorr} \sigma^a_{\alpha }= h^{a}{}_{\mu}d^{\mu\beta }\eta
_{\beta \alpha },
\end{equation}
where $ d^{\alpha \beta }$  is a symmetric matrix, `square root'
from the metric; more exactly,
\[ g^{\mu\nu}= \eta _{\alpha \beta } d^{\alpha \mu}d^{\beta \nu}.
\]
 It is easy to check that
\[ \eta _{ab}\sigma^a{}_\alpha \sigma^b{}_\beta
 =\eta _{\alpha \beta },\,
 \mbox{ i.e., }\, \sigma^a{}_\alpha  \in O(1,m). \]
 To ensure  uniqueness, the sings of eigenvalues of matrix
  $ d$ are taken in correspondence with the signature of
 metric $\eta^{ab}$; if, for example, matrix
$g^{\mu\nu}$ has a diagonal form, then
\[  g^{\mu \nu } = {\rm diag\,}
(-\lambda _0, \lambda _1,\ldots, \lambda _m) \; \mbox{ and }\;
d^{\alpha \beta } = {\rm diag\,}(- \lambda _{0}^{1/2} ,\lambda
_{1}^{1/2},\ldots,\lambda _{m}^{1/2}). \]
 In that case, \
 $\sigma^{a}{}_{\alpha }=\delta ^{a}_{\alpha }+O(\varepsilon )\,
\mbox{ on }\, R^{m}\backslash U^{m};\;\;\;
 \sigma^{a}{}_{\alpha }\in SO(1,m) $.

The localized field $\sigma^{a}{}_{\alpha }(x)$ can be further
deformed that to have \
 $\sigma^a{}_\alpha =\delta ^a_\alpha $ \  on\, $R^m/U^m$.
 Uniting all the peripheral points,
  $R^m\backslash U^m$, we obtain some
 mapping:  $S^m \rightarrow SO(1,m)$.
  The set of these maps consists of
 mutually disconnected components, homotopy classes,
  which form a
 group denoted as
$\pi_m(SO(1,m))$. There is a well known equality of homotopy
groups \cite{olsh}:
\[  \pi_m(SO(1,m))=\pi_m(SO_m) \ \ \ (SO(1,m)\approx SO_m) \]
(i.e., groups $SO(1,m)$ and $SO_m$ are homotopically equivalent).
 That is, one can further continue the deformation in order to
 remove boosts in Lorentz matrices $\sigma$,
and arrive to the next mapping:
\begin{equation}
\label{sso}
 \sigma : \ S^m \rightarrow SO_m \ \ (x_0 \mapsto 1^{m});
\end{equation}
here $x_0$ is a marked point on the sphere
 $S^{m}$, it corresponds
to  infinite points of the Cauchy surface (and maps into the
marked point of group $SO_{m}$, the unit matrix $1^{m}$).

So, any localised field  $h(x)$ can be characterised by  $p(h)$,
an element of  the homotopy group  $\pi_m(SO_m)$.
 It is clear that nor coordinate change on $R^m$
 (one should note that the indices of matrices $\sigma$ and $d$
  no more have  the usual covariant sense), nor
the equations of motion  (in absence of singularities)
 can change an integer value -- topological charge.
 Therefore, if $p(h)\neq 0$, we will say that the region
$U^m$ contains a particle, or topological soliton (or quantum).

The time reversal, $ x^0\mapsto -x^0$, does not change the element
 $p$; it acts as the  {\it right\/} transformation of field
 $h^a{}_\mu$,
on the Greek index, and this can be accompanied with the analogous
 {\it left\/} improper global rotation,
\[ s^a{}_b=\eta _{ab}\ \ (s^a{}_b\in O(1,m))\ ,\]
that to conserve the trivial asymptotics.
 If these transformations, left and right, are applied to all
 course of deformation
 $h(x) \rightarrow \sigma (x)\in SO_m$, one obtains the
  deformation
 of  transformed field $h^{*}(x)$ to the same `end point',
 $\sigma (x)$; this proves that the homotopy
  class does not change.

 Groups $\pi_m(SO_m)$ for all  $m\geq8$ can be found
  in \cite{solo},
 and all them are non-trivial, see Table~1; groups $\pi_m(SO_m)$
for $m<7$ are also  known -- from works by Mimura and Toda (and
Mimura)
 \cite{mito,mimu}; they are presented in Table~2.\\[2.5mm]
 {\normalsize {\bf Table 1.} Groups $\pi_{m}(SO_{m})$
  for \ $m>7$; $i\geq1$.}\\[2mm]
\begin{tabular}{c|c|c|c|c|c|c|c|c}
$m$&$\ 8i\ $&$8i+1$&$8i+2$&$8i+3$&$8i+4$&$8i+5$&$8i+6$&$8i+7 $ \\
\hline $\pi_m(SO_m)
$&$Z_2^3$&$Z^2_2$&$Z_4$&$Z$&$Z^2_2$&$Z_2$&$Z_4$&$Z$
\end{tabular} \\[6.5mm]
 {\normalsize {\bf Table 2.} Groups $\pi_{m}(SO_{m})$
for \ $m<8$.}\\[2mm]
\begin{tabular}{c|c|c|c|c|c|c}
 $   m       $&$\ 2\ $&$\ 3\ $&$\ 4\ $&$\ 5\
  $&$\ 6\ $&$\ 7\ $\\ \hline
$\pi_m(SO_m) $&$  0   $&$  Z   $&$ Z^2_2 $&$  Z_2 $&$  0   $&$  Z$
\end{tabular} \\[1mm]

It is interesting to note that only for two values of
  $m$ the group $\pi_m(SO_m)$
is trivial ($m=2$ or $m=6$). The infinite group  (integer group) $
Z$ corresponds, seemingly, to a soliton (`particle') of boson
kind,  while the cyclic second order group  $ Z_{2}$ (or
two-group) suits for solitons of fermion kind.

\newpage

\section*{3.2 \ Absolute, relative and $k-$ad
homotopy groups }
 This section contains some necessary information of mathematical
 sort which will be later used to answer the question:
 how the subsets of symmetrical solutions are situated on the
 whole set of localised solutions.

 One can start with several definition; more details can be found
 in textbooks
 \cite{dubr,fuks} (for initial familiarity with the subject,
 the book \cite{efre} is well suited, as well as
 the `A short guide to modern geometry for
physicists'
\cite{olsh}).\\
{\bf Homotopic mappings.} Continuous mappings (of space $X$ to
space $Y$)
$$ f:\;\,X\rightarrow Y  \mbox{\, and } \
 g:\;\,X\rightarrow Y \;\mbox{  are homotopic }(f\sim g),\,$$
 if there exists a continuous flock of mappings,
 $$ \varphi _t:\;\,X\rightarrow Y,\;\ 0\leq t\leq 1,  \,
\mbox{ such that } \varphi _0=f,\;\varphi _1=g.$$

 The subset of mappings homotopic (i.e., equivalent) to mapping
 $f$ forms a  {\it homotopy class} which is denoted  $[f]$.
 The set of all maps,
$${\cal C}(X,Y)=\{f:\ X\to Y\}\, ,$$
 becomes discretized, in this way, on homotopy classes
 which compose  a homotopy set denoted as $\pi(X,Y)$.

 If one has chosen in spaces $X, Y$ some {\it marked\/} points
 $x_0\in X$, $y_0\in Y$,
 then the subset ${\cal C}(X,Y;x_0,y_0)$ of mappings, which
 map $x_0$ into $y_0$ ($x_0\mapsto y_0$), can be picked out.
 If space $X $ is connected (single-component), then the space
$\pi(X,Y;x_0,y_0)$ does not depend on the choice of the point
$x_0$.

The case $X=S^{m}$ has (due to its importance) a special notation:
 \[ \pi_{m}(Y)\equiv \pi(S^{m},Y;x_0,y_0)
=\pi_0({\cal C}(S^m,Y;x_0,y_0));  \]
 in the case of localised fields
$\ \sigma ^{a}{}_{b}(x): \ S^{m}\to SO_m$ [see (\ref{sso})],
 the marked points, as was already mentioned, are the following
 ones: $x_0$ is the ``space
 infinity'', $y_0$ is the unit matrix of the group
  $SO_{m}$ (the marked
`point' of the set ${\cal C}(X,Y;x_0,y_0)$ is the trivial map
  $X \to y_0$).

 If  \ $Y$ is a group, the multiplication of maps,
 $f_3=f_1\cdot
f_2$ defines a group operation on the homotopy classes:
 $[f_3]=[f_1]\circ [f_2]$, that is, the set
 $\pi_{m}(Y)$ is also a group;
  moreover, it is an Abelian group (if $\,m\geq2$), because
 the `non-trivial regions' of two mappings can be moved away
 one
 from another, and interchanged
   (so, the group operation is just addition,
 and the trivial map relates
 to the zero element).

 If  \ $A\subset Y$ and a point $y_0\in A$ is marked,
 they name this a
  {\it
pointed pair\/} $(Y;A)$; a ball $D^{m+1}$ and its boundary $S^{m}$
form pair $(D^{m+1};S^{m})$. On the set of {\it relative
spheroids\/} (or pair mappings)
\[ (D^{m+1};S^{m})\to (Y;A) \ \ (D^{m+1}\to Y,
\ S^{m}\to A;\ x_0\mapsto y_0), \]
 it is also possible to pick out  classes of homotopic
 equivalence, which form a {\it relative\/}
(or diad) homotopy group $\pi_{m+1}(Y;A)$ (it is Abelian if $Y,A$
are groups). If $A=\{y_0\}$, the relative group reduces to the
{\it absolute\/} group $\pi_{m+1}(Y)$.

 Let $\Omega (Y,A)$ is the set of paths in $Y$
  (`path' is just a  map
of the unit segment $I=\{0\le t\le 1$\} to $Y$; $t_0=0$), which
start in the marked point $x_{0}\in Y\, (\in A)$ and end in the
subspace $A$. Then
\[  \pi_{m}(\Omega (Y,A))= \pi_{m+1}(Y;A)  \]
(the relative group is defined through the absolute one; this
equality, on the contrary, can serve as a definition).
 To prove
this equality, one can indicate
 the equivalence of the following maps:
\begin{equation}
\label{cha}
 S^{m}\to \{(I;\{1\})\to (Y;A) \}
\ \leftrightarrow\  (S^{m}\times I;S^{m}\times \{1\})\to (Y;A),
\end{equation}
 taking into account that
  $x_0\times I, S^{m}\times t_0 \mapsto y_0$.
 There is the (exact)
  {\it homotopic sequence of a pair}
\begin{equation}
 \label{gpp}  \cdots\to \pi_{r}(A)\to
 \pi_{r}(Y) \to \pi_{r}(Y;A)\to
\pi_{r-1}(A)\to\cdots,
\end{equation}
which is useful for calculation of a relative group. If pair
$(Y;A)$ makes a fiber bundle with a base $B=Y/A$, then
$\pi_{r}(Y;A)=\pi_{r}(B)$ and the sequence (\ref{gpp}) coincides
 with the homotopic sequence of a bundle \cite{post,dubr}.

 Even more general, {\it triad\/} homotopy groups are known
  \cite{blak,post}. More over, the transition from the case of
  relative group to the case of triad group, see
 \cite{post}, can be further extended with definition of $k$-ad
homotopy groups, for any (integer) $k$.\\
 Let $A_1,\ldots,A_{k-1} \subset Y;\  y_0\in A_i$,
i.e.\ $(Y;A_1,\ldots,A_{k-1})$  is a pointed $k$-ad \cite{post}.\\
{\it Definition }1 of $(k{+}1)$-ad homotopy group through $k$-ad
one (induction by $k$; $r{\geq} k$):
\[\pi_r(Y;A_1,\ldots,A_k) = \pi_{r-1}(\Omega (Y;A_1);
\Omega (A_2,A_2\cap A_1), \ldots,\Omega (A_k,A_k\cap A_1))\, ; \]
this induction leads also to the (exact) {\it homotopy sequence of
$(k{+}1)$-ad} (using the $k$-ad sequence and taking into account
the symmetry of  $k$-ad groups with respect to the lower elements
 $A_i$)
\[\cdots \to \pi_r(A_1;A_2{\cap} A_1,\ldots,A_k{\cap} A_1)\to
 \pi_r(Y;A_2,\ldots,A_k)\to \]
\begin{equation}
\label{gpk} \to  \pi_r(Y;A_1,\ldots,A_k) \to
\pi_{r-1}(A_1;A_2{\cap} A_1,\ldots,A_k{\cap} A_1) \to \cdots
\end{equation}
 The beginning of this induction is the relative and triad
 homotopy groups
\cite{post} and pair's homotopic sequence  (\ref{gpp}).

 The  equivalent {\it Definition }2  of $(k{+}1)$-ad
 group [it is manifestly symmetrical
 with respect to permutations
$A_i\leftrightarrow A_j$; the equivalence'
 prove goes by induction
 using a change similar to (\ref{cha})]:
\[\pi_{k+l}(Y;A_1,\ldots,A_k)=\pi_0({\cal C}_{k}^{l}), \]
where ${\cal C}_{k}^{l}$ is the set of maps ($(k{+}1)$-ad
spheroids, maps) of the following form:
\[  S^{l}\times I^{k}\to Y,
\ \ S^{l}\times I^{k-1}_{(i)}\to A_i; \ \ x_0\times I^{k},
S^{l}\times I^{k-1}_{(*)}\to y_0; \]
 here $I^{k-1}_{(i)}$ is the
$i$-th face of the unit cube $I^k$, which
 is adjacent to cube's
vertex $z_1$, $I^{k-1}_{(*)}$ is the collection of all other
cube's faces.

The set $\pi_{k+l}(Y;A_1,\ldots,A_k)$, evidently, will also be a
 group (evidently, Abelian, if $l>0$), if $Y$ is a group,
  and $A_i$  its subgroups.

\section*{3.3 \ Topological quasi-charges\\ for
symmetrical field configurations}
 Symmetrical solutions are of especial interest as they can be
  {\it stable}.
It seems that a discreet symmetry is too small for stability, so
we usually will assume that there is some continuous symmetry
subgroup.

 We will say that a solution $h(x)$ (configuration
 of co-frame, to be definite, field
 $h^a{}_\mu$) is {\it symmetrical\/} with respect to a
diffeomorphism, $x\mapsto y(x)$,
 if there exists a compensating
 global transformation; that is, a left-right transformation
(\ref{lrsy}) transfers this solution $h(x)$ into itself
 (a kind of stationary `point'):
\[  h^{*a}{}_\mu(y)=\kappa  s^{a}{}_b h^{b}{}_\nu(x)
\frac{\partial x^{\nu}} {\partial y^{\mu}}=h^a{}_\mu(y) ; \ \ s\in
O(1,m). \]

 A symmetry of a localized, island configuration is completely
 defined by the rotation $s$ -- there are no translational
 symmetries with  $s=1$ (replacing $y \to \kappa y $ one can
 always exclude the constant $\kappa $).
 Doubly repeated a transformation is also a symmetry, which
 corresponding rotation is
  $s^{2}$.
 It is clear that any possible  symmetry group  $G$
 is a subgroup of $O(1,m)$, and even  $G\subset O_m$
  (a boost is unable
 to be a symmetry).
 The presence of a symmetry does not depend on a choice of
 coordinates $x$, and they can be chosen in such a way that to
 make the corresponding diffeomorphism a global transformation:
  $\ {\partial
y}/\partial x=s={\rm const}$.

The deformation (\ref{dfm}) of frame field to a field of rotation
matrices  $\sigma (x)\in SO_{m}$ can be consistent with the
initial field's symmetry;
 that is, the problem of classification of $G$-symmetrical
 solutions in AP reduces to classification of
   $G$-symmetrical  configurations of the `chiral' $SO$-field
 (field of matrices $SO_m$).
 By the way, the symmetry group of  chiral models  (e.g., the
Scyrme model \cite{ryba}) includes both the transformations of the
ordinary space(time) and the rotations of `internal' (isotopic)
 space, but non-trivial symmetric solutions
  (as in AP) relate to {\it diagonal\/}
subgroups of the model's symmetry \cite{ryba}.

Repeat ones again the definition: a localized field
 $$  \sigma(x) :\
R^{m} \to SO(m),\ \ \ (\sigma (\infty)=1^{m})$$
 is $G$-symmetric
(i.e., symmetric with respect to group $G\subset O(m)$) if
 (in some appropriate coordinate system)
 \begin{equation}
\label{gsi}
 \sigma (sx)= s\sigma (x)s^{-1}\ \ \forall \ s \in G ;
\end{equation}
the set of such fields, ${\cal C}_G $, is divided on homotopy
classes which form group  $\Pi (G)$, that is, $ \Pi
(G)\equiv\pi_{0}({\cal C}_G)$.

 In addition to calculation of groups   $\Pi (G)$,
 one should also solve the other problem. Let
   $Sym1$ is greater than $Sym2$ (by the number of elements,
 or generators); that is, there is the natural
 embedding of the configuration sets:
 $$     i:\ {\cal C}(Sym1) \to {\cal C}(Sym2).$$
This embedding induces a homomorphism of homotopy groups,  $
\Pi(Sym)$:
$$     i_{*} :  \   \Pi(Sym1) \to  \Pi(Sym2), $$
so it is necessary to describe it (whether it is a monomorphism or
 a zero morphism, and so on; in other words, how `small' pieces of
 more symmetric solutions are situated with respect to `large'
 pieces of less symmetric solutions).

 Let us consider the simplest case -- the (discrete) group of
 one-coordinate (let, the first coordinate) inversion,  $P_1$:
 $$P_1 =\{1,p_{(1)}\},
\mbox{ \ where \ } p_{(1)}=
 {\rm diag\,}(-1,1,\ldots,1)=p_{(1)}^{-1}.$$
 It is enough to define the field   $\sigma (x)$  on the
 half-space $\frac1
2\,R^m=\{x_1\geq 0\}$,
 on the surface of stationary points, $R^{m-1}=\{x_1=0\}$,\,
matrix $\sigma $ should commute with the symmetry [see
(\ref{gsi})]:
\[p_{(1)}x=x\ \Rightarrow\ s(x)=p_{(1)}s p_{(1)}
\ \Rightarrow\ s\in 1 \times SO_{m-1}; \]
 therefore (taking into account the localization condition), we
 have here the next diad morphism (or relative spheroid):
\[(D^m;S^{m-1}) \to  (SO_{m}; SO_{m-1}), \mbox{ \ hence \,}
\Pi(P_1)=\pi_{m}(SO_{m};SO_{m-1})=\pi_{m}(S^{m-1}); \]
 the last equality follows due to the existence of
 fiber bundle $SO_{m}/SO_{m-1}=S^{m-1}$.

For the symmetry  $O_l$ (which acts on the first $l$ space
coordinates; $l\leq m$; the time index is omitted),
 matrix  $\sigma $ should obey the following
 restrictions [similar to the case of spherical symmetry
 (\ref{spsy})]:
\begin{equation}
\label{ol}
 \sigma _{pq}=\pmatrix{
e\,n_i n_j+d\,\Delta_{ij} & B\,n_i  \cr
 C\,n_j & A \cr},\  \ \  \sigma _{pq}\in O_{m};
\end{equation}
here $i,j=1,\ldots,l$.
 Again, it is necessary to fill the
 half-space
 $$ R^{m-l+1}_{+}=
\{x_i=0, i<l;x_l\geq 0\},$$
 the orbit set of the group $O_l$,
 where  $\sigma $ has the following  block-diagonal form [see
(\ref{ol})]:
\begin{equation}
\label{oll} n_i=(0,\ldots,0,1), \ \ \ \sigma =\pmatrix{ d & & & &
\vspace{-1mm}\cr
  & {\ddots} & & &\vspace{-1mm} \cr
  &  &   d & & \cr
  &  &  & e & B \cr
  &  &  & C & A \cr}
\end{equation}
(i.e., $d=1$, $\sigma \in 1^{l-1}\times SO_{m-l+1}$).
 On the
boundary $R^{m-l}=\{x_i=0\}$ (the set of stationary points)
 we have
$\sigma \in 1^{l}\oplus SO_{m-l}$
 -- these are matrices commuting
with this symmetry, $O_{l}$ ($B,C$ are uneven
 (vector)functions of
`radius', vanishing at $r=0$,
 see (\ref{oll}); $r^{2}=x_ix_i$).
Again, taking into account that  $\sigma (x)$ is localized, we
have the next diad  mapping:
\[(D^{m-l+1};S^{m-l}) \to  (SO_{m-l+1}; SO_{m-l}),\]
\[\mbox{ \ i.e., \ }
\Pi(O_{l})=\pi_{m-l+1}(SO_{m-l+1};SO_{m-l})=\pi_{m-l+1}(S^{m-l}).
\]
 If $l>3$, there is the equality  (isomorphism) $\Pi (O_l)=\Pi
(SO_{l})$, but for  $l=3$ or $l=2$ one should add to (\ref{ol})
the terms $e^*\varepsilon_{ijk}n_k $ or
$B^*(C^*)\varepsilon_{ij}n_j$, $ e^*\varepsilon_{ij}$,
respectively. Then, instead of  (\ref{oll}), we will have
(respectively for  $l=2$ or $l=3$)
\begin{equation}
\label{o23}
  \sigma =\pmatrix{
    d  &e^* &B^*\cr
  {-}e^* & e  & B \cr
   C^* & C  & A \cr}
\mbox{ \ or \ \ } \sigma =\pmatrix{
    d  &e^* & & \cr
   {-}e^*&  d & & \vspace{-1mm} \cr
       &    & e & B \cr
       &    & C & A \cr}.
\end{equation}
Taking into consideration that all terms with (single) $n_i$ are
uneven, we obtain the result:
\begin{equation}
\label{so3} \Pi(SO_{3}) = \pi_{m-2}(SO_{2}\times
SO_{m-2};SO_{m-3})= \pi_{m-2}(S^1\times S^{m-3}),
\end{equation}
\begin{equation}
\label{so2} \Pi(SO_{2}) = \pi_{m-1}(SO_{m};SO_{m-2}\times SO_{2})=
\pi_{m-1}(RG_+(m,2)).
\end{equation}

`Many-component' symmetries, looking  like
 $SO_{l1}\times\cdots\times SO_{l{\rm k}}$
 leads to
$(k+1)$-ad
 homotopy groups:
 every component $SO_{l{\rm a}}$
  has its own half-space of its orbits;
 their intersection forms the base of
${(}k{+}1{)}$-ad mapping (suspension cube $I^{k}$), which is
defined by joining up the conditions like  (\ref{oll}),
(\ref{o23}). For example,
\[\Pi(SO_2^{(1)}\times\cdots\times SO_2^{(k)})=
\pi_{m-k}(SO_{m};A_1,\ldots,A_k),\]
 where $A_i=SO_{m-2}^{(i)}\times
SO_{2}^{(i)}$ -- the subset of rotations commuting with
 the component $SO_{2}^{(i)}$
(i.e., with rotation of axes  $2i-1, 2i$; $2k\leq m$).
\newpage
\section*{3.4 \ Dimension
$1+4$; SO$_2$-symmetrical quasi-charges }
 In this section, we fix dimension
 $m=4$ and use the spherically-symmetrical
 cosmological model sketched in
   \S~2.6. Take a coordinate system  co-moving with
   the wave-guide, such that the coordinates
 $x_a=x_1,x_2,x_3$ go (tangentially) alone the wave-guide.
 The radial (the extra) coordinate $x_{0}$ has the origin in the
`center' of wave-guide, where $x_{0}=0$; the time coordinate is
marked in this section as   $x_{\emptyset}$ (or $t$). In `our'
 reference frame, the wave-guide's thickness, $L=\gamma \lambda $,
 is large enough ($L> 10^{10}\,$cm), but still much smaller
 than the wave-guide's radius ($R
>10^{28}\,$cm). It seems that the extra dimension  $x_{0}$ has no
 specific scale other than  $L$:
 `thermalized' waves (cosmological noise, 'zero-point
 oscillations') should move almost tangentially to the wave-guide
 to be trapped in it.%
\footnote{Residual escaping of noise waves might be useful
 (the horizon problem) at intermediate stages of expansion,
 serving as a kind of stochastic cooling, or rather smoothing
 of noise `temperature'.}
  As concerns coordinates $x_a$,
just the contrary, alongside with the large scale $R$, there
exists a small  typical scale $\lambda_{0}$ -- the spectrum edge
  (or temperature) of noise ($\lambda_0 <10^{-17}\,$cm); this scale
  can define the size (along the usual dimensions)
  of particle-like field configurations
  carrying a topological (quasi)charge.

 At large expansion times, the waveguide is scarcely
 distinguishable  (if
$x_a\ll R$) from a plane wave.
 One can assume that, along with the
 $O_{3}$-symmetry acting on $x_a$,
  there is an additional waveguide
symmetry, $P\{0\}$, which inverts the coordinate $x_{0}$ (time
reversion $T=P\{\emptyset\}$ is also an approximate symmetry).
 So, the highest possible symmetry of the waveguide reads
 \[Sym_0=O_3+P_1=O\{1,2,3\}+P\{0\}\subset O(4). \]

In the first place, let us consider symmetries which
 have continuous subgroup, and conform to the symmetry of
 waveguide:  $Sym\subset Sym_{0}$.

 Quasi-charge groups
  $\Pi(Sym)$ for several symmetries are already known to us (see
(\ref{so3}), (\ref{so2}); $RG_{+}(4,2){=}S^{2}{\times}S^{2}$,
\cite{fuks}):
\[ \Pi(1)=Z_2^2,\ \ \Pi(SO_{3})=\Pi(O_3)=0,
\ \  \Pi(SO_{2})=Z^2;\]
 this means that spherically symmetric
(along the usual coordinates $x_a$) quasi-solitons do not
 exist,\footnote{* Perhaps this means  that spin zero elementary
 particles are absent in the theory.}
 quasi-solitons of cylindrical symmetry are possible.

 The following will require the quaternion representation of
 group $SO_{4}$ \cite{dubr}, as well as some
 elementary information about quaternions
    \cite{fuks,pon}. We will use different ways to denote a
    quaternion $\rm x$\/:
\[ {\rm x}=x_{0}+x_1\mbox{i}+x_2\mbox{j}+x_3\mbox{k}=
x_{0}+x_a\mbox{i}{}_a=x_p\mbox{i}{}_p; \] i${}_{0}=1$.
 The conjugate quaternion is
 $\mbox{\={x}}=x_{0} - x_a\mbox{i}{}_a$ (there are the real part,
  $x_{0}$, and the imaginary part);
$|\mbox{x}|=(\mbox{x}\mbox{\=x})^{1/2}$ is quaternion length
(module). The set of unit quaternions   (which comply with
x${}^{-1}=\mbox{\=x}$) forms sphere $S^{3}$, whereas the set of
unit imaginary quaternions forms sphere $S^{2}$.

Matrix $\sigma \in SO_{4}$ can be bring in correspondence with
 a pair of unit quaternions, (f,\/g), following the next rule
\cite{dubr}:
\begin{equation}
\label{sfg} x^*_p=\sigma _{pq}x_{q} \leftrightarrow
\mbox{x}{}^{*}=\mbox{(f,\/g)\/x}=\mbox{f\/x\/g}{}^{-1};
\end{equation}
 Values $|$x$|$ and $|$x${}^{*}|$
 are the same. Indeed this is a representation, because
\[ \mbox{ (f\/,g)(\~f,\/\~g)\/x} = \mbox{(f\/\~f,\/g\/\~g)\/x}\,,\]
 and two-valued one, because the pairs (f,\/g) and (--f,\/--g)
act equally (they are equivalent), that is,
\begin{equation}
\label{slr} SO_{4}=S^{3}_{l}\times S^{3}_{r}/\pm.
\end{equation}
We distinguish these two spheres as  {\it left\/} and {\it
right\/} not only because the first quaternion in  (\ref{sfg})
 acts from the left, while the other -- from the right,
 but also because the inversion of one coordinate
  ($x_{0}$; the orientation change) changes the order of
  the elements of pair (f,\/g).

Pairs (f,\/f) do not change the real part of a quaternion x, see
(\ref{sfg}), that is, they correspond, to group $SO_{3}\subset
Sym_0$. In this quaternion representation,  angle $\phi $
rotations of coordinates \ 1) $x_2,x_3$, \ 2) $x_{0},x_1$ have,
respectively, the following form
 (they commute with each other):
\begin{equation}
\label{aab} 1) \ \ (\mbox{a,\/a})(\phi ),\ \ 2) \ \
(\mbox{a,\/a}{}^{-1})(\phi) ; \ \ \mbox{a}(\phi )=\cos\frac\phi
2+\mbox{i}\,\sin\frac\phi 2.
\end{equation}
 Inversion $p_{0}$ of coordinate $x_{0}$ does not relates
 to any pair because
 $p_{0}\not\in SO_{4}$.
 However, the transformation
   $\sigma \mapsto
p_{0}\sigma p_{0}^{-1}$
 keeps $\sigma $ in $SO_{4}$, and it corresponds to the
 permutation of  elements in the pair,  (f,\/g).
For example, pair (i,\/1) (it relates to the combination of
rotations  (\ref{aab}) with $\phi =\pi /2$) transforms into the
dual`mirror' pair  (1,\/i) (the rotation of axes $x_{0},x_1$
changes its sign).
 Inversion of another coordinate reduces to
 inversion  $p_{0}$ plus some rotation.

Therefore, the map $\sigma (x)$ can be substituted by [or divided
on, see (\ref{slr})] the next two maps:
 \[ \mbox{f}:\ R^4\to S^3_l,
\ \ \mbox{g}:\ R^4\to S^3_r;\ \ \mbox{f,\/g}(\infty)=1. \]
 After replacements $\sigma
\to $(f,\/g)$, x\to$x, the symmetry condition,  (\ref{gsi}), also
splits into two halves:
\begin{equation}
\label{fab}
 \mbox{f\/(a\/x\/b}{}^{-1})=\mbox{a\/f(x)\/a}{}^{-1},
\ \ \mbox{g\/(a\/x\/b}{}^{-1})=\mbox{b\/g(x)\/b}{}^{-1} \ \forall
\mbox{(a,\/b)\/} \in Sym \subset SO_{4}.
\end{equation}
Groups $\Pi (Sym)$ for $Sym\subset SO_4$  are split into two
(equal) parts, and at addition of inversion $P_1$ or $P_3$ (three
coordinates inverted), quasi-charge group $\Pi $ reduces in half
(two parts merge):
\[ \Pi(Sym)=\Pi_l(Sym)+\Pi_r(Sym),
\ \ \Pi(Sym+P_1)=\frac12\Pi(Sym) \ \ (1\mapsto 1_l+1_r);\]
 map g is defined through  f; for example, for symmetry
$P_1=P\{0\}$, (\ref{gsi}) leads to the simple relation:
 g$(x)=\mbox{f}(p_{0}x)$, or
 g\/(x)$=$f(--\=x).

Let us consider in more details the `structure' of a non-trivial
(left) map  f\/$\in 1_l\in \Pi _l(1)$. Fig.~2 shows a sphere of
unit quaternions with two antipodal, marked points, $1$ and $-1$,
${\rm f}{}^{-1}(1)=\infty$.
 If preimage set $M={\rm f}{}^{-1}(-1)$
is empty, than, making a pin-hole in the sphere in the point $-1$
and shrinking the sphere (along itself) to the antipodal point 1,
we obtain a homotopy which leads
 map f to the trivial map, f${}^{*}=1$. Therefore,
 the first requirement is just that  $M$ is not empty
  (one can take any other point of the sphere, but
  $M$ is a kind of the `center of a topological quantum').
 \\[11mm]
 \includegraphics[bb=0 0 155 182,scale=0.8]{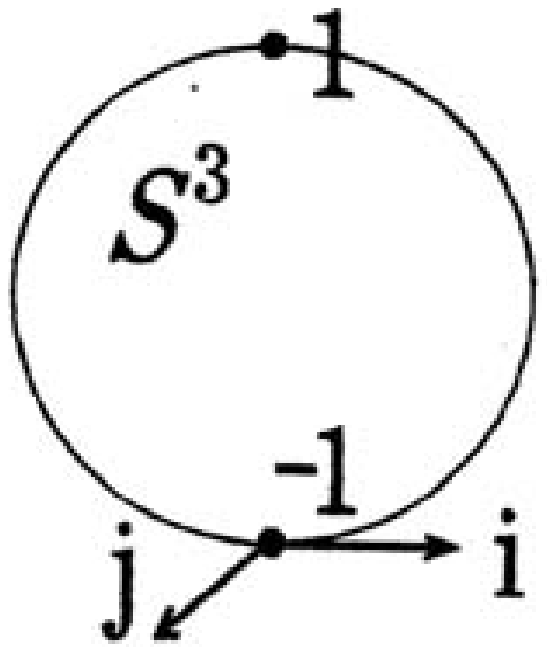} %
 \hspace{4.5cm}  
  \includegraphics[bb=0 0 206 138,scale=0.8]{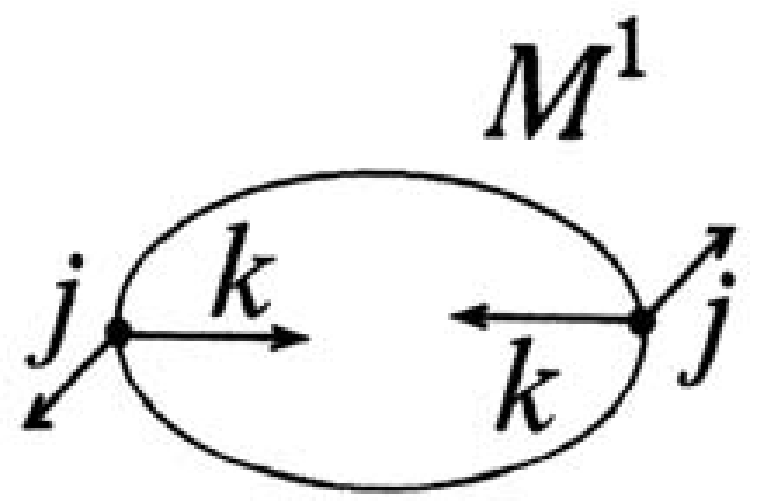}
 \\[-1mm]

 \noindent
Fig.\,2. The $S^3$-sphere of
 unit\hspace{4.5cm}Fig.\,3. A framed\\
\phantom{Fig.\,2. }quaternions\hspace{6.4cm}
1-manifold in $R^{4}$\\[3mm]

\noindent For any smooth map,  f, of {\em general position}, point
$-1$ is {\it regular\/} \cite{dubr,pon}: the tangent map f${}_{*}$
has (maximum possible) rank 3, i.e., any tangent vector  n in this
point, $-1\in S^3$, has, in every point
 $x\in M$, a vector-preimage  $n(x)$ orthogonal to
 $M$; at that, $M$ is a smooth one-dimensional
manifold;  Fig.\,2 and Fig.\,3 show not all (tangent or
orthogonal, respectively) vectors.

The connection between (smooth) maps and  {\it framed manifolds\/}
is reversible \cite{dubr},  and the notion of homotopic
equivalence can be transferred in a sense on the framed manifolds.
 Say, one can deform $M$, joining
disconnected parts, to a circle $S^{1}$. Moreover, the set of
vectors  $i_{a}$ ($i,j,k$) can be orthogonalized (using some
method  \cite{pon}), such that $i_{a}i_{b}=\delta_{ab}$. In this
case a point $y=x+\tau n$, where $x\in M$, $\tau\ll1$, maps on
$S^{3}$ to the point $-1 +\tau (ni_a)\,\mbox{i}{}_a
+O(\tau^{2})$.

A framed circle corresponds to a map  $S^1\to SO_3$, which defines
how the vectors of the orthogonal framing rotate relative to some
fixed framing (for example, the {\it trivial\/} one, where one
vector holds in the circle plane, looking along the radius, while
two other translate in parallel  during  circle tracing
\cite{dubr}); it follows from this that
\[\Pi_l(1)=\frac12\Pi(1)=\pi_{1}(SO_{3})=Z_2\ (=\pi_{4}(S^{3})).\]

A non-trivial framing can look as follows: one vector can
translate trivially, in parallel, while two others rotate (during
the circle's tracing) on  $2\pi$ (or on an uneven number of
complete revolutions). Therefore, the second condition for f (or
[f]) to be non-trivial is imposed on the map's differential (in
the image point f$=-1$).

 In the Faddeev model, the topological charge relates with
 the mapping of the `center of a quantum' (framed circle) to the
 orientable (`single-valued') manifold,
  $S^{1}\to SO_{2}$ \cite{ryba}; but in our case, the
  `soliton's center' maps to the non-orientable
 (`double-valued' in a sense) space $SO_{3}=S^{3}/\pm$;
 perhaps, in order to correctly write the topological current, one
 should introduce auxiliary, double-valued (spinor) fields.

In the case of axisymmetric maps f\/($x$), where, for
definiteness, $SO_{2}=SO\{1,2\}$, which is the  group of rotations
of coordinates $x_1,x_2$, eq-n (\ref{fab}) gives the following
condition:
\begin{equation}
\label{kk}
\mbox{f\/(a\/x\/a${}^{-1}$)}=\mbox{a\/f\/(x)\/a}{}^{-1}, \mbox{ \
\ where }\ \mbox{a}(\varphi)= \cos\frac\varphi
2+\mbox{k}\,\sin\frac\varphi 2.
\end{equation}
At that, manifold $M^{1}$ (if only it does not lie in the plane
$\{x_1=x_2=0\}$, the set of stationary points)
 is a circle intercepting in the semi-space
 \[R^{3}_{+}=\{x_2=0,\ x_1\geq0\} \]
  (it is the set of orbits of
this  $SO_{2}$-symmetry) one point. The condition (\ref{kk}),
translated on the map's differential, defines the framing: vector
$k$ translates trivially  (`trivial' vector), while vectors $i,j$
transform into rotated combinations ($i\to i\cos\varphi +
j\sin\varphi $), i.e., rotate on $2\pi$  during the circle tracing
  (the direction of rotation changes under the change of the
  orientation of pair  $i,j$, that is, under quasi-charge sign
  reversal, see Figs.\,3,4).

So, a framed circle of non-trivial class $1\in \Pi_l(1)$ admits
symmetry $SO_{2}$, i.e., there is the epimorphism,
 \[ \Pi _l(SO_2)=Z \stackrel{e}{\to}\Pi _{l}(1)=Z_2\
 \ (1\mapsto 1).\]
 Cylindrically symmetric `quasi-particle', in addition to a
  topological quasi-charge ($\in Z$; it would be prettily to
  try to relate it with the lepton charge),
  has the topological charge ($\in Z_2$; `left fermionity').

The already discussed equations of frame field has no dimensional
constants  (as distinct from, say, the Scyrme  model), which could
define the typical size, a stable `radius' of $SO_{2}-$soliton.
However, if a quasi-soliton is contained in a  `reservoir with
noise' (i.e., in our cosmological waveguide), there is a scale
$\lambda_0$, at which the expansion of a (quasi)soliton can stop
(or at least slow down).

\section*{3.5\ Phenomenology of quasi-solitons
 (topological quanta);\\ \hspace*{9mm}
secondary (proxy) fields and 'flavors'}

In this section we are enforced to resort to qualitative
descriptions and rough wordings; all `coincidences' of
quasi-solitons' names with the names of the observed elementary
particles have to be considered quite casual (or adequately
 conditional).
 The overall problem here is to outline a possible
 phenomenological description of topological quasi-solitons
 (their creation, evolution, and annihilation) in a `noise enough'
 (and quite degenerate along the extra dimension) environment
 of cosmological wave-guide; its the most symmetrical state, free
 of topological quasi-solitons, is a kind of `vacuum state'.

 We will proceed from a few assumptions and requirements in
 relation to this vacuum state; the main of them is the assumption
 that the vacuum state is energetically stable -- with respect to
 creation of topological (soliton) excitations; that is, the
 presence of topological quantum (quasi-soliton) in some region
  $U$ should increase the energy density in that region.

 It is assumed also that the noise is non-linear and, in average,
 of high symmetry; that is, the noise amplitude,
  $a_0$, is large so much that, at the scale $\lambda _0$, there
  are frame field fluctuations of order unit, which bring to
  creation of soliton--antisoliton pairs; at the same time, the
  amplitude of $f-$component of noise, which carry $D-$momentum,
  can be small, $a_f\ll a_0$ (for more precise conservation of
  energy-momentum).\footnote{*
   There is a very interesting phenomenon
  in this theory, the linear instability of the trivial solution:
  $f-$components cause linear growth of some polarizations, which
  do not contribute to $D-$momentum and angular momentum, i.e.,
  intangible in a sense; at that, the components of Riemann
  tensor, gravitational polarizations,
  do not grow (although intangible too); see
  \href{http://arxiv.org/abs/0812.1344}{arXiv: 0812.1344}.}

 It is not clear at what extent one can apply thermodynamic
 considerations to this vacuum state, with ensemble of `thermalized
 waves'; in any case, some scaling features in the spectrum of
 `steady-state thermalized noise', and its high symmetry,
 homogeneity and isotropy (in the natural coordinate system, where
 the waveguide itself is isotropic), seem to be quite possible.

 A tremendous number of noise harmonics enforce us to repudiate,
 to give up any attempts to keep track of an exact solution, of
 all frame field's fluctuations.
 Perhaps it is more important to keep an eye on the topological
 excitations of the `vacuum', quasi-solitons, which carry
 long-lived discrete information -- topological (quasi)charges.
 The natural `calling card' of a soliton is its framed manifold
 (the center of topological soliton, quantum); however, in
 the environment of non-linear noise, it is some inconvenient,
 rapidly changing an object.
  One can, however, make an averaging
over the scale $\lambda _{\star}$,  $\lambda _0\leq \lambda
 _\star\ll L$; consequently, one will have a more stable object,
 a `thickened framed manifold', which is very extended along the
 extra dimension($\sim L$; there are no other
 typical scales, excepting  $L$, on this
 dimension), and of size $\sim \lambda _0$ along the usual
 spatial dimensions; it is natural to expect that the averaged
 configuration (of quasi-soliton) is still symmetric,
 cylindrically symmetric in usual dimensions.

 This average object is a kind of a `tread with arrows'
  (*that not
 to say `string'), where the
 `arrow' is the overall name for the next set of
 parameters, which  define an
   $SO_{2}$-symmetrical configuration:\\
1) direction of the symmetry axis --
 $\,RP^{2}\, \ [=RG(3,2)\,]$; \\
2) `direction' of the framing -- $\,O_{3}$, see Fig.~4.
 \\[5mm]

\noindent
 \includegraphics[bb=0 0 182 178,scale=0.85]{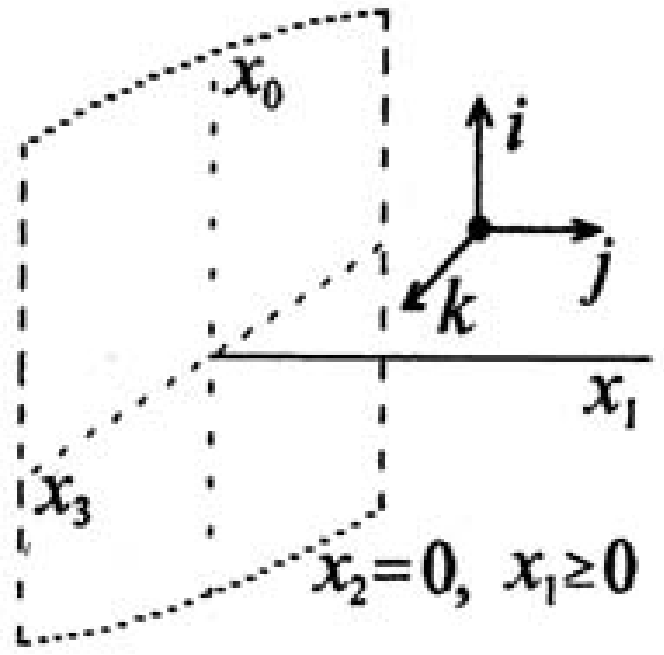}
 \hspace{3.5cm}
  \includegraphics[bb=0 0 191 186,scale=0.81]{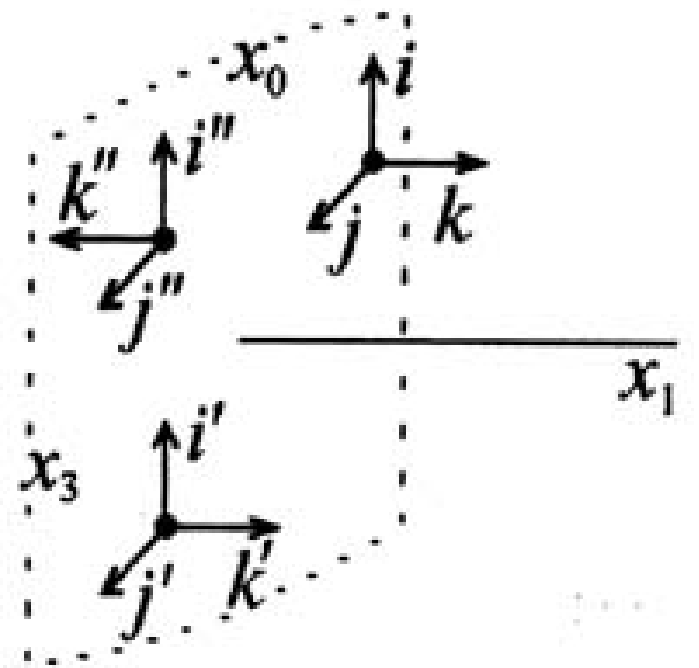}

 \noindent
Fig.\,4. {\normalsize Framed point
 in $R^{3}_{+}$}\hspace{3.55cm}
Fig.\,5. {\normalsize The action of inversion $P\{0,3\}$ (${}'$),\\
(on the set of orbits of symmetry $SO\{1,2\}$). \hspace{0.7cm}
and $P\{1,3\}$ (${}''$) on the framing vectors.}\\[0.5mm]

It may turn out that some of these parameters (which make
something different, such that degeneration is lacking) have
preferred, dedicated (stable or metastable, single or multiple)
values which do not change along the thread. This fixed part of
the `arrow' is to be called a `flavor' (in the case of
  $SO_2$-quanta, we have in mind an analogy with the lepton
  flavors). The variable part of the `arrow', in turn, can be
  divided on a spin part, which transforms at the coordinate
  rotations from  $SO_{3}\subset Sym_0$, and the rest part (if
  it remains), which is to be called
  `color'.\footnote{* The angles between the framing vectors
  are obviously not degenerate: small angles correspond to large
  derivatives; we arbitrarily assume that the most stable
  configurations correspond to orthonormal framing.}

  For the case of axisymmetrical solitons, the flavor set is
  connected, evidently, with the direction of the `trivial' vector
  of the framing -- in the case of symmetry  SO\{1,2\},
  it is vector $k$; Fig.\,4 and Fig.\,5 show examples of
  `flavors'.
 The direction of other two (orthogonal) vectors of framing
 (they rotate at tracing the circle) can differ in `phase';
 it seems that phase change, phase rotation,
  is a possible, natural for this symmetry,
 way of framing evolution (the possibility of
 a stable solution, with a stationary framing, seems not likely).

 It is important also to know
 how the `arrow components' transform at
 other transformations of group $SO_{4}\cap Sym_0$
 (they keep soliton `left'). Fig.\,5
shows how the inversion-rotation $P\{2,3\}$ and $P\{0,3\}$ --
 they change the direction of the third axis, the axis of
 symmetry -- act on $SO\{1,2\}-$soliton.
 The former transformation changes the sign of quasi-charge
 (the orientation of the framed point), while the letter does not.

 Evolution of this `thread with arrows' still has, seemingly,
 stochastic, irreproducible character, and one should seek
 for some even
 less detailed, but more stable, features of quasi-soliton's
 evolution.
 `Degeneracy' of vacuum noise along the extra dimension makes it
 natural and justified to average somehow along this dimension; so
 the question is just how (and why so) to do this averaging:
  should one  sum up `probabilities', or, instead, some
  `amplitudes, i.e., `arrows' themselves.

  If there was a `lantern', shining through the wave-guide along
  the radius $x_0$,
 then, in order to define a soliton's `shadow', one should sum up
 `probabilities'. However, such a lantern is absent, and the
 question is some different: how does a soliton thread scatters,
 disturbs the `electro-magnetic' component of noise? ($f-$waves,
  quite  the contrary, move tangentially, along the wave-guide)
  and how the positive (for energy stability of the vacuum)
  soliton's contribution into the (averaged
  along  $x_0$) energy density can be expressed through soliton's
  parameters. At raising this question, seemingly, one has to sum
  up `arrows',  $a(y_0)$; i.e., to sum up the amplitudes
  of scattered $f-$waves, contributions of different
  parts of a soliton thread;
  the `mean arrow'  turns out to be a secondary,
  proxy, $4D$ chiral field:
\[\psi_{(l)}(x_a)=\frac1L\int\! a_{(l)}(y_0)
\delta(x_a-y_a(y_0))\,dy_0\ \ (a=1,2,3), \]
 where $y_a(x_0)$ defines the thread location;
 at the change of orientation (with inversion $P_1$ or $P_3$),
a left thread becomes right, i.e., $\psi_{(l)}$ realises a
representation of group $SO_{3}$, but not $O_{3}$.

 The quanta's contribution to the energy density should start with
 terms quadratic in the secondary fields (linear terms, evidently,
are not possible due to symmetry considerations). Evolution of the
secondary fields, being a result of averaging along the huge
number of thread's parts (likewise the Feynman path integration
   \cite{fey}), perhaps obeys to some Lagrangian rules, with
   conservation of soliton's energy ($D-$momentum) as a
 Noetherian  consequence. The formalism of second quantization
 could be appropriate for phenomenological description of
 many-soliton excitations (indistinguishability of solitons of the
 same flavor is ensured  just by the chaos of vacuum).

 It is possible that the symmetry of the vacuum state is not
 restricted by rotations only, but there exists an additional
 approximate symmetry, Lorentz covariance along the coordinates
   $x_a$: if in the coordinate system moving, with velocity
 $u^{\mu}$ with regard to the `preferred'
system, the energy-momentum tensor of vacuum (doubly averaged --
over $x_0$ and over the scale $\lambda_\star$) looks like
\[ T^{\mu\nu}_{\star}=Ag^{\mu\nu}+Bu^{\mu}u^{\nu}
 \mbox{ \ and \ }
|B/A|\ll 1.\] In this case, the symmetry of the Lagrangian of
secondary fields should also include Lorentz-covariance.

 The position of soliton's center in the waveguide --
 central or peripheral -- is, perhaps, of `flavor kind';
 it may turn out that the latter is stable. However, the stability
 of the central position can be ensured by the additional
 symmetry,  $P\{0,3\}$; the framed point (flavor) located at the
 axis  $x_1$, see
 Fig.\,5, transforms into itself at this inversion (or rotation).

 If there exist stable quasi-solitons (topological quanta),
 corresponding to extended symmetry groups, they, of course,
 should behave (with their secondary fields) as independent
 channels of `topological excitation' of the vacuum, with their
 own contribution to the energy density and soliton Lagrangian.

In the case of extended symmetry
  $SO\{1,2\}+P\{0,3\}$, one need first to fill up the
quadrant $\{x_{0},x_{2}=0;\, x_{1},x_{3}\geq0$\}, which boundaries
should  obey to condition f\/($x$)\/k$=$k\/f\/($x$); such a
filling  relates to group $\pi_{2}(S^{3};S^{1})=\pi_{2}(S^{2})$.
Next, using $SO\{0,3\}$ rotations, fill half-space $R^{3}_{+}$
(Fig.\,4), obtaining some `reference' mapping f\/${}^{*}(x)$.
Then, the ratio F\/$(x)=$\/f\/$(x)/$\/f\/${}^{*}$ relates to an
element of the group  $\pi_{3}(S^{3};S^{1}_{(\mbox{k})})$:
 one has to fill the region
 $\{x_{2}=0;\, x_{0},x_{1}\geq0\}$, with $F(x_{0}=0)=1$,
$F(x_1=0)\in S^{1}_{(\mbox{k})}$.

 The results of classifications of proper symmetry groups
  ($\subset SO_4$), preserving (complying with) the waveguide
  symmetry, $Sym_0=O_3+P_1$, are brought to the Table 3.
  We use the names of the (Standard model's) elementary particles
  that not to concoct new names for different kinds of
  quasi-solitons.
 The next two circumstances are taken into account:
  reflection $P\{0\}$ is closely connected
   with the charge conjugation;
 solitons carrying the topological charge
   ($\in Z_2$) are fermions while others are
 bosons. The  last, `most symmetrical'
  $\Pi$-groups, and concomitant morphisms, also can be obtained
   through similar considerations: fill the orbit set,
   a quarter, or a half of space  $R_{+}^{3}$ on Fig.\,5
   [corresponding
 tetrad group, $\pi_{3}(S^{3};S^1_{(\mbox{i})}, S^1_{(\mbox{j})},
S^1_{(\mbox{k})})$, and triad group, through usage of exact
consequence (\ref{gpk}), reduce to $\pi_{3}(S^{3};S^{1})$]; then,
using discrete transformation, one can fill all space $R_{+}^{3}$.
The notation $\stackrel{0}{\to}$ is for the zero morphism  (image
consists of zero); notations $\,e,i,m2\,$ denote epi-, iso- and
2-monomorphism ($1\mapsto 2$), correspondingly.
 \vspace{3mm}

  \noindent
 {\bf Table 3.}
 The quasi-charge groups $\Pi_{l}(Sym)$
and their morphisms to the group of `previous' symmetry;
for $Sym\subset SO_{4}\cap Sym_0$.\\[4mm]
\hspace*{\parindent}
\begin{tabular}{c|c|c}
\hline
 $Sym $ &$ \Pi_l\to\Pi^*_l $&$ \mbox{`name-analogy'}$ \\ \hline
$1 $&$ Z_2 $&  \\[1mm] \hline
$SO\{1,2\} $&$ Z_{(e)} \stackrel{e}{\to}Z_2 $&$ e $ \\[1mm] \hline
$SO\{1,2\}+P\{0,3\} $&$ Z_{(\nu)}+Z_{(\gamma)}
\stackrel{i,m2}{\to}Z_{(e)} $&$ \nu^0;\
 \gamma^0\to e+e$ \\[1mm] 
$SO\{1,2\}+P\{2,3\} $&$ Z_{(W)}
\stackrel{0}{\to}Z_{(e)} $&$ W\to e+\nu^{0}$ \\[1mm]
$SO\{1,2\}+P\{0,2\} $&$ Z_{(Z)}
\stackrel{0}{\to}Z_{(e)} $&$ Z^0\to e+e$ \\[1mm]\hline
$SO\{1,2\}+P\{0,3\}+ $&$ Z_{(X)}
\stackrel{0}{\to}Z_{(\gamma)} $&$ X^0\to
\gamma^{0}+\gamma^{0}$ \\[0mm]
$ +P\{2,3\} $&$ \ \ \ %
\stackrel{0}{\to} Z_{(W)}$ &$ \ \ \ \to W+W$\\
 \hline
 \end{tabular}

 \newpage
 It is quite surprising that the Table 3 involves all
`flying out' (i.e., white) particles of the Standard
 Model;\footnote{* At this moment, I would prefer to
 interchange in this table $\gamma^0$ and $X^0$.}
however, there exist also   `confined' elementary particles --
quarks, gluons.

This section is already full of assumptions, so we should not be
afraid to add a few new ones.
 Up to this moment, we have been considering
  the symmetry groups which enter
 into the waveguide's symmetry; now let us discard this
 requirement.
 At a small enough scale  ($<\lambda _0$),
 there is the symmetry  $SO_{4}$ (flat space).

 Quark analogies give rise to the idea that a kind of
 `collective stability' is possible for quasi-solitons relating
 to symmetries which  break the wave-guide symmetry
  (confined,
`collectivist' quasi-solitons;
 *i.e., the hadron bag's background is of great
 `average' symmetry,  $SO_{4}$-like).
 In this connection, one can consider the
   `chiral'  one-parameter
 groups $SO_{2}^{+}$ (left) and $SO_{2}^{-}$ (right), which
 generators are, respectively, the sum and the difference of the
 generators of usual  $SO_{2}$-groups; for example,
\[ SO^{\pm}\{1,2\}: \ \{1,2\}\pm \{0,3\} .\]
 Taking into account (\ref{aab}), (\ref{kk}), one can easily show
 that
\[  SO^{+}\!\{1,2\}=SO^{+}\!\{0,3\}\ni
(\cos\varphi +\mbox{k}\,\sin\varphi ,1); \ \
 SO^{-}\!\{1,2\}\ni
(1,\cos\varphi +\mbox{k}\,\sin\varphi ).\]

 At the orientation change, these groups,
 $(+)$ and $(-)$, interchange.
 The set of stationary points of group  $SO^{+}_2$ consists of
 one point,\footnote{* This is some strange feature. So the other
 possibility is to use usual $SO_2-$groups, assuming that
 superpositions (plus and minus) of dual solitons are the real
 minima, i.e.,  more stable quanta.}
  $x_p=0$, f\/$(0)$\,k${}={}$k\,f\/$(0)$, while
g\/$(0)$  is free of restrictions.
 The set of orbits is a $3D$ manifold, therefore
 \[\Pi_{(l)}(SO_{2}^{\pm})=\Pi_{(r)}(SO_{2}^{\pm})=Z;\]
 perhaps, these quasi-charges can be concerned with
 the baryon charge.

 Let us prove that maps
$\Pi_{(l)}(SO_{2}^{\pm})\to \Pi_{(l)}(1)$ are epimorphism. Choose
 a framed  circle,  the central orbit  f${}^{-1}(-1)$,
 in the plane (1,2) ($x_0,x_3=0$).
Orbits shifted from it along its radius are also situated in this
 plane (and are circle too), and they correspond to the
 `trivial' framing vector. Orbits shifted along  $x_0$ or $x_3$,
 make one rotation around the central orbit, and they are
 concerned with the `phase' vectors (for $(-)$-group,  \ f \
 does not change along an orbit, so the framing
 vectors rotate together with orbits; for $(+)$-group, there is
  the additional rotation of the `phase' vectors
 on two turns).

 In the case of chiral groups, the `arrow' has more parameters
 (both dual planes should no longer lie in `our' space):
\[ RG(4,2)\, \mbox{(choice of 2-plane)} \times O_3\,
\mbox{(framing)};\]
 if, as for `leptons', degeneration is lacking only for two
  parameters (two-dimensional `flavor' space), then there is a
  room for `color'  (a space like $S^{2}$). Similar to
 Table\,3, one can consider extended symmetry groups
 $SO_{2}^{\pm}+P_2$, and so on, and, perhaps, obtain
 `color' bosons.

\newpage
\section*{Conclusion}
\addtocounter{chapter}{1}

Closing this work, it is worth to say a few words about the
 possible themes  for a further research.
 Of course, it would be desirable to obtain some quantitative
 predictions, for example, for spin-dependent post-Newtonian
 effects, to estimate the proportionality factor between
 `mathematical' field, $f_{\mu\nu }$), and
 `physical' electromagnetic field, $F_{\mu \nu }$,%
\footnote{In units  $c{=}1{=}\hbar$, the latter is defined by the
equality of photon's frequency and its `physical' energy
$\int\!F^{2}\,dV\,$.}
 to find a relation between the cosmological waveguide's parameter
  $L=\gamma \lambda $
 (or $\lambda $) and the gravitation
 constant,\footnote{* Motives of modified gravity
 are touched in
 \href{http://ru.arxiv.org/abs/0704.0857}{%
arXiv: 0704.0857} and
 \href{http://arxiv.org/abs/0812.1344}{arXiv:
0812.1344}.}
 to make estimations for parameters of the cosmological
 model.\footnote{* This relativistically expanding model
 (relativistic surfing model) corresponds, in FRW-framework,
 to anti-Milne model, with $a=H_0 t, k=+1$; it describes
 SNe Ia and GRB data quite well (without free parameters excepting
 the Hubble constant, $H_0$; see
 \href{http://arxiv.org/abs/0902.4513}{arXiv: 0902.4513}).}

 Numerical calculations of solutions of system
  (\ref{eq02}), (\ref{eq04}) can give some insights on the
  qualitative behavior of parameters  $R,\lambda ,\gamma ,a$
  at initial stages of expansion ($R/\lambda >1$).
 It is also important, however, to obtain some analytical
 estimations for their time dependence at more long stages, when
    $R/\lambda ,\gamma \gg1$. One can try to expand a solution
    of equations  (\ref{eq04}) as a series in small parameter
 $\lambda /R$.

 The formal test of possibility of co-singularities discussed in
 the second chapter, the checking that the symbol keeps
 being involutive at the degeneration of co-frame, is the
 necessary, but perhaps not sufficient condition; in most cases
 it holds trivially (gas dynamics); so much the more interesting
 was to find the opposite case (the AP equation which solutions
 are free of singularities).

There is the variant of gas dynamics equations
 (weakly nonlinear case) where, at some restriction on
 the {\sl lowest\/} terms of the system, the gradient catastrophe
 is nevertheless absent \cite{rozh};
 however, this is the case of
  {\sl weak\/} co-singularities. But in the case of {\sl
strong\/} co-singularities of AP (the lowest terms diverge at
co-frame degeneration), it seems impossible that some
 restriction imposed on the lowest terms can suppress
 singularities.
 Other kinds of singularities (in solutions
 of general position), which would differ from the considered ones,
 co- and contra-, seemingly are impossible: the choice of
 dependable variables should take into account
 the symmetry of equations
($LD-$covariance).

The test on co-singularities can be easily extended
 \cite{z5} on the equations of
 $R^{2}-$gravity  (forth order equations   $C_a
 D_{\mu\nu}^{(a)}$, see (\ref{d1})--(\ref{d2}), are irregular in
 the case $C_1=0$, which, however, can be found in literature
\cite{r2}).
  The similar analysis of forth (and higher) order equations of AP
  (which would give a well-posed Cauchy problem) is
  much more difficult problem, but still seemingly solvable.

 Yet one other problem is to derive covariant expressions
 of topological charges and quasi-charges, following, for example,
 the approach of \cite{ryba} to the Faddeev model;
 indeed, it would be desirable to find examples of solutions with
 non-trivial (quasi)charge.
 For dimension  $D=4$, the topological current
 (identically conserving, irrespectively of field equations)
 should seemingly reduce, at zero Riemannian curvature
($R_{ab\mu\nu}=0$), to the next expression (looking like the
 current of the Scyrme  model
  \cite{ryba}):\footnote{* This problem is  solved for
  $4D$-case; the $5D$-case is  more difficult; see
  \href{http://www.arXiv.org/gr-qc/0610076}{arXiv:
  gr-qc/0610076}.}
\[ J^{\mu}\sim \varepsilon^{\mu\nu\lambda\tau}
\gamma_{ab\nu}\gamma_{bc\lambda}\gamma_{ca\tau}, \] see
(\ref{gala}); perhaps, the exact expression for this current
should also
 include terms like  $R\gamma$.

The large scope of not so easy questions relates to the problem of
stability of both cosmological solution and particle-like
solutions with topological (quasi)charge -- in the `real
environment' of cosmological wave-guide filled with noise.

 Notwithstanding the large number of unclear questions and
 unsolved problems, it seems that the
 best variant of absolute parallelism is a
 very beautiful theory, full of great possibilities and plentiful
 pictorial means.
 The notion of `beauty' seems to be scarcely useful for a
 mathematical formalization; however, there is an opinion that
 a  {\it beautiful\/} theory should be  {\it  simple.}
 At this point some mathematical allusions become possible:
 for example, the set of real numbers is divided by mathematicians
 on {\it
complex}, or {\it random} numbers (composing overwhelming
majority) and {\it simple\/} numbers (rational numbers, numbers
like $\pi$, and so on), which can be exhaustively described by a
finite information, finite algorithm \cite{chir}.

 Analogously,  a physical theory can be called {\it simple\/} (or
{\it nonrandom}), if it is defined by a finite information, i.e.,
by a simple algorithm  (may be formulated exhaustively on a finite
number of pages, and so does not contain  {\it random\/} numbers
-- free parameters, or constants).
 In regard to solutions of a (physical) theory, just opposite, it
 would be desirable to have as more complexity as possible
(equations should be simple and beautiful, while their solutions
are intricate and interesting).

In this connection, let us add a few words on the choice of the
global topological class, as well as the topology of solution, or
topology of the Cauchy surface, using only quite qualitative
reasoning, without attempt at a more exact wording.
 Let us say a solution is {\it uninteresting\/} (or
vanilla), if it can be defined using a finite algorithm; for
example, the trivial solution. Let us call a solution
  {\it rather uninteresting} if, some finite time later, it
  approaches closer and closer to some
  {\it uninteresting\/} solution, which, hence, is a kind of
 {\it partly attracting} (time reversal gives a solution which
 is keeping away). A possible example is an expanding wave,
 of decreasing amplitude, which approaches the
 trivial solution.
 One can surmise that the presence of
 a {\it uninteresting, partly attracting\/} solution
 in a homotopy class
 means that all solutions in this class are rather uninteresting.
  On the other hand, the presence of rather uninteresting
  solutions of general position is quite sufficient reason that to
  reject this homotopy class of solutions.
  So, there is the problem how to reveal  such
 uninteresting (but simple and beautiful) solutions, and to
 find an `interesting' homotopy class where such solutions are
 absent (perhaps, in the same way one can also pick out
 a case of  {\it
 interesting\/} topology of Cauchy surface which should be
 a parallelizable manifold).

The variant of AP, which was discussed in  sections 3.4, 3.5
(trivial space topology, $D=5$, $\Pi =Z_{2}^{2}$), has three
 nontrivial homotopy classes of solutions;
 one of them, class  $1_l{+}1_r$, admits greater symmetry of
 solutions, i.e., perhaps, has more chances to be uninteresting.
 The choice between classes $1_l$ and
 $1_r$ (left and right) does not matter,
 because they change one another at the orientation change
 (on the Cauchy surface).

 The choice of positive time direction (the arrow of time)
 can be related to the cosmological expansion
 (of the `initial island state', of class  $1_l$),
 to the process of creation of
`discrete information' -- from infinite reservoir
 of `continuous information' (contained in the coefficients of
 Taylor or Fourier series), to that direction of this game
 of scales and symmetries, chaos and stability, where creation
 of new topological quasi-solitons (quanta) takes place.

 Note that there is one possible peculiarity of cosmological
 solution originating from the `left' initial state. At the stage
 of forming the sphere-like waveguide with weak noise waves
 trapped inside, the `left' component of `zero-point
 oscillations', chaos waves, could be dominating over the other,
 `right' component. In this case the `symmetry of noise' (averaged
 over a scale  $\sim
\lambda_0$) is not $O_{3}$, but $SO_{3}$, and interactions of left
and right quanta can be sufficiently different as well.


\end{document}